\documentclass[ a4paper,twocolumn,superscriptaddress,10pt, allowfontchageintitle, accepted=2017-09-04]{quantumarticle}
\pdfoutput=1
\usepackage[utf8]{inputenc}
\usepackage[english]{babel}
\usepackage[T1]{fontenc}
\usepackage{bbm, amsmath, amssymb, amsthm, bm,textcomp, nicefrac,geometry}
\usepackage{graphicx,epstopdf,verbatim,enumitem}
\usepackage[dvipsnames]{xcolor}
\usepackage{dsfont}

\usepackage[unicode=true,pdfusetitle, bookmarks=false,bookmarksnumbered=false,bookmarksopen=false, breaklinks=false,backref=false]{hyperref}

\usepackage[numbers,sort&compress]{natbib}

\DeclareGraphicsExtensions{.png,.pdf,.eps}
\graphicspath{ {./figs/} }

\newcommand{\eins}{\mathbbm{1}}

\newcommand{\ket}[1]{\left|#1\right\rangle}
\newcommand{\bra}[1]{\left\langle #1\right|}
\newcommand{\bracket}[2]{\left\langle #1|#2\right\rangle}
\newcommand\defn[1]{\emph{#1}}

\newcommand\ketbra[2]{|#1\rangle\langle#2|}
\newcommand{\proj}[1]{\ket{#1}\!\bra{#1}}
\newcommand\cH{{\mathcal H}}
\newcommand\cI{{\mathcal I}}
\newcommand\cB{{\mathcal B}}
\newcommand\cE{{\mathcal E}}

\newcommand\cF{{\mathcal F}}
\newcommand\cL{{\mathcal L}}

\newcommand\bigO{{O}}

\newcommand{\Vs}{{\bm \varsigma}}
\newcommand{\vs}{{\bm \sigma}}
\newcommand{\im}{\mathrm{Im}}
\newcommand{\re}{\mathrm{Re}}

\renewcommand{\t}[1]{\textrm{#1}}

\newcommand{\m}[1]{\mathsf{#1}}
\renewcommand{\v}[1]{{\bf #1}}
\newcommand{\Cos}[1]{\mathrm{c}_{#1}}
\newcommand{\Sin}[1]{\mathrm{s}_{#1}}
\newcommand{\aMSE}{$\textrm{AvMSE}\,$}

\renewcommand\d{\mathrm{d}}
\newcommand\e{\mathrm{e}}
\newcommand\ii{\mathrm{i}}
\newcommand\T{T}

\newcommand{\be}{\begin{equation}}
\newcommand{\ee}{\end{equation}}
\newcommand{\bea}{\begin{eqnarray}}
\newcommand{\eea}{\end{eqnarray}}
\newcommand{\al}[1]{\begin{align} #1 \end{align}}

\newcommand{\eref}[1]{\eqref{#1}}
\newcommand{\eqnref}[1]{Eq.~\eqref{#1}}
\newcommand{\eqnsref}[2]{Eqs.~\eqref{#1} and \eqref{#2}}
\newcommand{\figref}[1]{Fig.~\ref{#1}}

\newcommand{\secref}[1]{Sec.~\ref{#1}}
\newcommand{\appref}[1]{App.~\ref{#1}}

\newcommand{\refcite}[1]{Ref.~\cite{#1}}

\newcommand{\obsref}[1]{Res.~\ref{#1}}

\def\sone{\sigma_1}
\def\stwo{\sigma_2}
\def\sthree{\sigma_3}

\def\tr{\mathrm{tr}}
\def\QFI{\mathcal{F}}

\newtheorem{obs}{Result}

\begin{document}

\title{
{\fontsize{22}{22}\selectfont
Quantum metrology with full and fast quantum control 
}
}

\author{Pavel~Sekatski}
\affiliation{Institut f\"ur Theoretische Physik, Universit\"at Innsbruck, Technikerstr. 21a, A-6020 Innsbruck,  Austria}
\author{Michalis~Skotiniotis}
\orcid{0000-0001-6935-7460}
\affiliation{Institut f\"ur Theoretische Physik, Universit\"at Innsbruck, Technikerstr. 21a, A-6020 Innsbruck,  Austria}
\affiliation{F\'isica Te\`orica:~Informaci\'o i Fen\`omens Qu\`antics, Departament de F\'isica, Universitat Aut\`onoma de Barcelona, 08193 Bellatera (Barcelona) Spain}
\author{Jan~Ko{\l}ody{\'n}ski}
\orcid{0000-0001-8211-0016}
\affiliation{ICFO-Institut de Ciencies Fotoniques, The Barcelona Institute of Science and Technology, 08860 Castelldefels 
(Barcelona), Spain}
\author{Wolfgang~D\"ur}
\affiliation{Institut f\"ur Theoretische Physik, Universit\"at Innsbruck, Technikerstr. 21a, A-6020 Innsbruck,  Austria}

\date{\today}

\begin{abstract}
We establish general limits on how precise a parameter, e.g., frequency or the strength of a magnetic field,  can be estimated with the aid of \emph{full and fast quantum control}. We consider uncorrelated noisy evolutions of $N$ qubits and show that fast control allows to fully restore 
the Heisenberg scaling ($\sim\!1/N^2$) for all rank-one Pauli noise except dephasing.  
For all other types of noise the \emph{asymptotic} quantum enhancement is unavoidably limited to a constant-factor 
improvement over the standard quantum limit ($\sim\!1/N$) even when allowing for the full power of fast control. The latter holds both in 
the single-shot and infinitely-many repetitions scenarios.  However, even in this case allowing for fast quantum control helps to improve the asymptotic constant factor.
Furthermore, for frequency estimation with \emph{finite} resource we show how a parallel scheme utilizing any fixed number of entangled qubits but no fast quantum control can be outperformed by  a simple, easily implementable,  sequential scheme which only requires entanglement between one sensing and one auxiliary qubit.  
\end{abstract}

\maketitle

\section{Introduction}
\label{intro}

Precision measurements play a fundamental role in physics and beyond, as they constitute the main ingredient for many
state-of-the-art applications and experiments~\cite{Dowling2015}. When investigating the limits of known or speculative theories
experimentally one quickly enters into regimes where quantities and parameters need to be measured with unprecedented
precision. In this context, it is of utmost importance to know the ultimate limits nature sets on how precise any given quantity 
can be determined and how to achieve this.

These questions are at the focus of quantum metrology~\cite{Giovannetti2011}. Quantum mechanics is a
probabilistic theory and the intrinsically stochastic nature of measurements ultimately limits the achievable precision. 
When considering classical probes (or particles) independently sensing a physical parameter, such as phase or frequency,  the maximum attainable precision (as quantified by the Mean Squared Error---MSE) follows the 
\emph{standard scaling}, $1/N$, where $N$ is the number of probes~\cite{GLM04}. 
In turn it was shown that quantum entanglement allows one to achieve the so-called \emph{Heisenberg scaling} (HS) in precision, 
$1/N^2$,  a quadratic improvement as compared to classical approaches~\cite{Buzek1999, Giovannetti2006}.
These precision limits apply to both single-shot 
protocols~\cite{Sanders95,Berry00, Peres2001, Bagan2004, Chiribella2004a,Chiribella2004b,Chiribella2005, Higgins2007} 
as well as protocols utilizing many repetitions~\cite{Yurke86, GLM04, Giovannetti2006}.  Still, it remains unclear to what extent 
such an improvement can be harnessed in practice under non-idealized conditions, i.e., when taking unavoidable noise and 
imperfections into account~\cite{Huelga:97,Banaszek2009, Maccone2011}.

Due to the difficulty of obtaining exact precision limits in the presence of noise, several asymptotic lower bounds have been 
established for particular noise models~\cite{Fujiwara:08, Escher:11, Escher:12, Kolodynski:12,Kolodynski:13, Al14, Kn11, Knysh:14}.  
These lower bounds are often not only cumbersome to evaluate but also hard to optimize, relying on educated guesses,
numerical methods employing semi-definite programming or a combination of the two. Nevertheless, these bounds show that for typical 
uncorrelated noise processes the possible gain due to the usage of quantum resources is limited to a 
constant-factor improvement over the standard scaling, as opposed to a different scaling.  On the other hand, it was shown that for some 
types of noise---namely noise perpendicular to the Hamiltonian that encodes the parameter of interest---the restriction to standard scaling 
can be circumvented, and HS can be fully restored, by allowing for additional resources, such as   
perfectly protected auxiliary particles used to perform quantum error 
correction~\cite{Preskill2000,Dur:14, Kessler:14, Arrad:14, Ozeri:13,Lu2015, Herrera:2015}.

In this work we develop a general framework that provides us with analytic results for all types of noise processes 
described by a time-homogeneous master equation. In order to establish general limits, we account for the possibility of 
ancillary resources that do not take part in the sensing process together with \emph{full and fast quantum control} (FFQC) of 
the system and the ancillae\footnote{We remark that we do not allow for additional control over the environment as has been 
recently considered in~\cite{Gefen:15, Plenio:15}.}.
Such FFQC allows one to effectively modify the Lindblad superoperator that describes the noise process, a possibility which has 
hitherto not been considered\footnote{To the best of our knowledge, quantum control has been used to modify 
the overall system-environment dynamics but not the dissipative evolution itself~\cite{Arrad:14, SSD:15}.}. Our approach allows 
us to identify all types of qubit noise that can be fully corrected, and thus restore HS in precision.  
In addition, for all other types of uncorrelated noise we are able to provide fully analytic bounds applicable to the entire 
hierarchy of metrology schemes. In contrast to the available methods \cite{Escher:11,Kolodynski:12}, our bounds 
can be generally determined without need to explicitly solve the system dynamics.

Our main results can be summarized as follows:
\begin{itemize}[topsep=3pt,parsep=3pt,leftmargin=19pt]
\item[(i)] FFQC allows to restore HS by completely eliminating any Pauli rank-one noise (that is not parallel to the Hamiltonian) 
at the cost of slowing down the unitary evolution by a constant factor;
\item[(ii)] All other noise processes unavoidably limit the quantum gain to a constant factor improvement over standard scaling
despite full quantum control. We obtain analytic bounds for the achievable improvement factor.
\item[(iii)] For standard scaling-limited noise processes FFQC may yet allow for  significant improvement of precision in case of limited 
resources. We provide explicit examples demonstrating the advantage of FFQC-assisted schemes 
over ones without intermediate control for both continuous and discrete processes.
\end{itemize}

Let us already stress that our results do not pertain only to the frequentist approach to parameter estimation, which assumes an infinite 
number of protocol repetitions (i.e., sufficiently large statistics), but also apply to the Bayesian approach within which one considers a 
finite statistical data or, in the extreme case, even a single experimental run.  Moreover, the most general protocol involving FFQC 
lends itself naturally to adaptive strategies in which one is allowed to modify the protocol (measurements, control operations etc.) 
"on-the-fly"---basing on the record of the measurements already collected.

The paper is organized as follows. In \secref{sec:hierarchy} we recall some recent results in quantum metrology, 
and compare them through a hierarchy of quantum metrology schemes accounting for various levels of control. We show that 
the FFQC scheme, that tops the hierarchy, is the most powerful one allowed by quantum mechanics.
\secref{sec:background} introduces the model of noisy 
quantum processes described by a time-homogeneous master equation, and reviews the concept of quantum Fisher 
information (QFI).  In \secref{sec:qfi_noise}, we review upper bounds on the QFI for metrology schemes, accounting for 
intermediate control, and generalize them to incorporate FFQC~(\secref{sec:no_go}) 
demonstrating (i) and (ii). Our extension of the bounds to the limit of short evolution times is crucial to show (ii), and is the key 
to obtain analytic bounds (as it removes the necessity to solve system dynamics).   In \secref{sec:implications} we discuss the 
performance of FFQC-assisted metrology 
both in the single shot regime as well as the regime of asymptotically many repetitions.  \secref{sec:lim_res} considers protocols with limited 
resources, where we show that FFQC-assisted schemes outperform parallel schemes without intermediate control (see (iii)). 
This is demonstrated for noisy frequency estimation in~\secref{sec:sup_of_seq}, and for noisy phase estimation 
in~\secref{sec:phase}.  We summarize and conclude our results in \secref{sec:outlook}. A reader who is familiar with quantum 
metrology, 
or who is primarily interested in the results, can directly proceed to sections \ref{sec:rank1pauli} and \ref{sec:univ_bound}
where the first two of our main results ((i) and (ii)) are presented.
\begin{figure}[!t]
\centering
\includegraphics[keepaspectratio,width=\columnwidth]{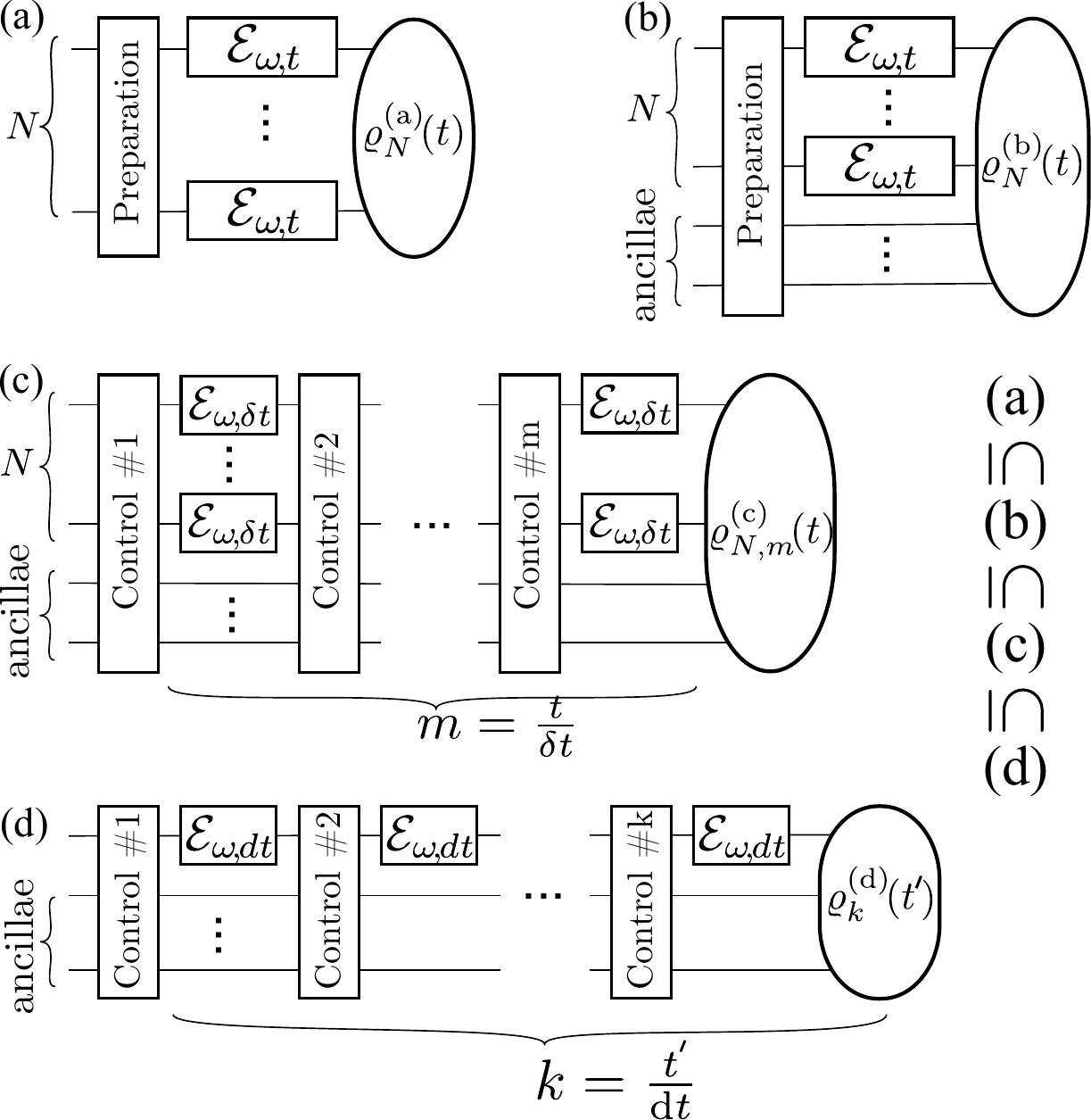}
\caption{Hierarchy of quantum metrology protocols  
for sensing a frequency-like parameter $\omega$ that is encoded on each probe
during its noisy evolution $\cE_{\omega,t}$. Within the standard schemes of type (a), $N$ probes independently sense the parameter for 
a time $t$. In (b), additionally an unlimited number of ancillae is allowed to perform 
a more general measurement strategy that may include a \emph{single error-correcting step} at the end of 
the protocol.  In (c), the sensing process is interspersed at time intervals, $\delta t$, with quantum control operations acting on 
both the sensing probes and ancillae. Finally in (d), control operations are further allowed 
to be of \emph{infinitesimally} small duration, i.e, $\delta t \to \d t$,
so that it is enough to consider a single probe that sequentially senses the 
parameter for the \emph{elongated} total time $t'\!=\!Nt$. As \emph{full and fast quantum control} 
(FFQC), applied frequently on the global state, may constitute swap operations any protocol
of type (c) can be simulated. Hence,
the schemes form the following hierarchy:~(a)$\subseteq$(b)$\subseteq$(c)$\subseteq$(d),
when ordered in terms of their ultimate power.}
\label{fig:hierarchy}
\end{figure}

\section{Quantum metrology protocols}
\label{sec:hierarchy}
In a standard quantum metrology protocol, a system consisting of $N$ probes (photons, atoms etc.)~is 
carefully engineered, so that each probe may be used to independently sense a parameter of interest 
$\omega$. As depicted in \figref{fig:hierarchy}(a), the experimentalist prepares then a suitably entangled 
state of the $N$ probes, which undergo the sensing process \emph{in parallel} for a time $t$, during which
the unknown parameter $\omega$ is imprinted on the state of the probes independently.
The final state of the system is measured in order to most precisely retrieve information regarding $\omega$.

If the sensing process is noiseless, the strategy (a) of \figref{fig:hierarchy} is known to optimally 
achieve the HS~\cite{Giovannetti2006}.
However, various uncorrelated noise types have been shown to constrain the precision to follow the standard scaling 
in the asymptotic $N$ limit, even if one optimally prepares the probes in an entangled but noise robust 
state~\cite{Huelga:97,Fujiwara:08,Escher:11,Kolodynski:12}. 

A more powerful metrological protocol is depicted in \figref{fig:hierarchy}(b).  Here, in addition to the $N$ 
sensing probes, the experimenter is equipped with ancillary particles that do not take part in the sensing process, and  may be used to 
implement single-step \emph{error correction} at the final measurement stage~\cite{Kolodynski:13}.  
Notice that the protocol of \figref{fig:hierarchy}(b) is more general than that of \figref{fig:hierarchy}(a);  indeed one recovers the 
latter by choosing not to entangle the sensing probes with the ancillae.  Consequently
any bound on precision valid 
for  scheme (b) 
also applies to scenario (a). 
Methods have been recently proposed that allow to derive bounds on the precision 
for protocols (b) from a Kraus representation of a single channel $\cE_{\omega, t}$ \cite{Kolodynski:12,Kolodynski:13}. 
Although these methods allowed to prove the asymptotic standard scaling for various noise types,
 the derived bounds 
cannot be guaranteed to be tight (even in the asymptotic $N$ limit). Nevertheless, in case of dephasing and 
particle-loss noise types, the corresponding bounds have been shown to be asymptotically achievable 
already within scheme (a)~\cite{Ulam2001, Kn11, Knysh:14}. For amplitude-damping 
noise, however,  scheme (b) has been shown to give a strictly better scaling than scheme \figref{fig:hierarchy}(a)~\cite{Do14}.

An interesting situation is the case of  the X-noise:~a noise similar to dephasing but with 
the generator perfectly transversal to the Hamiltonian encoding the parameter. For a fixed 
sensing time $t$ the bounds~\cite{Escher:11,Kolodynski:12} impose asymptotic standard scaling. However,  in frequency 
estimation one strongly benefits from decreasing the sensing time while increasing the number of probes $N$. Indeed, 
by optimizing the sensing time of the protocols (a) and (b) for each given $N$, it was shown that one may asymptotically beat the standard scaling and
achieve the $1/N^{5/3}$ precision scaling~\cite{Chaves:12}. 

Moreover, it has been shown that one can even restore the HS of precision for such 
strictly transversal noise after allowing for fast possibly multi-step error correction~\cite{Dur:14, Kessler:14, Arrad:14, Ozeri:13}. 
This is a particular case of the metrological scenario depicted in \figref{fig:hierarchy}(c).  Here, in addition to employing ancillary particles, the 
experimenter is capable of freely interjecting the sensing process with $m$ control pulses that may act on both probes and 
ancillae representing, e.g. 
error correction steps~\cite{Dur:14, Kessler:14, Arrad:14, Ozeri:13}, 
dynamical-decoupling pulses~\cite{Viola:98,Viola:99,Viola:03,Viola:09,Viola:10,West:10}  or any general adaptive feedback scheme~\cite{Wiseman2009,Chiribella2012}.
Notice that protocol (c) is more general than \figref{fig:hierarchy}(b) and one can 
obtain the latter from the former simply by allowing the intermediate operations to be the identity.  

The most general and powerful metrology protocol, and the main focus of the current work, is the one depicted in 
\figref{fig:hierarchy}(d). Here, the experimenter prepares a suitably entangled state between a \emph{single} 
probe and  many ancillae, and is capable of frequently interjecting the evolution with FFQC---an arbitrary number, $k$, 
of most general intermediate control operations acting on the overall state of probe-plus-ancillae.  Moreover, by choosing $k$ 
sufficiently large one can ensure 
the sensing time spent by the probe in between successive FFQC steps to be
\emph{infinitesimal}. It is this additional power of frequently interjecting the sensing process with \emph{fast} quantum control 
that we exploit throughout the remainder of this work to establish the ultimate bounds on precision. This intermediate control 
allows one in fact to modify the noise process, a possibility that has not been considered in previous approaches. 

To see that this is indeed the most general strategy, we note that for any protocol (c), with $N$ parallel probes 
and total sensing time $t$, there exists a protocol of type (d) with total sensing time $t'=Nt$ whose metrological performance is 
just as good as that of (c) or better.  We note in passing that protocols (c) and (d) are not equivalent; not only are the timesteps 
in protocol (d) assumed to be {\it infinitesimally} small, but protocol (d) allows us to use the same resources (namely $N$ and 
$t$) in a "sequentialized" fashion.

The above metrological schemes apply equally to the case of Bayesian parameter estimation, i.e., when considering only a 
finite number of experimental repetitions and a prior probability $p(\omega)$ representing our knowledge about the parameter 
$\omega$.  Examples of such Bayesian scenarios include reference frame alignment~\cite{Bagan2004, Chiribella2004a, 
Chiribella2004b, Chiribella2005}, adaptive Bayesian estimation~\cite{Berry00, Higgins2007, Sergeevich2011}, and single-shot 
estimation~\cite{Tsang:12, Gill:95}.

\subsubsection{A remark on error correction}

Let us briefly remark that the most general error correction step (encoding+syndrome readout+correction) only requires
the dimension of the space describing the ancillae  to be the same as the system one.
Indeed, given that the system containing the sensing probes has an overall dimension $d$ that
is smaller than the dimension of the ancillary space, any bipartite pure state of the two can be written using the Schmidt decomposition as
$\sum_{i=1}^{d}\!\lambda_i\ket{i}_{\t S}\!\otimes\!\ket{\psi_i}_{\t A}$. Here, $\{\ket{i}\}_{i=1}^d$ denotes 
a given fixed basis for the system, while $\{\ket{\psi_i}\}_{i=1}^{d}$ is a set of orthonormal states of ancillae 
that spans a $d$-dimensional subspace. Thus, since any error correction protocol may be implemented by restricting to states 
of the above form, it suffices to choose the overall space describing the ancillae to be of same dimension as the system.

\section{Background}
\label{sec:background}
In order to assess the performance of any of the protocols introduced in \secref{sec:hierarchy}, we review 
below the crucial tool of quantum metrology---the QFI. In addition, in \secref{sec:noise_models} 
we describe the general time-homogeneous qubit noise processes we consider throughout this work. 

\paragraph{Notation.}
Before doing so, let us first introduce the notation we utilize. In what follows, 
we denote the derivative with respect to the estimated parameter
$\omega$ (whenever applied to operators, vectors or matrices; differentiating
adequately the entries) as $\dot{\bullet}\!:=\!\frac{\d \bullet}{\d \omega}$; 
and shorten the trigonometric functions to $\Cos{\theta}\!:=\!\cos(\theta)$ and $\Sin{\theta}\!:=\!\sin(\theta)$.
$||A||\!:=\!\max_\psi | \bra{\psi} A\ket{\psi}|$ stands for the operator norm,
whereas $\eins$ and $\mathcal{I}$ for the identity operator and identity linear map respectively.
We use bold face to denote vectors, so that $\v{r}\!:=\! (x,y,z)^\T$ represents a 3-dimensional complex vector,
while $\v{x}$, $\v{y}$ and $\v{z}$ the Cartesian unit vectors.
We denote the standard Pauli 4- and 3-vectors 
as $\Vs\!:=\!(\sigma_0, \sone,\stwo,\sthree)^\T$ and $\vs\!:=\!(\sone,\stwo,\sthree)^\T$ 
respectively ($\sigma_0\!:=\!\eins$).
The scalar and outer products of vectors are then defined in the usual way irrespectively of the 
type of vector entries, i.e., $\v{v}^\dag\v{v}'\!:=\!\sum_i\! v_i^\dag v'_i$  and $[\v{v}'\v{v}^\dag]_{ij}\!:=\! v'_iv_j^\dag$. 
We write $\sigma_{\v{n}}\!:=\!\v{n}^\T\vs$ to denote 
a Pauli operator in the spatial direction ${\v{n}}=(s_\theta c_\phi, s_\theta s_\phi,c_\theta)^\T$.
We reserve the special font, $\m{M}$, for matrices whose entries, $\m{M}_{ij}$, are 
(complex) numbers, so that the multiplication of all types of vectors
can then be defined in the standard way:~$\m{M}\v{v}\!=\!(\sum_{j}\m{M}_{1j}v_j,\sum_{j}\m{M}_{2j}v_j,\dots)^\T$.
For example, any linear qubit map, $L$, can be unambiguously represented
by constructing its corresponding matrix $\m{L}$ (in the Pauli basis) such that
$L\!=\!\Vs^\dag\m{L}\Vs\!=\!\sum_{\mu,\nu=0}^3\!\sigma_\mu\m{L}_{\mu\nu}\sigma_\nu$.

\subsection{QFI and its role in quantum metrology}
\label{sec:par_est}
In all metrology schemes discussed in \secref{sec:hierarchy} the information about the estimated parameter of interest 
is encoded in the final state of the system consisting generally of both probes and ancillae.  
Thus, we denote the final state of the system in any of the schemes (a,b,c,d) of  \figref{fig:hierarchy} as 
$\varrho_\omega$. It is then $\varrho_\omega$ that is measured, in order to most precisely construct
an estimate of the parameter of interest, $\omega$.

How best to quantify the precision of this estimate depends 
crucially on how the information about $\omega$ has been collected.
If we are freely allowed to repeat the protocol sufficiently 
many times and collect statistics, then the best precision of estimation, as quantified by the \defn{mean squared error} (MSE), 
is determined by the well-known Cram\'{e}r-Rao bound (CRB)~\cite{Helstrom1976, H80}
\be\label{eq:CRB_def}
\delta^2 \omega \geq \frac{1}{\nu \QFI(\varrho_\omega)},
\ee
where $ \QFI(\varrho_\omega)$ is the QFI of the state and $\nu$ is the number of repetitions.
On the other hand, if the information about $\omega$ is inferred from the outcome of just a 
single repetition of the protocol, then the single-shot MSE has to be 
averaged over the prior distribution representing the knowledge we possess about $\omega$
prior to its estimation. 
Then, the \defn{average mean squared error} (\aMSE) can be lower-bounded by either the 
Bayesian Cram\'{e}r-Rao bound (BCRB)~\cite{Gill:95} (for well behaved prior distributions) or the 
Ziv-Zakai bound (ZZB)~\cite{Tsang:12}.   

Whether considering the CRB, BCRB, or ZZB, the crucial quantity of interest is the QFI:~$\QFI(\varrho_\omega)$. 
Given the spectral decomposition $\varrho_\omega\!=\!\sum_{i}p_i(\omega)\proj{\psi_i(\omega)}$, the QFI of the state 
$\varrho_\omega$ 
evaluated with respect to the estimated parameter $\omega$ generally reads
\begin{equation}
{\cF}(\varrho_\omega):=2\!\!\sum_{\underset{p_j+p_j\neq 0}{i,j}}\frac{1}{p_i(\omega)+p_j(\omega)}|\bra{\psi_i(\omega)}
\dot{\varrho}_\omega\ket{\psi_j(\omega)}|^2.
\label{QFI}
\end{equation}
On the other hand, the QFI may be equivalently defined by using its relation with the fidelity between 
quantum states \cite{BC94}, $F(\rho,\sigma)\!:=\!\tr\sqrt{\sqrt{\sigma} \rho \sqrt{\sigma}}$:
\be
\cF(\varrho_\omega) := 8 \lim_{\d\omega\to 0}\frac{1-F(\varrho_\omega,\varrho_{\omega+\d\omega})}{\d\omega^2}.
\label{eq:QFI_fidelity}
\ee

The QFI satisfies the following important properties. It is 
\emph{additive}, so that if the final state of the system is a \emph{product} state, i.e., 
$\varrho_\omega\!=\!\rho^{\otimes N}_\omega$, then its QFI satisfies
$\QFI(\rho^{\otimes N}_\omega)\!=\!N\QFI(\rho_\omega)$ and thus 
gives rise to standard scaling~\cite{GLM04}.  If the final state of the probes 
is \emph{pure}, $\varrho_\omega\!=\!\proj{\psi_\omega}$, then
\be
\QFI(\varrho_\omega)=4\left(\bracket{\dot{\psi}_\omega}{\dot{\psi}_\omega}-\left|\bracket{\dot{\psi}_\omega}{\psi_
\omega}\right|^2\right).
\label{QFIpure}
\ee
Moreover, if the parameter is also encoded via a unitary, so that $\ket{\psi_\omega}\!=\!\e^{-\ii \omega H}\ket{\psi}$ where 
$H$ is some Hamiltonian,  then $\QFI(\varrho_\omega)\!=\!4(\bra{\psi}H^2\ket{\psi}-\bra{\psi}H\ket{\psi}^2)=:4\,\mathrm{Var}_{\ket{\psi}}(H)$.  
In this case the QFI is maximized by preparing the probes in the state
$\ket{\psi}\!=\!\frac{1}{\sqrt{2}}\left(\ket{\lambda_{\t{max}}}\!+\!\ket{\lambda_{\t{min}}}\right)$, where $\ket{\lambda_\t{max/min}}$ are 
the eigenstates corresponding to the maximum/minimum eigenvalues of $H$.  Furthermore, if the $N$ probes are subjected to a 
unitary evolution generated by a \emph{local} Hamiltonian, e.g., $H=\frac{1}{2}\sum_{n=0}^N \sthree^{(n)}$, then such a state 
corresponds to the GHZ state and attains the HS by virtue of the CRB \eqref{eq:CRB_def}~\cite{GLM04}.
However, note that the QFI is important not only for metrology but also for entanglement 
detection~\cite{Pezze2009,Lucke2014,Strobel2014}; defines a natural geometric distance
between quantum states \cite{Pires2015}, thus serving as a tool to derive speed limits on quantum evolution 
\cite{Pires2015,Taddei2013}, as well as constitutes a measure of macroscopicity~\cite{Florian2012}.
The QFI is the primary focus of this work and we further discuss its role in quantifying
the precision of estimation in metrology protocols in more detail in \secref{sec:implications}.

\subsection{Time-homogeneous qubit evolution}
\label{sec:noise_models}

The interaction dynamics via which a probe senses the parameter is generally 
described by a quantum channel~\cite{MikeIke} (see \figref{fig:hierarchy}), $\cE_{\omega,t}$, that 
encodes the parameter $\omega$ onto the state of the probe over the interrogation time $t$:~$\rho_\omega(t)\!=\!\cE_{\omega,t}(\rho)$.
The channel consists of a unitary evolution part, encoding the parameter of interest, and a non-unitary (noise) part describing additional system-environment interactions.  In this work we shall consider that the non-unitary part  arises due to uncorrelated noise processes 
described by a \emph{time-homogeneous master equation of Lindblad form}---often referred to as the semigroup 
dynamics \cite{Alicki1987}%
---which provides an appropriate description for most physically relevant noise processes. 
In particular, in such a typical setting the environment is assumed to be sufficiently large such that
it disturbs the system in the same fashion independently of the system state and the time instance \cite{Breuer2002}.
Thus, the time-homogeneity of evolution allows us to unambiguously apply the FFQC techniques as any operation applied 
on the system does not affect the environment. Moreover, one may then freely swap the sensing particles 
at any time with the ancillary ones, which experience afterwards the same fixed noisy dynamics. 
In contrast, noise processes described by time-inhomogeneous master equations have 
been recently considered within the context of quantum metrology in~\cite{Matsuzaki2011,Chin2012,Macieszczak2015,Smirne2015}, where
time-inhomogeneity has been shown to be beneficial at short time-scales--- the so-called Zeno regime---where the bath can
no longer be assumed to be uncorrelated from the system. Let us stress that such a regime, 
however, does not allow for the general control operations considered here to be unambiguously 
applied without explicitly modelling the environment \cite{Addis2016,SSD:15}.

We describe the evolution of a single qubit probe by following time-homogeneous master equation
\be
\frac{\d \rho_\omega(t)}{\d t} = -\ii \frac{\omega}{2}[\sthree,  \rho_\omega(t)] +\cL( \rho_\omega(t)),
\label{eq:ME}
\ee
where $\omega$ is the parameter to be estimated and the noise is given by the Lindblad super-operator, $\mathcal{L}$.
The Liouvillian is then generally defined as
\begin{equation}
 \cL(\rho):=\frac{1}{2}\sum_{\mu,\nu=0}^3 \m{L}_{\mu\nu}\left(\left[\sigma_\mu \rho,\sigma_\nu\right]+
\left[\sigma_\mu, \rho \sigma_\nu\right]\right),
\label{eq:L}
\end{equation}
where $\m{L}$ is the matrix representation of the Lindblad superoperator -- an Hermitian positive semi-definite matrix whose  entries are independent of 
$t$ and $\omega$.   We note that  all the noise terms $\m{L}_{0i}$ and $\m{L}_{i0}$  can be straightforwardly corrected within FFQC
as they lead to an additional  Hamiltonian evolution term in the \eqnref{eq:ME}, sometimes referred to as Lamb-shift. 
This $\omega$-independent Hamiltonian term can be cancelled by continuously applying the inverse unitary rotation.
In what follows we assume that the control operations already incorporate such unitary and will, thus, 
be concerned with the restriction of $\m{L}$ to the subspace spanned by $\{\sigma_1, \sigma_2,\sigma_3 \}$, which we denote as $\bar{\m{L}}$.

We can group all the relevant noise processes into three important families: 
\begin{itemize}[topsep=3pt,parsep=3pt,leftmargin=10pt]
\item \emph{Rank-one Pauli} noise of strength $\gamma$
\be
\mathcal{L}^\t{1P}_\v{n}(\rho) :=\frac{\gamma}{2} ( \sigma_\v{n} \rho \sigma_\v{n} - \rho ),
\label{eq:rank1pauli}
\ee
with the particular case of \emph{dephasing} noise $\mathcal{L}^\t{1P}_{\bf z}$ 
for which $\sigma_\v{z}=\sigma_3$.

\item \emph{General rank-one} noise $\mathcal{L}^\t{1G}_\v{r}$ defined by the matrix
\be
\bar{\m{ L}}^\t{1G}_\v{r} :=\v{r}\,\v{r}^\dagger,
\label{eq:rank1gen}
\ee
where $\v r = (x,y,z)^T$ and the strength of the noise is denoted $\gamma/2\!=\!|\v{r}|^2$. 
The special case $\re(\v{r})\!\times\!\im(\v{r})\!=\!0$ corresponds to the rank-one Pauli noise of
\eqnref{eq:rank1pauli}, whereas whenever $|\re(\v{r})|\!=\!|\im(\v{r})|$ and $\re(\v{r})^\T\im(\v{r})\!=\!0$ 
an \emph{amplitude damping} channel is recovered, which represents spontaneous emission 
along some particular direction.

\item \emph{Rank-two Pauli} noise $\mathcal{L}^\t{2P}_{\bf \Omega}$ defined by the matrix
\bea
\bar{\m{ L}}^\t{2P}_{\bf \Omega} := \frac{1}{2}\,
R_{\bf \Omega}^T\begin{small} \left(\begin{array}{ccc}
\gamma_1 & &\\
& \gamma_2 &\\
&&0
\end{array}\right)
 \end{small} R_{\bf \Omega},
\label{eq:rank2pauli}
\eea
where $\v{\Omega}\!\in\!\mathrm{SO}(3)$ and $R_{\bf \Omega}\!=\!R_z(\varphi) R_y(\theta) R_z(\xi)$ 
is its matrix representation written in the Euler form with angles $(\varphi,\theta,\xi)$. 
The special case of $R_{\bf \Omega}\!=\!\eins$ corresponds to \emph{asymmetric X-Y noise}, which 
we conveniently parametrize with $\gamma_1\!=\!\gamma\,p$ and $\gamma_2\!=\!\gamma(1-p)$ 
for $0\!\le\!p\!\le\!1$ to define
\bea
\mathcal{L}_\t{X-Y}(\rho) := \frac{\gamma}{2} (p \, \sone \rho \, \sone + (1-p) \stwo \rho \, \stwo - \rho ).
\label{eq:xynoise}
\eea
\end{itemize}

In order to explicitly determine the form of the quantum channel 
describing the qubit evolution, one has to integrate \eqnref{eq:ME}~\cite{Anderson:07}.
The resulting dynamics can then be expressed, e.g, with help of the corresponding 
dynamical matrix, $\m{S}$ (specified in the Pauli operator basis), or via a Kraus representation as follows:
\bea
\cE_{\omega,t}(\rho)
 &=& \sum_{\mu,\nu=0}^3 \m{S}_{\mu\nu}(\omega,t) \, \sigma_\mu\, \rho\, \sigma_\nu \label{eq:Srep}\\
&=&\; \sum_i K_i(\omega,t) \,\rho\, K_i(\omega,t)^\dag \label{eq:Krep}.
\eea
Note that the Kraus representation of \eqnref{eq:Krep} is not unique as starting from a set of $r$ linearly 
independent Kraus operators, $\v{K}\!=\!(K_1,\dots,K_r)^\T$, we may simply construct 
another valid set $\v{K}'\!=\!\m{u}\v{K}$ choosing $\m{u}$ to be any (potentially $\omega$-,$t$- dependent) 
$r\!\times\!r$ unitary matrix.

\section{Noisy metrology with full and fast control}
\label{sec:qfi_noise}

In this section we derive the main results of our work. 

In \secref{sec:rank1pauli} we outline the optimal FFQC protocol suitable for correcting rank-one Pauli noise that is 
not parallel to the Hamiltonian \eqref{eq:rank1pauli}. In \secref{sec:SQL} we introduce the necessary tools required to bound the QFI for FFQC (\figref{fig:hierarchy}(d)), and show that any other noise unavoidably leads to a linear scaling of the QFI.  We exemplify our findings by deriving analytical bounds for the QFI for several physically relevant noise-types. Finally, in \secref{sec:comparison} we discuss different aspects in which FFQC allows to outperform scenarios without fast control, using the cases of X-Y noise \eqref{eq:rank1pauli} and transversal rank-one Pauli noise \eqref{eq:rank2pauli} as examples.

\subsection{Removing rank-one Pauli noise}
\label{sec:rank1pauli}

Here we construct a general FFQC strategy that allows one to correct for 
\emph{any} rank-one Pauli noise $\mathcal{L}^\t{1P}_{\v{n}}$ introduced in \eqnref{eq:rank1pauli}
with $\v{n}\!\neq\!\v{z}$. For convenience, we perform a change of basis so that the noise is generated by $\sone$ 
($\v{n}\!\!=\!\!\v{x}$), whereas $\omega$ is encoded via a rotated Hamiltonian 
$\sigma_\theta\!:=\!\Sin{\theta}\sthree + \Cos{\theta}\sone$. 
We consider the qubit probe to be aided by an ancillary qubit and define the two-qubit \emph{code space} as a subset of 
$\cH_\t{S}\otimes\cH_\t{A}\supset \cH_\t{C}\!:=\!\mathrm{span}\{\ket{00},\ket{11}\}$, with 
corresponding projector $\Pi_\t{C}\!:=\!\proj{00}+\proj{11}$, 
and the \emph{error space}, $\cH_\t{E}\!:=\!\mathrm{span}\{\ket{01},\ket{10}\}$, with $\Pi_\t{E}\!:=\!\eins-\Pi_\t{C}$. 

Let the probe-plus-ancilla be prepared in a pure state $\varrho\!=\!\proj{\psi}$ with 
$\ket{\psi}\!=\!\alpha \ket{00}+\beta \ket{11} \in \cH_\t{C}$. The dynamics on the probe-plus-ancilla
is then described by \eqnref{eq:rank1pauli} with only the probe system evolving
\be
\frac{\d \varrho}{\d t} = -\ii \frac{\omega}{2}[\sigma_\theta\otimes\eins, \varrho] +\mathcal{L}^\t{1P}_{\bf x}\!\otimes\mathcal{I}_\t{A}\,(\varrho).
\label{eq:qubit_ancilla_evol}
\ee
Integrating \eqnref{eq:qubit_ancilla_evol} over an elementary timestep $\d t$, 
we may write the probe-plus-ancilla state, up to first order in $\d t$, as
\bea
\varrho(\d t) &=& \varrho (1-\frac{\gamma}{2} \d t)-
\ii\frac{\omega}{2}[\sigma_\theta\otimes\eins, \varrho]\d t 
\label{eq:qubit_ancilla_int_dt} \\
&&+\,\frac{\gamma}{2}(\sone\otimes\eins) \varrho (\sone\otimes\eins)\d t+ \bigO(\d t^2).
\nonumber
\eea
Projecting $\varrho(\d t)$ onto the code and error subspaces yields the unnormalized states 
$\varrho_\t{C}(\d t)\!=\! \Pi_\t{C}\varrho(\d t) \Pi_\t{C}$ and $\varrho_\t{E}(\d t)\!=\!\Pi_\t{E}\varrho(\d t) \Pi_\t{E}$ given by 
\begin{align}
\varrho_\t{C}(\d t) &= \varrho (1-\frac{\gamma}{2} \d t) - \ii\,\Sin{\theta}\frac{\omega}{2}[\sigma_3\otimes\eins,\varrho] \d t +O(\d t^2),
\nonumber\\
\varrho_\t{E}(\d t) &= \frac{\gamma}{2} (\sone\otimes\eins) \varrho (\sone\otimes\eins)\,\d t+O(\d t^2)
\end{align}
respectively.

If an error is detected, we simply apply $\sone$ on the sensing probe (so after correction
$\bar\varrho_\t{E}(\d t)\!:=\!(\sone\otimes\eins)\varrho_\t{E}(\d t)(\sone\otimes\eins)$), and otherwise do nothing. Hence, 
the state of probe-plus-ancilla after an infinitesimal-timestep followed by fast error correction is the mixture
\bea
\bar \varrho(\d t)&:=& \varrho_\t{C}(\d t) + \bar\varrho_\t{E}(\d t) \nonumber\\
&=&\varrho -\ii\,\Sin{\theta}\frac{\omega}{2}[\sigma_3\otimes\eins,\varrho] \,\d t+O(\d t^2),
\label{eq:corrected_state}
\eea
which, to first order in $\d t$, is equivalent to a unitary evolution under the 
projected Hamiltonian $\frac{\omega}{2} s_\theta \sigma_3$. 
As the measurement may always be adjusted 
to compensate for a known rotation of the Hamiltonian encoding the parameter, 
the above strategy perfectly corrects rank-one Pauli noise \eref{eq:rank1pauli} at the price of 
slowing the evolution down by a factor $s_\theta=\sqrt{1-({\v{n}}^\T \v{z})^2}$.

We summarize the above result in the following observation:

\begin{obs}
\label{obs:rank_1_pauli}
A general FFQC strategy allows one to completely eliminate the impact of any rank-one Pauli noise 
$\mathcal{L}^\mathrm{1P}_{\v{n}}$ \eref{eq:rank1pauli} that is not exactly parallel to the parameter-encoding Hamiltonian.  
The resulting dynamics of the system are then described by a noiseless evolution 
with a rotated Hamiltonian and the estimated parameter being rescaled to 
$\sqrt{1-(\v{n}^T \v{z})^2}\,\omega$, which still yields HS in precision.
\end{obs}

Note that in the above derivation nothing forbids us from replacing the scalar parameter $\omega$ with any operator $B$ acting on an additional system described  by the Hilbert space $\cH_\t{B}$.
Consequently, one can effectively modify the noisy dynamics of any $\varrho_\t{BS} \in \mathcal{B}(\cH_\t{B}\otimes \cH_\t{S})$:
\be
\frac{\d \varrho_\t{BS}}{\d t} = -\ii \frac{1}{2}[B\otimes\sigma_\theta, \varrho_\t{BS}] +\mathcal{I}_\t{B}\!\otimes\!\mathcal{L}^\t{1P}_{\bf x}\,(\varrho_\t{BS}),
\label{eq:qubit_ancilla_B_evol}
\ee
by adding an ancillary qubit and implementing the FFQC strategy described above. Then, the dynamics of the error-corrected $\bar \varrho_\t{BS}=\tr_\t{A}\{\bar\varrho_\t{BSA}\}$ 
is governed by
\be
\frac{\d \bar \varrho_\t{BS}}{\d t} = -\ii \frac{\sqrt{1-({\v{n}}^\T \v{z})^2}}{2}[B \otimes\sigma_3, \bar \varrho_\t{BS}],
\label{eq:qubit_ancilla_B_evol_eff}
\ee
with the rank-one Pauli noise removed. Hence, our strategy can be directly used for the implementation of unitary gates $U=e^{\ii g \,B\otimes \sigma_3}$ in the presence of rank-one Pauli noise on the qubit.
 
%

\subsection{Other noise-types:~unavoidable linear scaling}
\label{sec:SQL}

In this subsection we prove our main no-go result, $\QFI_\mathcal{L}\leq 4 \,\alpha_\mathcal{L}\, t'$, showing that even with FFQC the QFI is bound to a linear scaling for all uncorrelated noise-types other than rank-one Pauli noise, and to minimize the constant $\alpha_\mathcal{L}$ appearing in this bound for several practically relevant noise-types introduced in \secref{sec:noise_models}. These results are presented in \secref{sec:univ_bound} and \secref{sec:examples} respectively. The preceding sections are devoted to elaborate the method that allows us to derive these results. In \secref{sec:no_go} and \secref{sec:bound_opt} we describe the channel extension (CE) 
method used to upper bound the QFI~\cite{Fujiwara:08, Kolodynski:12,Kolodynski:13, Do14} and then review the  bounds it yields for strategies (a), (b) and (c) of \figref{fig:hierarchy}.  
However, the CE method fails to yield a meaningful bound when considering the limit in which the sensing process lasts for an 
infinitesimal time-step. In \secref{asymptoticbounds} we resolve this issue by amending the CE method 
in order to obtain an upper bound on the QFI for strategies that employ FFQC.  

We stress that we are able to derive analytic bounds by extending the CE method to the limit $\d t \to 0$, in which the problem  actually simplifies: one does not have to solve the master equation and determine the full dynamics! This is even more remarkable as these bounds apply to any metrological scheme (see the hierarchy in \figref{fig:hierarchy})

\subsubsection{Upper bounds on the QFI}
\label{sec:no_go}
We would like to compare the maximum achievable QFI in the various metrology schemes introduced in 
\secref{sec:hierarchy} and, in particular, quantify the improvement (if any) between protocols  (b), (c) and (d). To this end we use 
variations of the
CE method~\cite{Fujiwara:08, Kolodynski:12,Kolodynski:13, Do14} that allow one to upper-bound the QFI depending on the 
scenario considered. 
Crucially, they constrain the QFI using solely the properties of the quantum channel responsible for encoding the parameter.

\paragraph{CE bound.}\label{sec:CE_bound}
The QFI of a state representing the output of a quantum channel
encoding the parameter may always be upper-bounded 
purely by the channel properties after performing a channel extension (CE). 
Given a probe in a state $\rho$ and a quantum channel $\cE_{\omega,t}$ one may always construct the 
\emph{CE bound} on the QFI of the output state $\cE_{\omega,t}(\rho)$~\cite{Fujiwara:08}:
\be
\QFI( \cE_{\omega,t}(\rho))
\;\leq\;
\max_\varrho
\QFI( \cE_{\omega,t} \otimes \mathcal{I}_\t{A}(\varrho))
\;\leq\;
4\,||
\alpha(t)
||,
\label{eq:CE_bound}
\ee
where $\varrho\!\in\!\cB(\cH_\t{S}\otimes\cH_\t{A})$ represents a joined (\emph{extended}) 
state of the probe and an ancilla, and the operator
\bea
\alpha(t)&:=&\sum_i \!\dot K_i^\dag \dot K_i =\dot{\v{K}}^\dag \,\dot{\v{K}}
\label{eq:alpha}
\eea
for any of the Kraus representation \eqnref{eq:Krep} of the probe channel. Here, $\dot K_i$
are the derivatives of the Kraus operators with respect to~$\omega$ (in what follows, we write explicit 
dependencies of the Kraus operators on $t$ and $\omega$ only if necessary)%

Although the first inequality in \eqnref{eq:CE_bound} may not be tight---entangling the probe 
and ancilla may increase the achievable QFI---the second is. A valid Kraus 
representation of $\cE_{\omega,t}$ for which the second inequality is saturated  
is guaranteed to exist as long as $\dim(\cH_\t{A})\!\ge\!\dim(\cH_\t{S})$~\cite{Fujiwara:08}.
In consequence, the CE bound \eref{eq:CE_bound} directly determines the maximal attainable QFI  for the ancilla-assisted parallel scheme 
 depicted in \figref{fig:hierarchy}(b) with $N\!=\!1$~\cite{Kolodynski:13}.

\paragraph{Parallel CE bound.}\label{sec:parallelCE}
The CE bound \eref{eq:CE_bound} may be directly applied to upper-bound the 
ultimate attainable QFI of any parallel scheme of \figref{fig:hierarchy}(b)
that employs $N$ probes and ancillae. In scenario (b),
the final state generally reads 
$\varrho^{\t(b)}_{N}(t)\!=\!\cE^{\otimes N}_{\omega,t}\otimes\mathcal{I}_\t{A}(\varrho_N^\t{(b)})$.
Hence, applying the CE bound \eref{eq:CE_bound} to channel $\cE^{\otimes N}_{\omega,t}\otimes\mathcal{I}_\t{A}$  
that describes the overall evolution and fixing a tensor product Kraus representation such that
\be
\cE^{\otimes N}_{\omega,t}(\rho^N) 
= \!\!\!\!
\sum_{i_1,\dots,i_N} \!\!\!\!
K_{i_1}\!\otimes...\otimes K_{i_N}\;\rho^{N}\,K_{i_1}^\dag\!\otimes...\otimes K_{i_N}^\dag, 
\label{eq:krausrep_par}
\ee
one obtains the \emph{parallel CE bound}~\cite{Fujiwara:08,Kolodynski:12,Kolodynski:13}:
\al{
\QFI\Big(\varrho^\t{(b)}_{N}(t)\Big)
&\leq
\max_{\varrho_N^\t{(b)}}\,\QFI\Big(\cE^{\otimes N}_{\omega,t}\otimes\mathcal{I}_\t{A}(\varrho^{(\t b)}_N)\Big) \nonumber\\
&\leq
4 N || \alpha(t) || + 4 N (N-1)  || \beta(t)||^2,
\label{eq:CE_par}
}
where $\varrho_N^\t{(b)}\!\in\!\cB(\cH_\t{S}^{\otimes N}\otimes\cH_\t{A})$ describes
the initial probes-plus-ancillae state and
\be
\beta(t) := \ii\sum_i \!\dot K_i^\dag  K_i =\ii\,\dot{\v{K}}^\dag\, {\v{K}}
\label{eq:beta}
\ee
which depends solely on Kraus operators and their derivatives. 
 In order to obtain \eqnref{eq:CE_par} from \eqnref{eq:CE_bound} note that due to the tensor product structure the term $\dot K_{\v i}^\dag \dot K_{\v i} = \frac{\d}{\d\omega}(K_{i_1}^\dag\otimes...\otimes K_{i_N}^\dag)\, \frac{\d}{\d\omega}(K_{i_1}\otimes...\otimes K_{i_N})$ in \eqnref{eq:CE_bound}  contains $N$ terms for which in both parenthesis the derivative applies on the Kraus operator acting on the same probe and $N(N-1)$ terms where it applies on the Kraus operators acting on two different probes~ \cite{Fujiwara:08}. 
Yet, in contrast to \eqnref{eq:CE_bound}, the parallel CE bound \eref{eq:CE_par} is then not 
generally attainable, even after optimizing \eqnref{eq:CE_par} over all Kraus representations 
of the single-probe channel. This is because the Kraus decomposition for the map 
$\cE^{\otimes N}_{\omega,t}\otimes\mathcal{I}_\t{A}$ that makes \eqnref{eq:CE_bound} saturable 
might not be of a tensor-product form, as assumed in \eqnref{eq:krausrep_par}.

\paragraph{Sequential CE bound.}
\label{par:seq_bound}
Now, we would like to apply the CE methods to the two strategies involving control, i.e., 
(c) and (d) of \figref{fig:hierarchy}. However, as argued in \secref{sec:hierarchy}, any 
scheme of type (c) employing $N$ probes with $m$ control steps (each lasting $\delta t$) can 
always be mimicked by a scheme \figref{fig:hierarchy}(d) employing a single probe but lasting 
$N$-times longer, i.e., with $t'\!=\!Nt$ and involving more $k\!=\!Nm$ steps 
each still of duration $\delta t$. Hence, we may always upper-bound the QFI of  \emph{any} 
scheme depicted in \figref{fig:hierarchy}(c) as
\be
\QFI\Big(\varrho^\t{(c)}_{N,m}(t)\Big) \leq 
\max_{
\underset{k=Nm,\,t'=Nt}{\t{schemes\,(d)}}
}
\QFI\Big(\varrho^\t{(d)}_{k}(t')\Big)
\label{eq:QFI(c)vs(d)}
\ee
where the maximization is over all sequential schemes of type (d) 
with step duration $\delta t\!=\!N/m$. Thus, in what follows, 
we focus on applying the CE methods to the scenario 
of \figref{fig:hierarchy}(d). Note that by decreasing 
$\delta t$ further, or equivalently by raising the number of control steps $k$, 
we may only increase the right hand side of \eqnref{eq:QFI(c)vs(d)}. 
Hence, the optimal FFQC scheme yielding the maximum in \eqnref{eq:QFI(c)vs(d)}
must correspond to the limit of $k\to\infty$, or equivalently, $\delta t\!\to\!0$.
It is so, as any FFQC scheme of step duration $\delta t$ can be mimicked by a protocol 
with shorter $\delta t'=\delta t/k$ (and any $k\ge2$) and setting some of the intermediate 
control operations to be trivial, i.e., the identity.

For sequential strategies employing intermediate control a CE-based upper bound 
on the corresponding QFI has recently been derived in \refcite{Do14}.
Let the output state for the sequential protocol of \figref{fig:hierarchy}(d) be given as
\be
\varrho^\t{(d)}_{k}(t')=\sum_{{\bm i}^{(k)}} K_{{\bm i}^{(k)}}\!(t')\,\varrho^\t{(d)}\,K^\dag_{{\bm i}^{(k)}}\!(t')
\label{eq:rho^(d)}
\ee
with $\varrho^\t{(d)}\!\in\!\cB(\cH_\t{S}\otimes \cH_\t{A})$ being the initial state. 
However, in contrast to \eqnref{eq:krausrep_par}, the overall Kraus operators
\be
K_{{\bm i}^{(k)}}\!(t)\!:=\!(K_{i_k}\!(\delta t)\otimes\eins_\t{A})U_k \dots (K_{i_1}\!(\delta t)\otimes\eins_\t{A})U_1
\label{eq:kraus_seq}
\ee
have a composition rather than tensor-product structure, 
where $U_\ell$ stands for the a 
control operation acting on both probe and ancillae at the $\ell$th step , while $K_{i_\ell}$ are the 
Kraus operators of the channel $\cE_{\omega,\delta t}$ acting solely on the probe at the $\ell$th step. 
An application of the general CE bound \eqref{eq:CE_bound} to this case was derived in~\refcite{Do14}, leading to 
the \emph{sequential CE bound} on the QFI attainable using \emph{any} sequential scheme with FFQC (with timestep $\delta t$):
\al{
&\QFI\Big(\varrho^\t{(d)}_{k}(t')\Big)
\leq
4 k || \alpha(\delta t)||+ \label{eq:CE_seq}\\
&\qquad+4 k (k-1)  ||\beta(\delta t)||(||\alpha(\delta t)||+||\beta(\delta t)||+1),
\nonumber
}
which applies to \emph{any} protocol (d) of \figref{fig:hierarchy} with time-steps of duration $\delta t$,
and hence also to \emph{all} schemes (c) of \figref{fig:hierarchy} upon substituting $k\!=\!Nm$.

Notice that, just as in the case of the parallel CE bound \eref{eq:CE_par}, there 
may not exist a Kraus representation of the single-probe channel such that the sequential CE 
bound of \eqnref{eq:CE_seq} is guaranteed to be tight. This holds also for the optimal sequential strategy (d)
that maximizes the corresponding QFI (i.e., the one maximizing the right hand side of \eqnref{eq:QFI(c)vs(d)}).

\subsubsection{Optimization of the CE bounds}
\label{sec:bound_opt}

All the CE bounds presented this far:~\emph{standard} (\eqnref{eq:CE_bound}), \emph{parallel} (\eqnref{eq:CE_par}) 
and \emph{sequential} (\eqnref{eq:CE_seq});~rely solely on the structure of the single-probe channel $\cE_{\omega,t}$ encoding the parameter 
and its particular Kraus representation \eref{eq:Krep}. Thus, in order to obtain the tightest versions of these bounds one 
should minimize them over all equivalent Kraus representations, $\v{K'}\!=\!\m{u}\v{K}$ (with $\m u^\dag \m u \!=\!\eins$), 
describing the single-probe dynamics. Although such a gauge freedom is generally parameter dependent, i.e., 
$\m u\!=\! \m u(\omega)$,  the operators $\alpha$ \eref{eq:alpha} and $\beta$ \eref{eq:beta} 
depend solely on $\m h\!:=\!\ii\, \m u^\dag \dot {\m u}$.  Hence, it always suffices to search only through Kraus 
representations satisfying $\v{K}'\!=\!\v{K}$ and $\dot{\v{K}}'\!=\!\dot{\v{K}}\!-\!\ii\m{h}\v{K}$ or,
in other words, any CE-based bound considered may always be optimized by simply performing its minimization over all 
Hermitian matrices $\m{h}$~\cite{Fujiwara:08}.

In contrast to the sequential CE bound \eref{eq:CE_seq}, the standard \eref{eq:CE_bound} and 
parallel \eref{eq:CE_par} CE bounds are quadratic (convex) in $\m{h}$, and thus may be minimized 
numerically in a systematic manner by means of semi-definite programming (SDP)
given some $\v K$~\cite{Kolodynski:12,Kolodynski:13}. 
On the other hand, when considering t (or $\delta t$ in case of \eqnref{eq:CE_seq})
to be fixed and the asymptotic limit of $N$ \eref{eq:CE_par} ($Nm$ \eref{eq:QFI(c)vs(d)}, or $k$ \eref{eq:CE_seq}),
it is always optimal, if possible (contrary to the unitary case), to set $\beta(t)\!=\!0$ in the CE bounds, 
so that they \emph{asymptotically} become linear in $N$ ($Nm$, or $k$ respectively).
In such a regime, as the constraint $\beta(t)\!=\!0$ is linear in $\m{h}$, one may always apply 
the SDP methods to minimize the remaining coefficient $4||\alpha(t)||$, which
is then emergent in all the bounds~\cite{Kolodynski:12,Kolodynski:13}.

Equivalently, the same argumentation can be made starting from the dynamical matrix $\m S$ of 
the channel $\cE_{\omega,t}$, defined in \eqnref{eq:Srep}, instead of its Kraus representation. 
The two are directly related (see also \appref{app:1}), as any valid vector of Kraus operators ${\bf K}$
can always be expressed in a complete operator basis, e.g. the Pauli basis in case of a single-qubit 
maps, via a matrix $\m{M}$ satisfying
\bea
{\bf K} = \m{M} \Vs \quad \text{and} \quad \m{M}^\dag \m{M} = \m{S}^\T.
\label{eq:conditionsM}
\eea
As a result, one may rewrite the operators $\alpha$ \eref{eq:alpha} and $\beta$ \eref{eq:beta} 
with help of the matrix $\m{M}$ and a general Hermitian matrix $\m{h}$ to be optimized as
\bea
\alpha(t) &=& \Vs^\dag \, (\dot{\m{M}}^\dag + \ii  {\m{M}}^\dag {\m{h}}) \,(\dot {\m{M}}- \ii {\m{h}}{\m{M}})\, \Vs, 
\label{eq:alphaM}\\
\beta(t)  &=& \Vs^\dag \, (\dot{\m{M}}^\dag + \ii  {\m{M}}^\dag {\m{h}})\, {\m{M}}\, \Vs \label{eq:betaM}.
\eea
In what follows, we always choose $\m M\!=\!\m M^\dag\!=\!\sqrt{\m S}^T$ and thus
unambiguously fix the starting Kraus representation $\v K$ in \eqnref{eq:conditionsM}.
We then make use of \eqnsref{eq:alphaM}{eq:betaM} in order to \emph{analytically} optimize 
over $\m{h}$ the infinitesimal-timestep version of the sequential CE bound \eref{eq:CE_seq}
as we now explain.

\subsubsection{Infinitesimal-timestep CE bound}
\label{asymptoticbounds}

In the most powerful FFQC setting of \figref{fig:hierarchy}(d), one allows for the control operations
to be \emph{arbitrarily fast} and thus assumes the steps of the protocol to last an \emph{infinitesimal} 
duration $\d t$. Hence, one deals then with a sequential strategy of infinitesimal step duration, for which 
the sequential bound of \eqnref{eq:CE_seq} can be directly applied after setting $k\!\to\!\frac{t'}{\d t}$ and taking the limit 
$\delta t\!\equiv\!\d t\!\to\!0$ (see \appref{app:2}). 
In order for the bound to remain meaningful, we adjust the derivation of~\refcite{Do14} in \appref{app:2},
so that it now reads:
\bea
&&\QFI\Big( \varrho^\t{(d)}_{t'/\d t}(t')\Big)\leq4 \frac{t'}{\d t} || \alpha(\d t) ||
\label{eq:CE_inf}\\
&& \;\;+4\, \Big(\frac{t'}{\d t}\Big)^2 ||\beta(\d t)||\!
\left( \frac{1}{\sqrt{\d t}}||\alpha(\d t)|| \!+\! ||\beta(\d t)|| \!+\!\sqrt{\d t}\right).
\nonumber
\eea
We refer to \eqnref{eq:CE_inf} as the \emph{infinitesimal-timestep CE bound}.

In order to derive the ultimate precision bounds valid in the presence of noise and FFQC,
we compute the infinitesimal-timestep CE bound \eref{eq:CE_inf} after substituting the expressions for $\alpha$ and $\beta$
given by \eqnsref{eq:alphaM}{eq:betaM}. 
We explicitly minimize \eqnref{eq:CE_inf} over Kraus representations as described in \secref{sec:bound_opt}, by
expanding the Kraus operators, $\v{K}(\d t)$, and the Hermitian matrix, $\m{h}(\d t)$, up to small orders of $\sqrt{\d t}$. 
As we are interested in the regime of \emph{asymptotic resources}, i.e., the limit
$t'\!\gg\!\d t$ (or equivalently $Nt\!\gg\!\delta t$ in case of the scheme (c)), we may assume without loss of generality that it is always optimal
to make the second term in \eqnref{eq:CE_inf} vanish by setting $\beta(\d t)\!=\!0$. We outline the methods we use below, 
with detailed calculations deferred to 
\appref{app:3}.

Consider the dynamical matrix $\m S$, defined in \eqnref{eq:Srep}, of the infinitesimal time channel $\cE_{\omega,\d t}$. Expanding it  in $\d t$ around $t\!=\!0$ gives $\m{S}(\d t)\!:=\!\m{S}^{(0)}\!+\!\m{S}^{(1)}\d t\!+\!O(\d t^2)$.
The evolution of the probe after an infinitesimally short time  is then given by
\bea
\cE_{\omega,\d t}(\rho) 
&=& 
\sum_{\mu,\nu=0}^3 \m{S}_{\mu\nu}(\d t)\; \sigma_\mu\, \rho\, \sigma_\nu. \\
&=&
\sum_{\mu,\nu=0}^3 (\m{S}^{(0)}_{\mu\nu}+\m{S}^{(1)}_{\mu\nu}\d t)\, \sigma_\mu\, \rho\, \sigma_\nu + \bigO(\d t^2).
\nonumber
\eea
The above expansion is unambiguously specified 
up to $\bigO(\d t^2)$ by the master equation in \eqnref{eq:ME}
\be
\cE_{\omega,\d t}(\rho) = \rho +\left(-\ii \frac{\omega}{2}[\sthree,  \rho]+\cL( \rho)\right)\d t + \bigO(\d t^2),
\ee
so we may directly relate the short-time expansion of $\m S$ to the Liouvillian 
of \eqnref{eq:L} without explicitly integrating \eqnref{eq:ME}:
\al{
\m{S}^{(0)}&= \mathrm{diag}(1,0,0,0), \label{eq:S(0)}\\
\m{S}^{(1)}&=
{\small \left(\begin{array}{c|ccc}
-\sum_{i=1}^3 \bar{\m{L}}_{ii} & \im\bar{\m{L}}_{23} & \im\bar{\m{L}}_{31} & \im\bar{\m{L}}_{12}+ \ii\frac{\omega}{2}\\ \hline
 \im\bar{\m{L}}_{23}& & &\\
 \im\bar{\m{L}}_{31}& & \bar{\m{ L}} &\\ 
 \im\bar{\m{L}}_{12} - \ii\frac{\omega}{2}
\end{array}\right)}\!, 
\label{eq:S(1)}
}
where $\bar{\m{L}}$ is just the restriction of the Liouvillian $\m{L}$ 
\eref{eq:L} to the subspace spanned by the Pauli operators.
We remark that in the limit $\d t\! \to\!0$ 
these are the only meaningful terms, as higher orders in the expansion do not affect the 
final state (which is precisely the reason why the master equation formalism is valid). From \eqnsref{eq:S(0)}{eq:S(1)} we directly obtain the the expansion of $\m M$ in $\sqrt{\d t}$, using the convenient choice $\m M\!=\!\m M^\dag\!=\!\sqrt{\m S}^T$.

Next, we expand the infinitesimal-timestep CE bound \eref{eq:CE_inf}
in orders of $\sqrt{\d t}$ by defining the expansions for operators
$\alpha(\d t)\!:=\!\sum_{\ell\geq 0} \alpha^{(\ell)} (\d t)^{\ell/2}$ and $\beta(\d t)$ (similarly).
The definitions \eqnsref{eq:alphaM}{eq:betaM} of $\alpha(\d t)$ and $\beta(\d t)$ allow one to relate their expansions  to the ones of the matrix $\m{M}$ (fixed by the Liouvillian via \eqnsref{eq:S(0)}{eq:S(1)}) and the Hermitian matrix $\m{h}$ (which is still free) order by order. In the last step one optimizes the matrix $\m{h}$ in order to obtain the tightest infinitesimal-timestep CE bound \eref{eq:CE_inf} in the limit $\d t\!\to\!0$.

%

We include the full details in \appref{app:2}, yet let us summarize here that a non-trivial bound in \eqnref{eq:CE_inf} 
can only be obtained if the expansion terms satisfy $\alpha^{(0)}\!=\!\alpha^{(1)}\!=\!\beta^{(0)}\!=\!\beta^{(1)}\!=\!0$, 
these condition are always satisfied if the low-order terms in the expansion of the matrix $\m h$ are constrained. If it is possible to choose $\m h$ such that $\beta^{(2)}=\beta^{(3)}=0$ also hold then the infinitesimal-timestep CE bound \eqref{eq:CE_inf}  
becomes
\be
\QFI\Big( \varrho^\t{(d)}_{t'/\d t}(t')\Big)
\leq
4 ||\alpha^{(2)}||\,t'
+(t')^2 O(\sqrt{dt}).
\ee
bounding the QFI to a linear scaling in $t'$ in the limit $\d t\!\to\! 0$ with $k=t'/\d t$. Note that $\beta^{(3)}$ can always be set to zero without affecting lower order terms, while the existence of a solution for $\beta^{(2)}=0$ depends on the noise, in particular it is impossible for rank-one Pauli noise as implied by the result of \secref{sec:rank1pauli}. If there exists a matrix $\m h$ such that $\alpha^{(0)}\!=\!\alpha^{(1)}\!=\!\beta^{(0)}\!=\!\beta^{(1)}\!=\!\beta^{(2)}\!=\!0$ we obtain 
\begin{align}   
\QFI\Big( \varrho^\t{(d)}_{t'/\d t}(t')\Big)
&\underset{\d t \to 0}{\leq}
4 \,\alpha_\cL\, t' \label{eq:CE_inf_lin} \qquad\t{with}\\
\alpha_\cL & := \min_{\m h \,\,\t{such that}\, \alpha^{(i)}=\beta^{(j)}=0} ||\alpha^{(2)}||,
\label{eq: alphaL}
\end{align}
for $i\leq 1$ and $j\leq 3$.

Note that as the sequential scheme with FFQC is the most powerful one (see \secref{sec:hierarchy}), 
\eqnref{eq:CE_inf_lin} holds for all protocols  depicted in \figref{fig:hierarchy} after accordingly setting $t'\!=\!Nt$.
In particular, \eqnref{eq:CE_inf_lin} also applies in case of control with finite time-step $\delta t$ (scheme (c)) and
in the absence of quantum control when $\delta t\!=\!t$ (schemes (a) and (b)).
Finally, let us remark that although in the case of parallel strategies with ancillae of \figref{fig:hierarchy}(b)
it is the parallel CE bound \eref{eq:CE_par} that must yield tighter limits on precision for finite 
$t$, one may show (following exact argumentation as in the previous paragraphs) that the linearly scaling 
parallel CE bound must always converge to its infinitesimal-timestep equivalent of \eqnref{eq:CE_inf_lin}
when fixing $t'\!=\!Nt$ and considering the $t\!\to\!0$ limit, in which the parallel protocol becomes infinitesimally 
short but involves an infinite number of probes.

\subsubsection{Universal asymptotic linear bound}
\label{sec:univ_bound}
Thanks to the infinitesimal-timestep CE bound \eref{eq:CE_inf}
yielding \eqnref{eq:CE_inf_lin}, 
we may formulate a general observation about the scaling of the QFI
applicable to any of the schemes depicted in \figref{fig:hierarchy}:

\begin{obs}
\label{obs:qfi_lin}%
For all noise processes described by \eqnref{eq:L} 
except rank-one Pauli noise-types that are not parallel to 
the parameter-encoding Hamiltonian, the QFI in any scheme
of \figref{fig:hierarchy} is upper-bounded by 
\be
\QFI_\mathcal{L}\leq 4 \,\alpha_\mathcal{L}\, t',
\label{eq:generalboundqfi}
\ee
where $\alpha_\mathcal{L}$ is a constant that solely depends on the 
particular form of the Liouvillian $\mathcal{L}$ defined in \eqnref{eq:L},
while $t'$ is the effective protocol time as defined via the most powerful
scheme (d) with FFQC in \figref{fig:hierarchy}.
\end{obs}

We explicitly prove \obsref{obs:qfi_lin} in \appref{app:3}, but outline the derivation here.
In \appref{app:3.1}, for any qubit Liouvillian 
$\mathcal{L}$ of \eqnref{eq:L}, except the rank-one Pauli noise of \eqref{eq:rank1pauli} with $\sigma_{\bf n} \neq \sigma_3$, we give the form of the hermitian matrix $\m h$ for which $\alpha^{(0)}\!=\!\alpha^{(1)}\!=\!\beta^{(0)}\!=\!\beta^{(1)}\!=\!\beta^{(2)}\! =\!\beta^{(3)}\!=\!0$. Hence, there exists some finite $\alpha_\cL$ for which  \eref{eq:CE_inf_lin} holds.
We first show that dephasing noise, as well as any rank-one noise which is not of Pauli type, 
imposes a linear scaling of the QFI due to \eqnref{eq:CE_inf_lin}. Next, we consider noise processes described by 
Liouvillians whose rank is strictly greater than one.  Any such noise process corresponds to a matrix representation of the Lindblad superoperator \eref{eq:L} 
that may be written as a direct sum of matrices of lower rank. We show that the QFI of noise processes of 
rank greater than one is always smaller than the QFI of any of its orthogonal components. Thus, the only rank-two noise 
processes that, in principle, could still allow for quadratic scaling of the QFI are the ones corresponding to a direct sum of 
two orthogonal rank-one Pauli noises.  
For this case, however, we explicitly show that the QFI scales linearly with $t'$. 
Moreover, this also settles the case of the rank-three noise types as any such noise can be expressed as a direct sum that 
contains either a rank-one non-Pauli or a rank-two Pauli noise, or both.
Finally, in \appref{app:5} we derive explicit forms of $\alpha_\cL$ for the exemplary noise-types
discussed below.

\subsubsection{Exemplary noise-types}
\label{sec:examples}
Although we have shown the validity of \eqnref{eq:generalboundqfi} for most types of noise processes, it still remains to 
compute $\alpha_{\cL}$ for a general Liouvillian (see \eqnref{eq:L}) by adequately minimizing \eqnref{eq: alphaL}.  In 
\appref{app:5} we perform such minimization in a completely analytic manner. In particular, we minimize
$||\alpha^{(2)}||$ in \eqnref{eq:CE_inf_lin} over Hermitian matrices $\m{h}$ (under the constraint of $\beta^{(2)}\!=\!0$ and 
$\alpha^{(\ell)}\!=\!\beta^{(\ell)}\!=\!0$ for $\ell\!\le\!1$) for all the exemplary Liouvillians stated in \secref{sec:noise_models}.
The corresponding CE bounds for dephasing, general rank-one, and rank-two Pauli noise types respectively read
\begin{subequations}
\label{eq:exemplary_Fbounds}
\al{
\label{dephasisg bound}
\QFI^\t{1P}_\v{z}&\leq \frac{t'}{2\gamma}\\
\QFI^\t{1G}_{\v{r}}&\leq \frac{t'}{4}\,\frac{\max\{|x + \ii y|,|x - \ii y|\}^2}{(|\re(\v{r})|^2 |\im(\v{r})|^2 -(\re(\v{r})^\T \im(\v{r}))^2 )}
\label{eq:qfi_rank1gen}\\
\QFI^\t{2P}_{\bf \Omega}&\leq \frac{t'}{2 \gamma_1 \gamma_2}\Big( c_\theta^2 (\gamma_1+\gamma_2) + s_\theta^2 
(\gamma_1 s_\varphi^2 + \gamma_2 c_\varphi^2) \Big) \label{eq:rank2pauli}\\
& \implies\; \QFI_\textrm{X-Y}\leq \frac{t'}{2 \gamma \, p (1-p)} \label{eq:Fb_xynoise}
}
\end{subequations}
We note that in the case of general rank-one noise, yielding $\QFI^\t{1G}_{\v{r}}$ in \eqnref{eq:qfi_rank1gen}, we analytically 
find the optimal $\m{h}$ for the cases where one of the components of $\v{r}$ ($x$, $y$, or $z$) vanishes (see \appref{app:5}). 
This allows us to use the structure of the optimal $\m{h}$ as an ansatz for the case of general rank-one noise $\cL^\t{1G}$. 
However, although the minimization of $||\alpha^{(2)}||$ subject to the constraint $\beta^{(2)}=0$ can be done for any fixed 
Liouvillian, it generally contains a large number of parameters to be optimized. Hence, we derive a valid, but not provably tightest, analytical bound for the case of general noise.

\subsection{The gain allowed by fast control}
\label{sec:comparison}
We now discuss the impact of fast control, in particular the gap 
in the achievable QFI between the scheme (b) and the schemes with FFQC, i.e., 
(c) and (d) (with $\delta t \!\to\! 0$ throughout this section) in 
\figref{fig:hierarchy}. To do the comparison we focus on the example of 
the X-Y noise $\cL_\t{X-Y}$ in \eqnref{eq:xynoise},  and  the rank-one transversal 
Pauli noise $\cL_\t{\v x}^\t{1P}$ in \eqnref{eq:rank1pauli}  with $\sigma_{\v n}=\sone$, 
which is also the limiting case of  X-Y noise for $p=1$. For  X-Y noise the master equation \eqref{eq:ME} 
can be analytically solved, hence the parallel CE bound \eqnref{eq:CE_par} 
allows one to upper-bound the QFI  of the parallel scheme $\QFI^\t{(b)}(t)$ (see \appref{app:4}).

\subsubsection{Transversal rank-one Pauli noise}

We begin with the rank-one transversal Pauli noise $\cL_\t{\v x}^\t{1P}\!=\!\cL_\t{X-Y}|_{p=1}$. 
From \obsref{obs:rank_1_pauli} and from previous results~\cite{Dur:14, Kessler:14, Arrad:14, Ozeri:13}, 
we know that this noise can be completely removed by FFQC without harming 
the evolution. Consequently, the strategies with FFQC attain the QFI given 
by $\QFI^\t{(d)}(t')\!=\!(t')^2$ and $\QFI^\t{(c)}(t)\!=\!(N t)^2$ respectively. Remarkably, 
in order to attain such QFI within (d) one only requires one qubit and one ancilla in parallel:~this 
is enough to implement the FFQC strategy of \secref{sec:rank1pauli} and get rid of the noise, and it is known that in the noiseless case a sequential single qubit strategy is optimal~\cite{Giovannetti2006}. For the parallel strategy (b) the CE bound reads $\QFI^\t{(b)}\!\leq\!4 N  \alpha_{\omega, \gamma, p=1}^\t{(b)}(t)$ 
(see \appref{app:4}). We plot $4 \, \alpha_{\omega,\gamma,p}^\t{(b)}$ 
for $p\!=\!\omega\!=\!\gamma\!=\!1$ in \figref{FFQCgain} (thin solid line). 
This clearly shows the advantage offered by FFQC: for any fixed $t$ in strategy (b) the attainable $\QFI^\t{(b)}$ is bound to a linear scaling in $N$. Moreover, for any $\gamma$ and $\omega$ as $t$ increases the bound  eventually starts to decrease (actually it has been shown that $\QFI^\t{(b)}(t)/{t}$ is maximized for $t \propto 1/ N^{\frac{1}{3}}$~\cite{Chaves:12}). Interestingly, the bound $\alpha_{\omega,\gamma,0}$ diverges when $t$ approaches zero. This is to be expected, as in this limit a super-linear scaling in $N$ is known to be possible already within scenario (a) \cite{Chaves:12,Brask:15}.

\subsubsection{X-Y noise}
\label{sec:XYnoise}

\begin{figure}[!t]
\vspace{5 mm}
\includegraphics[keepaspectratio,width=\columnwidth]{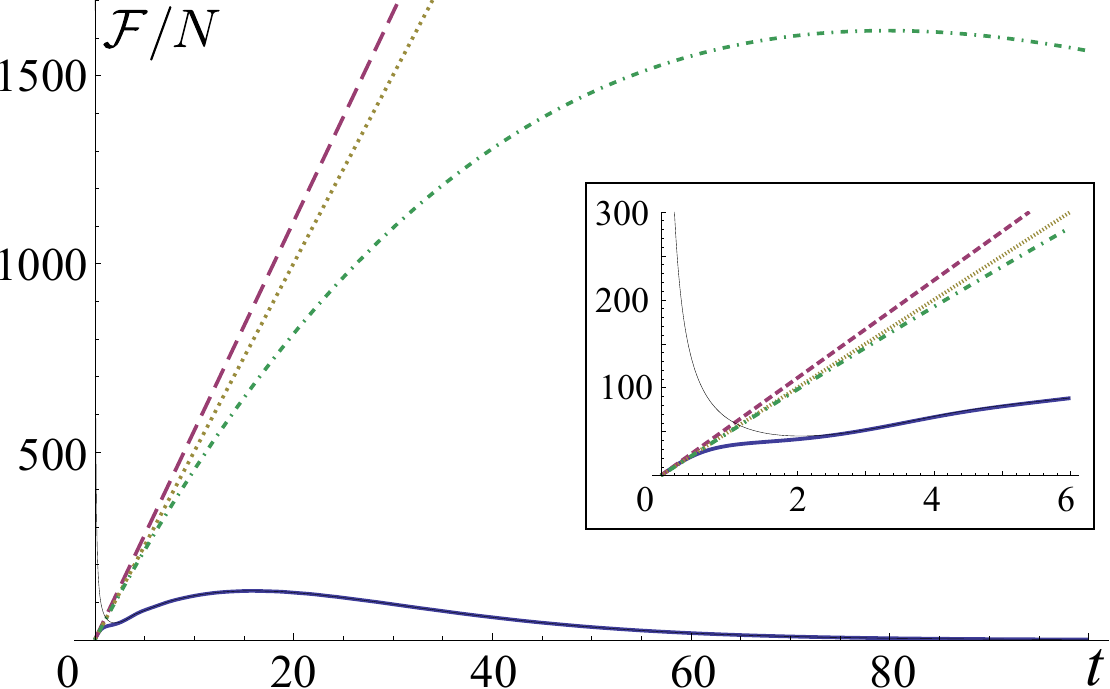}
\caption{X-Y noise: Scaling of the QFI as function of $t$ for various protocols and $\omega =1$, $\gamma=0.1$ and $p=0.1$. The solid line corresponds to the CE bound on $\QFI^\t{(b)}(t)/N$ for strategy (b) without FFQC (the original X-Y noise). The dashed straight is the ultimate upper-bound $\frac{t}{2 \gamma p (1-p)}$ \eqnref{eq:rank2pauli}. The dotted straight line corresponds to the attainable bound for the particular FFQC strategy of \secref{sec:XYnoise} where one corrects the dominant noise but discards the error record, it corresponds to $\frac{t}{2\gamma p}$  from \eqnref{dephasisg bound}. The dot-dashed line is the attainable bound for the same strategy but with fixed number of ancillae equal to $N$ and no intermediate measurements. X-noise: Finally, the thin line is the CE bound on $\QFI^\t{(b)}(t)/N$ for the X-noise with $p\!=\!\gamma\!=\!\omega\!=\!1$, note that it diverges as $t\to0$ since strategy (b) allows to super-classical scaling in $N$ for vanishing $t$. The inset is a zoom on the region close to the origin.}
\label{FFQCgain}
\end{figure}

Now let us turn to X-Y noise $\cL_\t{X-Y}$ (with $0\!<\!p\!<\!1$). 
For the parallel strategy (b), the same bound $\QFI^\t{(b)} \leq 4 N  \alpha_{\omega, \gamma, p}^\t{(b)}(t)$ applies
and the function $ \alpha_{\omega, \gamma, p}^\t{(b)}(t)$ is now well-behaved 
(it starts at zero for $t=0$). 
We plot  $\alpha_{\omega, \gamma, p}^\t{(b)}(t)$ for $\omega\!=\!\gamma\!=\!1$ and $p=0.1$ in \figref{FFQCgain} (thick solid line). 
The situation for FFQC schemes is a bit more subtle, and allows to nicely illustrate different aspects in which fast control is helpful. 
First there is the ultimate bound of \eqnref{eq:rank2pauli} which holds for any FFQC 
strategy and reads $\QFI^\t{(d)}_\t{X-Y}\leq \frac{t'}{2\gamma p(1-p)}$ ($\QFI^\t{(c)}_\t{X-Y}\leq \frac{N t}{2\gamma p(1-p)}$ for scheme (c)) 
but might not be attainable (dashed line in \figref{FFQCgain}).

Now let us consider a particular FFQC strategy, which is described in details 
in \secref{app:XYffqc}. Using the error correction code of \secref{sec:rank1pauli} one can detect if an error 
happens at a given timestep. However, as there is no way to tell if the error corresponds 
to a $\sigma_1$ or $\sigma_2$ the noise cannot be fully corrected. Still, one can chose to 
correct the most probable error term, say $p\leq1/2$ and it is  $\stwo$. Hence, if the guess 
was right the error is corrected $\stwo \stwo =\eins$, while the other case leads to a 
z-error  $\stwo \sone = -\ii \sthree$. This strategy modifies the X-Y noise process to an 
announced dephasing noise (one knows how many correction steps were performed), 
for which the same bound \eqnref{eq:Fb_xynoise} holds as all our control operations are unitaries.

Next, consider a suboptimal variant of the above scheme (we will show that this is 
strictly suboptimal in \secref{sec:lim_res})  where we forget the error register, i.e.,
the knowledge about the number of correction steps. In this case the resulting effective 
noise process is the usual dephasing ($\cL_{\v z}^\t{1P}$ of \eqnref{eq:rank1pauli}) 
with strength $\gamma_{\v z} = p\, \gamma$, and the bounds $\QFI^\t{(d)}\leq\frac{t'}{2 \gamma p}$ 
and $\QFI^\t{(c)}\leq\frac{t N}{2  \gamma p}$ of \eqnref{dephasisg bound} apply (dotted line in \figref{FFQCgain}). 
Notice that these bounds are attainable; it is known that the CE bound for dephasing may be 
asymptotically achieved already within strategy (a) when considering the limit of short dynamics $t \!\to\! 0$  
and large number of probes $N\!\to\! \infty$~\cite{Ulam2001,Kn11,Knysh:14}. 
Crucially, when considering the more powerful strategies (c) and (d), such a regime can always be mimicked 
by employing an unbounded number of ancillary qubits. These are then continuously swapped with the sensing probes and measured 
so that the required protocol of type (a) is indeed recovered.

A natural question to ask then is what happens if we keep the number of ancillary qubits constant, and equal 
to $N$ in (c) or (d). In this case, the FFQC strategy described above allows one to mimic 
the scheme (a) with dephasing noise $\cL_{\v z}^\t{1P}$ of strength $\gamma_{\v z} = p\, \gamma$ 
and a fixed number of probes $N$. For such an effective noise-type, the parallel 
CE bound is known to yield $\QFI^\t{(a)}(t)\leq  N\frac{t^2 \eta^2}{(1-\eta^2)}$ with 
$\eta=\e^{-p \gamma  t}$ \cite{Kolodynski:13}, which is attainable when $N$ is large 
(dot-dashed line in \figref{FFQCgain}). Hence, it is also attainable by strategy (c) with 
the number of ancillary qubits equal the number of probes $N$, or by strategy (d) 
with $N+1$ qubits in total (only one qubit is needed as an ancilla to run the error correction protocol, 
while the others can be used for consecutive preparation of the required initial entangled states). 

\subsubsection{Advantages of FFQC}

In summary, we see that FFCQ offers several advantages. First, it allows to modify the noise process particularly rank-one transversal Pauli noise which can be completely removed, while X-Y noise can be transformed to a weaker dephasing noise (dot-dashed line in \figref{FFQCgain}). In addition to this, if the number of available ancillary qubits is large FFQC offers the possibility to prolong the short-time dynamics by ``parallelizing the evolution'' (dotted line in \figref{FFQCgain})---mimicking a parallel evolution with short running time but large $N$ and the modified noise.  Finally, it also gives the possibility to keep the register of errors that happened during the evolution (dashed line in \figref{FFQCgain}).  To get a feeling of how the latter improves the QFI, think of the frequentist scenario where for the construction of the final estimator one values more the runs where less or no errors happened (we explore this in more depth in \secref{sec:lim_res}). In any case, FFQC undoubtedly outperforms the parallel scheme (b) (solid line in \figref{FFQCgain}) by a large amount.

We showed how the introduction of FFQC leads to an improved scaling of the QFI. Yet an interesting question is if there exists a gap between the two strategies (c) and (d). If the frequency with which one applies the controls in (c) is limited, i.e. the duration of each step $\delta t$ is fixed, it is easy to see that there exists a gap. The bound derived in the \appref{app:4} also applies to this case and yields $\QFI^\t{(c)}\leq 4 N m \, \alpha^\t{(b)}_{\omega,\gamma,p}(\delta t)$, and we just demonstrated that it can be outperformed by FFQC. However, it is not clear if there is a gap between (c) and (d) when $\delta t \!\to\! 0$ and $t=t'/N$. Furthermore, it is not clear  these strategies outperform scheme (b) for $t\!\to\! 0$  with $N=\frac{t'}{t}$, as all the CE bounds coincide in this limit. On the other hand, the fact that the CE bounds coincide does not disprove the existence of a gap, as none of the bounds is necessarily tight. Finally, we note that  protocol (d)  has following important advantage: one can read the error syndrome at any time-step before deciding on the optimal strategy for the future, whereas for protocols (b) and (c) some decisions have to be made beforehand. We will come back to this particular point, and show for a simple phase estimation example how this difference allows to boost the attainable QFI  in \secref{sec:phase}.

\section{Implications on attainable precision in metrology schemes}
\label{sec:implications} 

Hitherto our analysis focused on the scaling of the QFI evaluated on the final 
system state of the metrology protocols depicted in \figref{fig:hierarchy} in presence of
general time-homogeneous Liouvillian noise. We now discuss the implications of 
FFQC on the precision with which one can estimate the parameter $\omega$.  
As already mentioned in \secref{sec:par_est},  how to 
best quantify the estimation precision depends crucially on how information about $\omega$ is obtained.  
In \secref{sec:frequentist} we consider the implications of FFQC 
on precision for the case of asymptotically many repetitions, whereas \secref{sec:bayesian} deals with single-shot estimation. 
 Regardless of how one chooses to quantify the precision, whenever a noiseless estimation scenario exhibits a 
quantum improvement in precision scaling then this improvement may always be maintained in the presence of rank-one Pauli noise by employing the FFQC protocol described in \secref{sec:rank1pauli}.   We note again that the ability of our scheme to ameliorate for any rank-one Pauli noise finds applications beyond metrology, i.e., in the design of high fidelity gates, or indeed any other protocol where such noise terms may appear.

The achievable precision in estimating $\omega$ in any metrological protocol is dictated by the resources at hand.  
In what follows we shall consider two different ways of quantifying the resources of any metrological protocol:~%
time-particles
and number of probes.  In the \emph{time-particles} approach the total resource \emph{per experimental run} is defined 
as the product between the number of probes $N$ and the protocol duration $t$. Thus, for all the types of protocols depicted 
\figref{fig:hierarchy}  the time-particle resource is equal to $t'=Nt$.
In the \emph{number of probes} approach one is primarily interested in how estimation precision scales with $N$ only.  
In particular, this is how resources are defined in \emph{frequency estimation}~\cite{Huelga:97, Escher:11, Kolodynski:13, Chaves:12}, 
where the time for each experimental run, $t$, is bounded from above so that one can consistently consider the limit of many 
experimental repetitions by letting the overall experiment last sufficiently long.
In all cases we show how our results yield noise dependent upper bounds on the attainable precision and restrict it to a classical scaling in terms of the resource.

\subsection{Precision in the presence of free repetitions}
\label{sec:frequentist}
We now consider precision bounds for estimating parameter $\omega$ using FFQC for the case where the protocol is repeated 
asymptotically many times so that CRB is applicable.  We first consider the time-particles picture
before moving onto the scenario in which the number of probes $N$ is the resource.    

\subsubsection{Time-particles}
Given $\nu$ repetitions of the protocol and using the CRB, we bound the ultimate MSE attainable by the most general FFQC 
strategy in the presence of rank-one Pauli noise  as 
\bea
\delta \omega^2\,\nu &\geq& \frac{1}{(1- (\v{n}^\T\v{z})^2)}\,\frac{1}{{t'}^2}.
\label{eq:timeparts_pauli1}
\eea
On the other hand, stemming from \eqnref{eq:generalboundqfi} the MSE for all other noise processes may be lower-bounded 
as follows
\bea
\delta \omega^2 \,\nu &\geq& \frac{1}{4 \alpha_\mathcal{L}}\,\frac{1}{t'}.
\label{eq:timeparts_nogo}
\eea
Yet, in contrast to \eqnref{eq:timeparts_pauli1}, \eqnref{eq:timeparts_nogo} is not guaranteed to be achievable in the limit 
$\nu\to\infty$.

\subsubsection{Frequency estimation}
We now turn to the setting of frequency estimation in which 
the resource of interest is the number of probes $N$. 
Notice that the FFQC protocol in this case is the one depicted in \figref{fig:hierarchy}(c). Yet, its QFI 
can always be upper-bounded by that of scheme (d) after setting $t'\!=\!Nt$. In order to ensure a large number of 
repetitions we fix the time of a single run $t$ and demand that the total time $T\!:=\!\nu t\gg t$.  Hence, we can 
again use the CRBs of \eqnsref{eq:timeparts_pauli1}{eq:timeparts_nogo} and obtain the precision bounds:
\bea 
\delta \omega^2\, T &\geq& \frac{1}{ (1- (\v{n}^T \v{z})^2) } \frac{1}{N^2 t}, \label{eq:freq_pauli1}\\
\delta \omega^2\, T &\geq& \frac{1}{ 4 \alpha_{\mathcal{L}}} \frac{1}{N} \label{eq:freq_nogo}.
\eea
Indeed, \eqnref{eq:freq_pauli1} proves that HS is achievable for rank-one Pauli noises.  On the other hand, 
for all other noise processes, thanks to \eqnref{eq:generalboundqfi} being linear in $t'$, the bound of \eqnref{eq:freq_nogo}
is independent of the single-run duration $t$.  This suggests that FFQC allows us to indefinitely maintain the ultimate precision 
normally exhibited only at very short time scales (see \secref{sec:comparison}).

\subsection{Single-shot precision bounds}
\label{sec:bayesian}
We now consider precision bounds applicable in the single-shot scenario, i.e., when $\nu\!=\!1$.  As a result we are forced to consider 
the \aMSE as the figure of merit, where the average is with respect to the probability distribution
$p_0(\omega)$ describing our prior knowledge about the parameter.  The \aMSE can then be lower-bounded  with the help of 
BCRB and ZZB (see \secref{sec:par_est}).  However, note that  the FFQC-assisted sequential strategies of \figref{fig:hierarchy}(d) with 
sufficiently large $t'$ also incorporate all many repetition protocols. This is because one can always set the intermediate 
control gates to implement measurements and state re-preparations.  Hence, we expect all the single-shot bounds to reproduce 
the ones derived in the regime of many repetitions when considering $t'\to\infty$.

\subsubsection{Pauli rank-one noise}
As by repeating we may only improve the precision, whenever the precision is forced to follow the standard scaling in the many-repetitions regime,
it must be standard scaling-bounded in case of the single-shot estimation. Thus, it is most important to ask whether the HS can still be 
attained in presence of rank-one Pauli noise within strategies (c) and (d) with only a single run. In particular any protocol that achieves the maximum attainable precision in the absence of noise can maintain this precision in the presence of rank-one Pauli noise by utilizing FFQC.  For example, by 
incorporating FFQC in the protocol in~\cite{Sergeevich2011} one can maintain its 
optimal performance for any rank-one Pauli noise process; the number of 
measurements $N$ in~\cite{Sergeevich2011} correspond to the control-steps $k$ in our protocol (d).
In \appref{app:7}, we provide an explicit strategy which for a uniform prior---a prior distribution that is 
constant over a finite interval, and zero elsewhere--- leads to the \aMSE satisfying
\bea
\langle\delta \omega^2\rangle_\textrm{FFQC} \leq \frac{\pi^4 \kappa}{4(1- (\v{n}^\T\v{z})^2)} \, \frac{1}{t'^2}
\label{eq:bayesianbound}
\eea 
for (d) and (c) (after substituting  $t'=Nt$), where $\kappa=6.74$. This shows that, at least for this prior, quadratic scaling of precision in the 
\emph{total resource} ${t'}^2$ (or $(Nt)^2$) is attainable by a single shot strategy. In the context of frequency estimation 
of \secref{sec:frequentist}, this demonstrates that if one can decide whether to divide the total time $T$ in  many runs of 
a fixed duration $t$ or perform a single shot strategy with FFQC, one should choose the latter as it achieves precision that 
scales quadratically $\langle \delta \omega^2 \rangle \sim 1/(NT)^2$ in $N$ and $T$, and not only in $N$, as in \eqnref{eq:freq_pauli1}.

\subsubsection{Bayesian Cram\'{e}r-Rao bound} 
\label{BCRB}
The BCRB is defined as~\cite{Gill:95} 
\be
\langle\delta \omega^2\rangle
\geq 
\frac{1}{\langle\QFI_\mathcal{L}\rangle + F\big(p_0(\omega)\big)},
\ee 
where $F\big(p_0(\omega)\big)\!=\!\langle (\partial_{\omega}\log p_0(\omega))^2\rangle$ is 
the classical Fisher information of the prior distribution and $\langle\dots\rangle$ denotes averaging with respect to the prior 
distribution $p_0(\omega)$.  Hence, using \eqnref{eq:generalboundqfi}, we obtain the following lower bound on the \aMSE:
\be
\label{eq:BCRB}
\langle\delta \omega^2\rangle
\geq\frac{1}{4 \alpha_\mathcal{L} t'+F\big(p_0(\omega)\big)}.
\ee
Notice that if $F\big(p_0(\omega)\big)$ is finite, by substituting $t'\!\to\!\nu t'$ above we adequately recover   
the many-repetition bound of \eqnref{eq:timeparts_nogo}. 

\subsubsection{Ziv-Zakai bound} 
A more general bound that also allows for irregular priors is the
ZZB introduced in~\cite{Tsang:12}.  In \appref{app:4} we generalize the bound of~\cite{Tsang:12} so that 
it can be applied also to mixed states. As a result we can lower bound the \aMSE for large enough $t'$ by 
\bea
\langle\delta \omega^2\rangle \geq \frac{1}{12 \alpha_\mathcal{L} t'}
\eea
for any prior that fulfills some very mild regularity conditions (see \appref{app:4}). Again, by substituting $t'\to\nu t'$ 
we adequately recover the many-repetition bound of \eqnref{eq:timeparts_nogo}.

\section{Importance of quantum control for metrology with limited resources}
\label{sec:lim_res} 

In this section we further analyse the benefits of continuous quantum control for metrology in presence of noise that can not be completely removed. In addition, we adopt a practical perspective where the resources available for the implementation of the scheme are severely limited. This allows us to keep the analysis simple, but also makes the proposed protocols easily implementable in real experiments. In \secref{sec:sup_of_seq} to cope with an unbalanced X-Y noise we propose a simple FFQC scheme that only requires one sensing and one ancillary qubit, but can still outperform a parallel scheme with a large number of entangled probes and ancillae (but no continuous quantum control). In \secref{sec:phase} we show that also in the case of phase estimation the possibility to perform intermediate control (applied between the two successive applications of the channel on the probe qubits as shown in \figref{strategies}) leads to an improvement in the attainable QFI.

%
%
\subsection{Frequency estimation with X-Y noise}
\label{sec:sup_of_seq}

\begin{figure*}[!t]
\includegraphics[keepaspectratio,width=\textwidth]{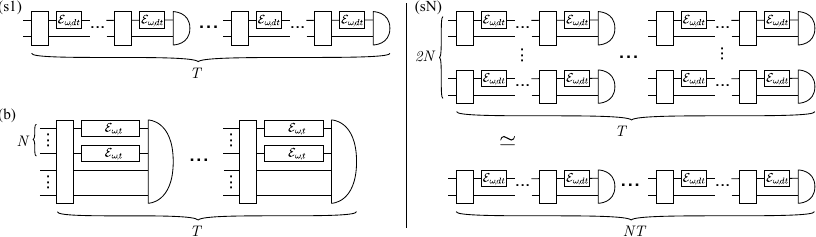}
\caption{(s1): A control-assisted sequential protocol that only involves one probe and one ancillary qubit.
(b): The most general parallel protocol of \figref{fig:hierarchy}(b) that requires to manipulate entangled states of $N$ probe and $N$ ancillary qubits.  
(sN): A control-assisted sequential protocol that employs $N$ independent probe-plus-ancilla entangled pairs running in parallel.}
\label{fig:XY_schemes}
\end{figure*} 

We study the setting of \emph{frequency estimation} in which the probes
sensing the parameter are also affected by the \emph{X-Y noise} specified in \eqnref{eq:xynoise}. 
In accordance to Eq.~(\ref{eq:Fb_xynoise}, \ref{eq:freq_nogo}) the attainable precision for this noise is asymptotically 
constrained to a constant improvement over standard scaling, even when considering the most powerful FFQC scheme of \figref{fig:hierarchy}(d) 
(with the exception of the limiting cases $p\!=\!0$ or $p\!=\!1$, where one of the components vanishes and one 
recovers the correctable perpendicular dephasing noise~\cite{Chaves:12,Dur:14,Kessler:14,Arrad:14,Ozeri:13}). Also in these cases strategies with FFQC can outperform those without yielding a large improvement factor as we showed in \secref{sec:comparison}.
In this section we strengthen this point by demonstrating how the most general parallel scheme
of \figref{fig:hierarchy}(b) with $N$ qubits and $N$ ancillae is outperformed by
a simple and experimentally tangible strategy, which only requires entanglement 
between one sensing and one ancillary qubit but employs fast control. 

More precisely, we compare the most general parallel strategy \figref{fig:XY_schemes}(b)
with simple sequential strategies \figref{fig:XY_schemes}(s1) and \figref{fig:XY_schemes}(sN), where one is only allowed to entangle probes and ancillae two by two rather then manipulate a global entangled state as required by (b). The two strategies (s1) and (sN) are actually the same, but correspond to two different ways of treating the total resources. In (s1) the experimentalist only manipulates one qubit and one ancilla and does the experiment for a total duration $T$. This protocol has the same duration as (b) but uses $N$ times less ``calls of the master equation'', i.e., one effectively applies the single-qubit evolution for a time that is $N$ times shorter as compared to (b). The scheme (sN) is equivalent to (s1) upon replacing $T$ with $N T$, such that it uses the same amount of time-particles as (b). Equivalently (sN) corresponds to running $N$ protocols (s1) simultaneously. In this respect it has the same number of qubits and the same time duration as (b). However, in contrast to (b), the scheme (sN) is granted with fast control on the one hand, but only requires two-qubit entanglement on the other. We demonstrate that, even with such restrictions, for any fixed $N$ both strategies (sN) and (s1) outperform any parallel schemes (b) given that the asymmetry in the X-Y noise  is high enough.

Our results show that the use of control and error correction techniques allow one to attain resolutions with a two-qubit (probe-plus-ancillae) setup
which outperform ones reached when considering systems containing large-scale entanglement. 
We believe that our results may support current state-of-art quantum metrology experiments 
with nitrogen-vacancy (NV) centres, in which the error correction protocols have already been implemented with 
great success \cite{Taminiau2014,Waldherr2014}.
In particular, as the dominant noise is such systems
has been argued to be nearly transversal \cite{Kessler:14,Arrad:14} and, hence, highly asymmetric \cite{Brask:15}, 
our work proves that such systems are indeed capable of attaining resolutions unreachable by the entanglement-based 
schemes employing comparable resources.

\subsubsection{Parallel strategy with N qubits and N ancillae}

A general parallel frequency estimation scenario 
is depicted in \figref{fig:hierarchy}(b) and \figref{fig:XY_schemes}(b). For a regular prior knowledge and sufficiently large $T$ (see \secref{BCRB}), or with the explicit assumption of  sufficiently many repetitions
($T/t\!\gg\!1$),  the ultimate attainable precision is determined by the CRB:
\be
T \delta^2 \omega
\;\geq\; 
\min_t \frac{t}{ \QFI\Big(\varrho_N^\t{(b)}(t)\Big)}.
\label{eq:CRB_freq}
\ee
The right hand side in \eqnref{eq:CRB_freq} is the inverse of the \emph{maximal QFI rate} 
for a given scheme of type (b) employing $N$ sensing qubits.
It is the optimized for the single-run duration that we dub $t_\t{opt}^\t{(b)}$. The optimal time
in general depends on the system size (the probe number $N$) and the form of the noise 
(the Liouvillian in \eqnref{eq:L}) \footnote{Note that in the absence of noise it is optimal to set 
$t\!=\!T$ as large as possible, which makes the CRB \eref{eq:CRB_freq} not applicable.}.
Following the methods described in \secref{sec:no_go}
(see also \appref{app:4}), we can upper-bound the
maximal QFI rate by employing the parallel CE bound \eref{eq:CE_par}:
\bea 
\frac{\QFI\Big(\varrho_N^\t{(b)}(t_\t{opt}^\t{(b)})\Big)}{t_\t{opt}^\t{(b)}} \;\le\;
 N\,\mathfrak{f}^{\t{(b)}\uparrow}_{N}, \qquad \t{where} \qquad \nonumber\\
 \mathfrak{f}^{\t{(b)}\uparrow}_{N}:= 4\max_t\,\min_{\m{h}(t)}\frac{||\alpha(t)||+(N-1)||\beta(t)||^2}{t}
\label{eq:freq_CE_par}
\eea
is the so-obtained upper bound on the \emph{maximal QFI rate per probe}.
We adequately label then as $t_{\t{opt}}^{\t{(b)}\uparrow}$ the optimal $t$ maximising \eqnref{eq:freq_CE_par}.
As a result, we may generally lower-bound the precision dictated 
by the CRB \eqref{eq:CRB_freq} as follows:
\be
\text{(b):}\qquad T \delta^2 \omega \geq
\frac{1}{N \mathfrak{f}^{\t{(b)}\uparrow}_{N}},
\label{eq:freq_par_CRB}
\ee
at the price of the saturability that cannot be guaranteed any more, even in the
$T/t_\t{opt}^{\t{(b)}\uparrow}\!\gg\!1$ limit.

In order to perform the minimization in \eqref{eq:freq_CE_par}, we resort to the SDP-based methods 
of \refcite{Kolodynski:13} reviewed in \secref{sec:bound_opt}. 
To this end we solve the master \eqnref{eq:xynoise} for the X-Y noise, 
and compute the Kraus representation of the resulting channel, 
$\cE_{\omega,t}^\t{X-Y}$, given in \appref{app:4}. For any fixed $N$ and $t$ this allows us to perform 
the minimization over the Hamiltonian $\m h(t)$ numerically,  while for $N\!=\!1$%
\footnote{For which the CE bound \eref{eq:freq_CE_par} is guaranteed to be tight, see \eqnref{eq:CE_bound}.} 
and $N\!\to\!\infty$ we obtain the corresponding expressions analytically for all times. 
Finally, to obtain $\mathfrak{f}^{\t{(b)}\uparrow}_{N}$  we numerically search for the 
optimal  $t_\t{opt}^{\t{(b)}\uparrow}$ in \eqnref{eq:freq_CE_par}. The detailed analysis on the solution of the master equation and derivation of the bounds may be found in \appref{app:4}. Note that the balanced X-Y noise ($p\!=\!1/2$) is phase-covariant, i.e., 
it commutes with the Hamiltonian, so that for this particular case 
the bound \eref{eq:freq_CE_par} can be obtained analytically \cite{Smirne2015}.

\subsubsection{Simple sequential strategy}

\label{sec:Finite_sequential}

For the sequential strategy, all the probe plus ancilla pairs  behave independently, and since the QFI is additive in this case in order to establish 
the global performance it is sufficient to consider a single pair up to the first measurement in \figref{fig:XY_schemes}(s).
We denote the state of a probe-plus-ancilla at any time by $\varrho(t)\!\in\!\cB( \mathds{C}^2\!\otimes\!\mathds{C}^2)$.

The  maximal QFI rate per probe then reads 
\bea
\mathfrak{f}^\t{(s)}\!:=\!\max_t\frac{\QFI(\varrho(t))}{t}=\frac{\QFI(\varrho(t_\t{opt}^\t{(s)}))}{t_\t{opt}^\t{(s)}}
\label{eq:QFI_pp_seq}
\eea
with $t_\t{opt}^\t{(s)}$ being now determined solely by the noise.

The sequential FFQC strategy was already sketched in \secref{sec:XYnoise} and is formally described in \secref{app:XYffqc}. It consists of continuously checking if the probe+ancilla state is in the code $\cH_\t{C}$ or the error subspace $\cH_\t{E}$. At each time step the probability of detecting an error is $\frac{\gamma \d t}{2}$, and if an error is detected the most probable error term, say $\sigma_1$, is canceled by applying the same unitary on the probe $\sigma_1 \sigma_1 =\eins$. This maps the state back into the code space but slightly degrades it with a residual dephasing coming from the other noise term $\sigma_1 \sigma_2 =\ii \sigma_3$.

As one keeps track of the number of errors $m$ that happened during such an FFQC assisted evolution, but disregards the exact times the errors occurred, the final state at time $t$ possesses a direct-sum structure 
\be\label{eq:stateDS}
\bar \varrho(t)\!=\!\bigoplus_{m=0}^\infty p(m;t)\bar \varrho_m(t).
\ee
The probability of errors follows a Poissonian distribution $p(m;t)\!=\!\e^{-\gamma t/2}(\gamma t/2)^m/m!$, while the state $\bar \varrho_m(t)$ conditional to the detection of $m$ errors is given in \eqref{eq:effective stateXY}.

The corresponding QFI may be straightforwardly evaluated:
\be
\QFI(\bar\varrho(t))= \sum_m p(m; t) \QFI\big(\bar \varrho_m(t)\big) 
= 
t^2\,\e^{-\gamma t\,2p(1-p)},
\label{eq:qfi_seq}
\ee
and yields the following maximal QFI rate per probe \eref{eq:QFI_pp_seq}:
\be
\mathfrak{f}^\t{(s)}=
\frac{1}{2\gamma\e}\frac{1}{p(1- p)},
\label{eq:f_seq}
\ee
which is attained after optimally  setting ${t_\t{opt}^\t{(s)}}\!=\!1/(2\gamma p(1-p))$.

In the regime of long experimental duration $T/t_\t{opt}^\t{(s)}\!\gg\!1$ that we are interested in,  
the CRBs \eref{eq:CRB_freq} for the two strategies \figref{fig:X-Y_noise}(s1) 
and \figref{fig:X-Y_noise}(sN) respectively read:
\begin{align}
\text{(s1):}& \qquad T\delta^2 \omega \geq \frac{1}{\mathfrak{f}^\t{(s)}}\\
\text{(sN):}& \qquad T\delta^2 \omega \geq \frac{1}{N \mathfrak{f}^\t{(s)}},
\end{align}
and are guaranteed to be attainable.
  
Before we proceed to the comparison of strategies (s) and (b), 
let us briefly comment on the role of the error register, and how it allows to 
boost the attainable performance. As we argued in \secref{sec:XYnoise} discarding the 
error register is equivalent to effectively modifying the X-Y noise to a dephasing noise of strength $p\, \gamma$. 
Concretely, if the error register is discarded, the state at time $t$ is a mere mixture  
$\bar \varrho(t)\!=\!\sum_{m=0}^\infty p(m;t)\bar \varrho_m(t)$ rather than a direct sum as 
in \eqnref{eq:stateDS}. The corresponding maximal QFI rate per probe is 
then given by $\mathfrak{f}^\t{(s)}\!=\!1/(2 \gamma \e p)$, which is reduced by a factor $1/(1-p)$ as compared 
to the strategy including the error register and described by \eqnref{eq:f_seq}. The intuition behind this gap is rather simple. 
When inferring the value of $\omega$ after a certain number of runs it is helpful to know how noisy each run was. 
This information, contained in the error register, allows to properly ponder the data 
obtained in each run in order to optimally construct the global estimator.

\subsubsection{Comparison of the strategies}

\begin{figure}[!t]
\includegraphics[width=\columnwidth]{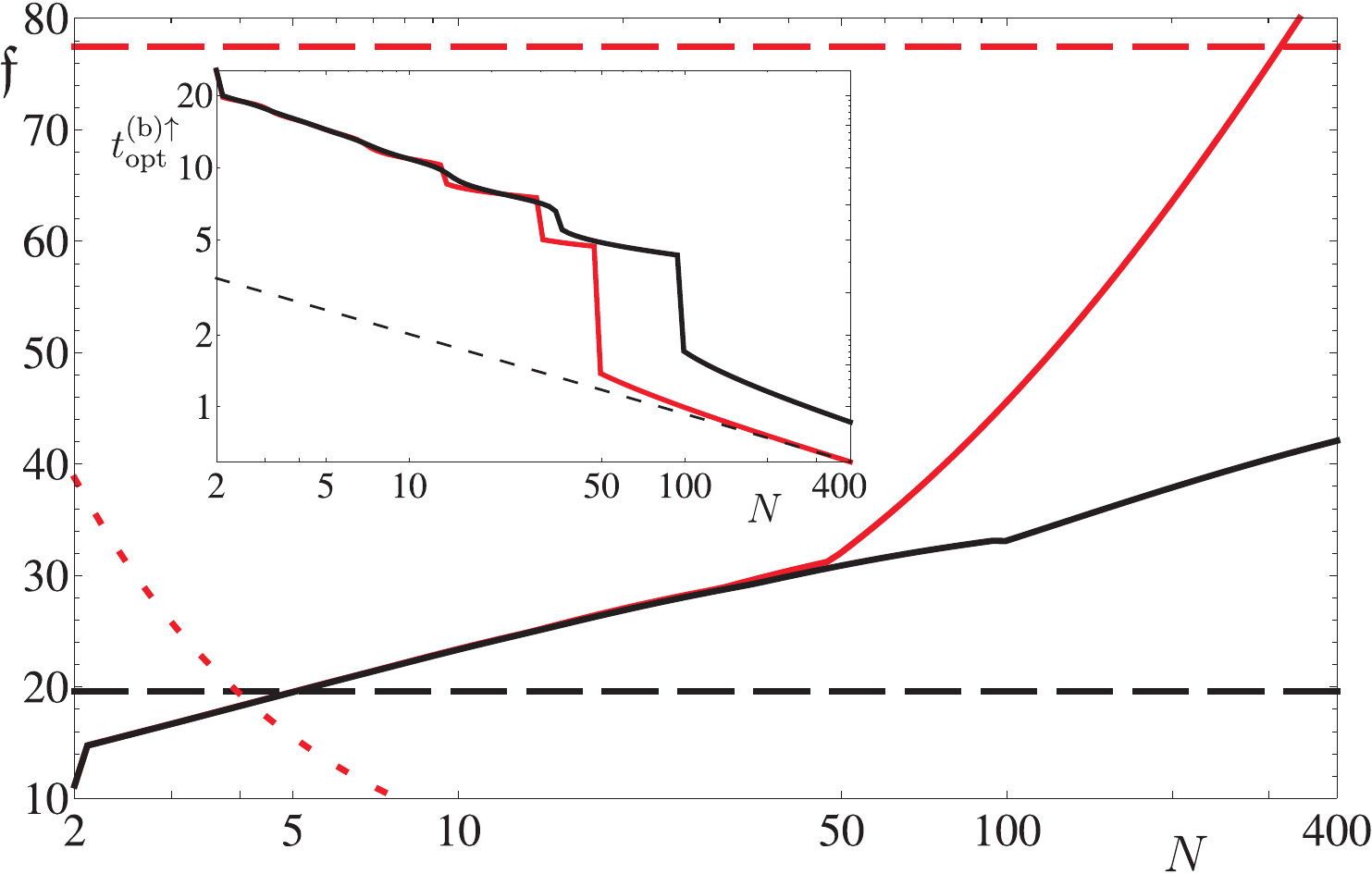}
\caption{%
Maximal QFI rate per probe $\mathfrak{f}^\t{(s)}$ attainable by the sequential strategies 
(\emph{dashed lines}) of \figref{fig:XY_schemes}(s1) and (sN), as compared 
to the upper bound, $\mathfrak{f}^{\t{(b)}\uparrow}_{N}$ (\emph{solid curves}), 
valid for all parallel protocols of \figref{fig:XY_schemes}(b) in presence of 
X-Y noise ($\omega\!=\!1$, $\gamma\!=\!0.05$).
For $p\!=\!0.25$ (\emph{black}) the curves indicate the sequential strategy (sN) to be superior 
for $N\!<5$, whereas for $p\!=\!0.05$ (\emph{red}) up to $N\!<\!316$ (while {(s1)} is superior 
up to $N\!<\!4$---\emph{dotted red curve}).
The inset depicts dependence on the probe number of $t_\t{opt}^{\t{(b)}\uparrow}$ that 
maximizes \eqnref{eq:freq_CE_par}. Despite not being smooth at finite $N$,
$t_\t{opt}^{\t{(b)}\uparrow}$ asymptotically follows $1/N^{1/3}$ scaling (\emph{dashed line}).}
\label{fig:X-Y_noise}
\end{figure}

In \figref{fig:X-Y_noise}, we explicitly present the maximal QFI rate per probe attained by the sequential 
strategy $\mathfrak{f}^\t{(s)}$ of \eqnref{eq:f_seq}, and compare it to the bound limiting 
the performance of any parallel scheme $\mathfrak{f}^{\t{(b)}\uparrow}_N$ of \eqnref{eq:freq_CE_par},
which we plot as a function of the number of probes $N$.
This corresponds to a comparison of strategies (sN) and (b) described in \figref{fig:XY_schemes}. 
In particular, after fixing $\omega\!=\!1$, $\gamma\!=\!0.05$, we show that 
when the noise generated by $\stwo$ is three times more dominant than
the one generated by $\sone$, i.e., $p\!=\!1/4$, the sequential strategy is guaranteed to be 
superior for $N$ less than $5$. However, with the increase of noise asymmetry 
the superiority of the sequential strategy drastically improves, 
so that for $p\!=\!0.05$ only if the probe number $N$ is greater than 
$316$ the parallel schemes (b) can potentially beat the sequential one (sN).
In this case, in order to also compare (b) to the strategy (s1), which employs only
a \emph{single} (rather than $N$) probe-plus-ancilla qubit pair for the same duration $T$, 
we also plot $\mathfrak{f}^\t{(s)}/N$. Strikingly, the parallel strategies (b) are still outperformed
by (s1) as long as $N\!<\!4$.

\begin{figure}[!t]
\includegraphics[width=\columnwidth]{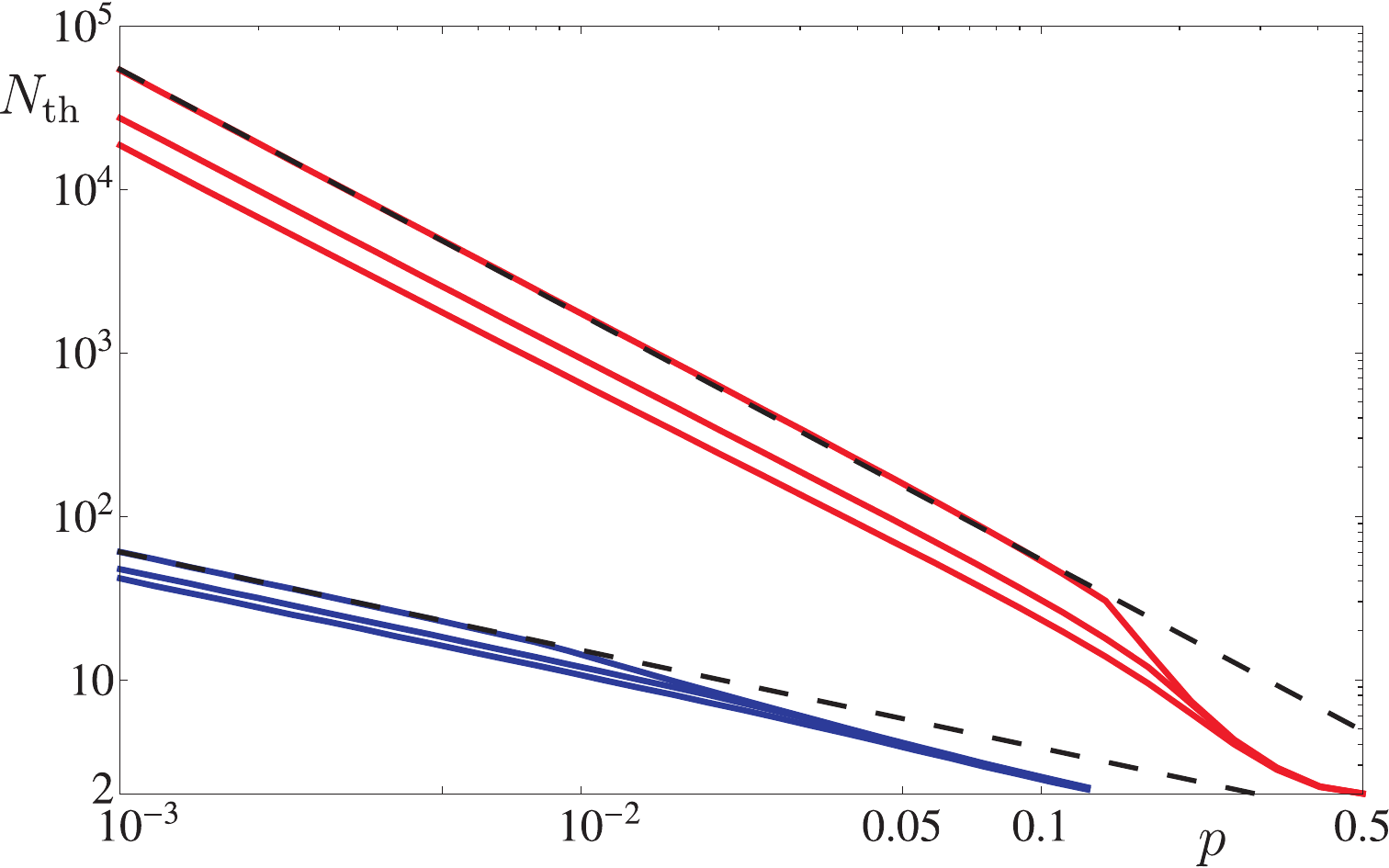}
\caption{%
Threshold numbers of probes, $N_\t{th}^\t{(s1)}$ and $N_\t{th}^\t{(sN)}$, 
below which the two sequential strategies (s1) and (sN) are superior to all parallel protocols (b) in \figref{fig:XY_schemes}.
as a function of the X-Y noise asymmetry $p$ for $\omega\!=\!1$ and $\gamma\!=\!\{0.1,0.2,0.3\}$ (\emph{top to bottom}). 
\emph{Red lines} are plotted assuming sequential protocol performance to be 
evaluated per probe (sN), while \emph{blue lines} for the sequential protocol  employing only a \emph{single} probe-plus-ancilla pair (s1). 
In the former case, $N_\t{th}^\t{(sN)}\!\ge\!2$ with equality always occurring strictly at $p\!=\!1/2$, while in 
the single-pair scenario there exists a range of $0.13\lessapprox\!p\!\le0.5$ for which trivially $N_\t{th}^\t{(s1)}\!=\!1$.
In both cases, $N_\t{th}$ diverges with noise asymmetry, as $p\!\to\!0$, and eventually
follows the $p^{-3/2}$, $p^{-3/5}$ scalings respectively (\emph{dashed lines}).}
\label{fig:Nth}
\end{figure}

To compare the strategies more directly we explicitly present in \figref{fig:Nth} 
the threshold probe numbers below which the sequential strategies are guaranteed to be superior over all parallel schemes.
Concretely, to compare (sN) with (b) we plot $N_\t{th}^\t{(sN)}$ such that 
{
$\mathfrak{f}^\t{(s)}\!>\!\mathfrak{f}^{\t{(b)}\uparrow}_{N}$ for all $N\!<\!N_\t{th}^\t{(sN)}$ (the top curves), and to 
compare (s1) with (b) we plot $N_\t{th}^\t{(s1)}$ such that 
$\mathfrak{f}^\t{(s)}/N\!>\!\mathfrak{f}^{\t{(b)}\uparrow}_{N}$ for all 
$N\!<\!N_\t{th}^\t{(s1)}$ (the bottom curves).} 
Note that $N_\t{th}^\t{(sN)}\!\ge\!2$ for any value of $p$, as in the worst case 
of balanced noise ($p\!=\!1/2$) one may explicitly show that 
{
$\mathfrak{f}^{\t{(b)}\uparrow}_{N=2}\!=\!\mathfrak{f}^\t{(s)}$, see \appref{app:4}.} 
Regarding the other threshold, although $N_\t{th}^{\t{(s1)}}\!=\!1$ in the range $0.13\lessapprox\!p\!\le0.5$, 
meaning that the sequential strategy (s1) only outperforms parallel strategies with a single {probe-plus-ancilla pair}; 
$N_\t{th}^{\t{(s1)}}$ still diverges to infinity when $p\!\to\!0$ and the noise approaches 
the perpendicular dephasing. We observe that the thresholds scale as $N_\t{th}^\t{(sN)}\propto p^{- 3/2}$ 
and $N_\t{th}^\t{(s1)}\propto p^{-3/5}$. This {proves that both sequential strategies
(s1) and (sN) can outperform the most general parallel one of type (b)} for any $N$, given sufficient noise asymmetry.

\subsection{Phase estimation with bit- and bit-phase-flip noise}
\label{sec:phase}
\begin{figure}[!t]
\includegraphics[keepaspectratio, width=\columnwidth]{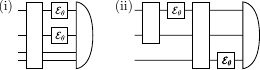}
\caption{Two different strategies for phase estimation involving two applications of the noisy channel $\cE_{\theta}$ presented in \secref{sec:phase}. In strategy (i) the two probe and ancillary qubits are prepared in a suitably entangled state and the channels are applied to the sensing systems in
parallel before they are finally measured. This strategy intrinsically includes error-correction but only at the final step (see
strategy (iii) in~\cite{Do14}). (ii) The sequential strategy from \secref{sec:phase}, after the application of the first channel one reads out the error syndrome and decides on the optimal strategy for the next step  (see
strategy (iv) in~\cite{Do14}).}
\label{strategies}
\end{figure}

We now return to the canonical phase estimation scenario~\cite{Giovannetti2006}, which may yet be interpreted 
as the frequency estimation considered before but with the evolution time $t$ being fixed, such that the parameter of interest is $\theta\!=\!\omega t$. 
In this case the sensing interaction of a single probe of duration $t$ constitutes the elementary block, drawn as a square box in \figref{strategies}. Each box corresponds to a CPTP map $\cE_\theta$ incorporating the noise~\cite{Do14}. 
In this context one is free to use ancillary particles to benefit from error correction techniques,  but also to combine the boxes in different ways.
The two schemes depicted in \figref{strategies}(i) and \figref{strategies}(ii) are the elementary phase estimation equivalents of  the schemes depicted in \figref{fig:hierarchy}(b) and \figref{fig:hierarchy}(d) respectively. 

We provide a simple example 
demonstrating that a sequential strategy outperforms the parallel one.  Specifically, consider the  
situation where each line in \figref{strategies} corresponds to a qubit and the map $\cE_\theta$ is a combination of the unitary evolution 
given by  $U_\theta=e^{-i \frac{\theta}{2}  \sthree}$ and of the noise which consists of a random application of either a bit flip 
($\sone$) or bit-phase flip ($\stwo$) with equal probability $\frac{1-p}{2}$. Accordingly, $\cE_\theta$ has a Kraus representation 
of the form $\{K_0=\sqrt{p} \, U_\theta,\, K_1=\sqrt{1-p}\, \ketbra{1}{0},\,K_2=\sqrt{1-p}\,\ketbra{0}{1} \}$. 
Note that this noise channel corresponds to a Pauli $X-Y$ noise channel which, however, differs from the solution of the master
equation for the balanced $X-Y$ noise described in \secref{sec:sup_of_seq}.

In the parallel strategy of \figref{strategies}(i) one starts with a suitably entangled state of two sensing qubits and two 
ancillae. The channel is applied on both qubits in parallel, and the final state is measured (this includes error-correction at the final 
stage see strategy (iii) in~\cite{Do14}).  Without loss of generality the input two-qubit plus two-ancilla state is given by
$\ket{\psi}= \sum_{i,j=0}^1 a_{ij} \ket{i,j}_Q \ket{i,j}_A$ with $a_{ij}\in\mathbb{R}$. This encoding is optimal as it allows us to 
distinguish between all the different branches of the noisy evolution. Moreover, within every branch the state remains pure, as 
the channel corresponds to either leaving the sensing qubit untouched---with probability $p$---or projecting it onto the 
$\{\ket{0},\,\ket{1}\}$ basis (followed by an irrelevant flip of the qubit) which can be detected by our error correction code. As all 
the branches can be detected, and within every branch the state remains pure, the QFI of the final state is given by the mean 
variance of $H=\frac{1}{2}(\sthree \otimes \eins +\eins \otimes \sthree)$ over all the branches. The mean variance can be straightforwardly maximized by a brute force numerical optimization over the three dimensional 
manifold spanned by the coefficients $a_{ij}$. It turns out that the optimal states are symmetric under permutations of the two 
qubits ($a_{01}=a_{10}$) and, moreover, satisfy $a_{00} = a_{11}$. Using these conditions at the outset one can analytically 
determine the optimal state to be 
\bea\nonumber
\label{optstate}
\ket{\psi}_{\mathrm{opt}}&=\sqrt{\frac{x_\t{opt}}{2}}(\ket{00}_Q\ket{00}_S +\ket{11}_Q\ket{11}_S)\\
&+ \sqrt{\frac{1-x_\t{opt}}{2}}(\ket{01}_Q\ket{01}_S +\ket{10}_Q\ket{10}_S),
\eea
with $x_\t{opt}=\min(1,\frac{2-p}{4 (1-p)} )$. The corresponding maximal QFI is 
\bea \label{eq:QFI_phase_a}
\QFI^\t{(i)} = 
 \left\{  
\begin{array}{cc}
 -\frac{(p-2)^2 p}{2 (p-1)} & p\leq \frac{2}{3} \\
 4 p^2 & \text{otherwise.}
\end{array}\right.
\eea

Now consider the following sequential strategy with intermediate error correction involving two sensing qubits and a single ancilla, see Figure~\ref{strategies}(ii). 
The first sensing qubit and the ancilla are prepared in the state $\frac{1}{\sqrt{2}}\left( \ket{00}_{Q_1 A} + \ket{11}_{Q_1 A}\right)$. After the application of the first box on the first sensing qubit but before the second one we perform our error detection scheme.  If an error is detected the state contains no information about the parameter, hence one discard $Q_1$ and prepare the second qubit and the ancilla in the same state  $\frac{1}{\sqrt{2}}\left( \ket{00}_{Q_2 A} + \ket{11}_{Q_2 A}\right)$. If no error is detected after the first application of the channel the second sensing qubit is introduced and entangled with the first qubit such that the overall  state is given by
\begin{align}\nonumber
\sqrt{\frac{1}{2}}\left( e^{-i\frac{\theta}{2}}  \ket{0}_{Q1} (\sqrt{y}\ket{00}_{Q2, A}+\sqrt{1-y}\ket{11}_{Q2, A})\right.\\
\left.+ e^{i\frac{\theta}{2}}  \ket{1}_{Q1} (\sqrt{y}\ket{11}_{Q2, A}+\sqrt{1-y}\ket{00}_{Q2, A})\right).
\label{statestratc}
\end{align}
One easily checks that the state that optimizes the final QFI (after the application of the second box) is the one with $y= y_{opt}=\min(1,\frac{1}{2(1-p)})$.  Finally the overall QFI of this scheme is given by  
\begin{equation}
\QFI^\t{(ii)}=
 \left \{
\begin{array}{cc}
 p (3 p+1) & p\geq \frac{1}{2} \\
 \frac{p \left(p^2-2 p+2\right)}{1-p} & \text{otherwise.} 
\end{array}\right.
\label{eq:QFI_phase_b}
\end{equation}

\begin{figure}[!t]
\vspace{5 mm}
\centering
\includegraphics[keepaspectratio,width=0.95\columnwidth]{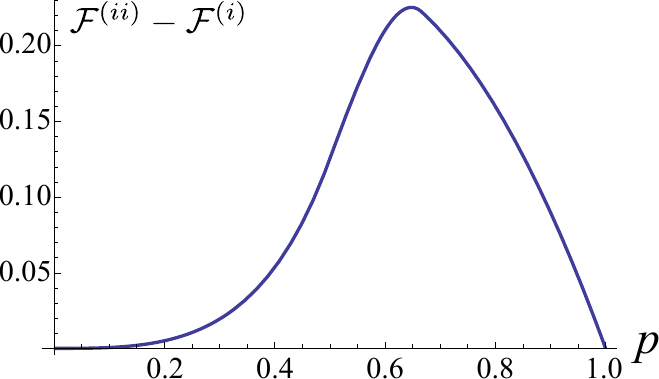}
\caption{Comparison of the QFI's for strategies (i) and (ii). The curve corresponds to the difference of 
QFI between the strategies (ii) in \eqnref{eq:QFI_phase_b} and (i) \eqnref{eq:QFI_phase_a}.}
\label{comparison}
\end{figure}

In Figure~\ref{comparison}, we plot the difference between QFI's of the strategies (ii) in \eqnref{eq:QFI_phase_b} and (i) in \eqnref{eq:QFI_phase_a}.
 One clearly sees that the strategy with intermediate control \figref{strategies}(ii) performs better than the parallel 
strategy \figref{strategies}(i). Moreover there is a gap for all values of $p$ except the trivial cases: the noiseless case $p=1$
and the case where no information about $\theta$ is contained in the channel $p=0$.

This simple example shows that intermediate control (FFQC) allows for an improvement in the achievable accuracy.  
Notice that such an advantage can be maintained for any finite number of blocks as one can choose to introduce new sensing 
qubits, appropriately entangled with the already existing ones, only after one has obtained information about the errors from 
the intermediate error correcting steps.  In the parallel strategy however, all the qubits have to be initialized in an entangled 
state, which is highly fragile to subsequent errors. However the question whether this improvement vanishes in the asymptotic case---as conjectured 
in~\cite{Do14}---or persists, as one might suspect from the finite $N$ results shown here, remains open.\\


\section{Summary and Outlook}
\label{sec:outlook}

We have considered the general limits of quantum metrology where one is, in principle, equipped with a full-scale quantum 
computer to assist in the sensing process. We have shown that one can use techniques from quantum error correction to 
detect or correct for certain kinds of errors while maintaining the sensing 
capabilities of the system.  In particular, we discover that the use of FFQC allows to restore the Heisenberg
scaling of precision for all all rank-one Pauli noise processes, except the Pauli noise that is identical to the Hamiltonian, 
at the cost of a slowing down the evolution by a constant factor. For all other noise-types, we have shown that the QFI is limited to a linear scaling 
with the number of resources. As we demonstrated, this result forbids the Heisenberg scaling of the estimation precision and provides noise dependent upper bounds for all metrological scenarios, local or Bayesian. Remarkably, considering the ultimate metrological scheme (FFQC) actually lead to a simplification of the problem, as in the 
limit of short evolution steps $\d t \to 0$ the CE method does not require to solve the master equation, this allowed us to obtain fully analytic bounds on the achievable precision for relevant noise processes.

However, even in the cases where FFQC does not allow for full restoration of the Heisenberg scaling, one can achieve a significant improvement 
over parallel strategies that operate on limited resources, and do not utilize FFQC.  We have demonstrated this for the example of  
asymmetric X-Y noise where the improvement over parallel schemes employing the same number of resources but do not 
use FFQC can be significant. Moreover, we have put forward simple sequential protocols that make use of only 
a single sensing and auxiliary system, but that nevertheless outperform parallel entanglement-based strategies for limited 
number of sensing systems. From a practical perspective, the existence of simple practical schemes that operate with only a single sensing and auxiliary system, are of high relevance, and may open the way for practical noisy metrology with a significant quantum advantage.
This demonstrates that fast quantum control and error correction are powerful tools in quantum metrology and may enable 
practical sensing with a significant quantum enhancement even in the presence of noise.

An important question one may ask is how our results generalize to higher dimensional systems.  Note that in higher dimensions the parameter encoding Hamiltonian may have support only on a subspace of the Hilbert space of the probe system, which may allow for more involved error correcting codes.


\paragraph{Note added.}
After making this work available online, the error correction protocol presented in \secref{sec:rank1pauli} capable of removing any 
rank-one Pauli noise has been successfully implemented in an NV-centre experiment to compensate for transversal dephasing 
noise \cite{Unden2016}. On the other hand, a dissipation-based scheme has been proposed to correct for such noise in 
ultra-cold ion sensing experiments \cite{Reiter2017}. Lastly, the general results presented in \secref{sec:rank1pauli} and \secref{sec:univ_bound} have been very recently generalized \cite{Demkowicz2017,  zhou2017} beyond the case of qubit probes, considered in this paper.

\section*{Acknowledgements}
This work has been supported by 
Austrian Science Fund (FWF:~P24273-N16, P28000-N27),  
Swiss National Science Foundation Grant (P2GEP2\_151964),
Spanish Ministry National plans FOQUS and MINECO (Severo Ochoa Grant No.~SEV-2015-0522),
Spanish MINECO  FIS2013-40627-P, Generalitat de Catalunya CIRIT  2014 SGR 966
Fundació Privada Cellex, Generalitat de Catalunya Grant (No.~SGR875),
as well as received funding from the European Union's Horizon 2020 research 
and innovation programme under the Marie Sk\l{}odowska-Curie Grant:~Q-METAPP (No.~655161).

\bibliographystyle{unsrtnat}
\bibliography{ffqc}

\begin{thebibliography}{75}
\providecommand{\natexlab}[1]{#1}
\providecommand{\url}[1]{\texttt{#1}}
\expandafter\ifx\csname urlstyle\endcsname\relax
  \providecommand{\doi}[1]{doi: #1}\else
  \providecommand{\doi}{doi: \begingroup \urlstyle{rm}\Url}\fi

\bibitem[Dowling and Seshadreesan(2015)]{Dowling2015}
Jonathan~P. Dowling and Kaushik~P. Seshadreesan.
\newblock {Quantum Optical Technologies for Metrology, Sensing, and Imaging}.
\newblock \emph{J. Lightwave Technol.}, 33\penalty0 (12):\penalty0 2359--2370,
  June 2015.
\newblock ISSN 0733-8724.
\newblock \doi{10.1109/JLT.2014.2386795}.

\bibitem[Giovannetti et~al.(2011)Giovannetti, Lloyd, and
  Maccone]{Giovannetti2011}
Vittorio Giovannetti, Seth Lloyd, and Lorenzo Maccone.
\newblock Advances in quantum metrology.
\newblock \emph{Nature Photon.}, 5:\penalty0 222--229, 2011.
\newblock \doi{10.1038/nphoton.2011.35}.

\bibitem[Giovannetti et~al.(2004)Giovannetti, Lloyd, and Maccone]{GLM04}
Vittorio Giovannetti, Seth Lloyd, and Lorenzo Maccone.
\newblock Quantum-enhanced measurements: Beating the standard quantum limit.
\newblock \emph{Science}, 306\penalty0 (5700):\penalty0 1330--1336, 2004.
\newblock \doi{10.1126/science.1104149}.

\bibitem[Bu\ifmmode~\check{z}\else \v{z}\fi{}ek
  et~al.(1999)Bu\ifmmode~\check{z}\else \v{z}\fi{}ek, Derka, and
  Massar]{Buzek1999}
V.~Bu\ifmmode~\check{z}\else \v{z}\fi{}ek, R.~Derka, and S.~Massar.
\newblock Optimal quantum clocks.
\newblock \emph{Phys. Rev. Lett.}, 82:\penalty0 2207--2210, Mar 1999.
\newblock \doi{10.1103/PhysRevLett.82.2207}.

\bibitem[Giovannetti et~al.(2006)Giovannetti, Lloyd, and
  Maccone]{Giovannetti2006}
Vittorio Giovannetti, Seth Lloyd, and Lorenzo Maccone.
\newblock Quantum metrology.
\newblock \emph{Phys. Rev. Lett.}, 96:\penalty0 010401, 2006.
\newblock \doi{10.1103/PhysRevLett.96.010401}.

\bibitem[Sanders and Milburn(1995)]{Sanders95}
B.~C. Sanders and G.~J. Milburn.
\newblock Optimal quantum measurements for phase estimation.
\newblock \emph{Phys. Rev. Lett.}, 75:\penalty0 2944--2947, Oct 1995.
\newblock \doi{10.1103/PhysRevLett.75.2944}.

\bibitem[Berry and Wiseman(2000)]{Berry00}
D.~W. Berry and H.~M. Wiseman.
\newblock Optimal states and almost optimal adaptive measurements for quantum
  interferometry.
\newblock \emph{Phys. Rev. Lett.}, 85:\penalty0 5098--5101, Dec 2000.
\newblock \doi{10.1103/PhysRevLett.85.5098}.

\bibitem[Peres and Scudo(2001)]{Peres2001}
Asher Peres and Petra~F. Scudo.
\newblock Entangled quantum states as direction indicators.
\newblock \emph{Phys. Rev. Lett.}, 86:\penalty0 4160--4162, Apr 2001.
\newblock \doi{10.1103/PhysRevLett.86.4160}.

\bibitem[Bagan et~al.(2004)Bagan, Baig, and Mu\~noz Tapia]{Bagan2004}
E.~Bagan, M.~Baig, and R.~Mu\~noz Tapia.
\newblock Quantum reverse engineering and reference-frame alignment without
  nonlocal correlations.
\newblock \emph{Phys. Rev. A}, 70:\penalty0 030301, Sep 2004.
\newblock \doi{10.1103/PhysRevA.70.030301}.

\bibitem[Chiribella et~al.(2004{\natexlab{a}})Chiribella, D'Ariano, Perinotti,
  and Sacchi]{Chiribella2004a}
G.~Chiribella, G.~M. D'Ariano, P.~Perinotti, and M.~F. Sacchi.
\newblock Efficient use of quantum resources for the transmission of a
  reference frame.
\newblock \emph{Phys. Rev. Lett.}, 93:\penalty0 180503, Oct 2004{\natexlab{a}}.
\newblock \doi{10.1103/PhysRevLett.93.180503}.

\bibitem[Chiribella et~al.(2004{\natexlab{b}})Chiribella, D'Ariano, Perinotti,
  and Sacchi]{Chiribella2004b}
Giulio Chiribella, Giacomo~Mauro D'Ariano, Paolo Perinotti, and Massimiliano~F.
  Sacchi.
\newblock Covariant quantum measurements that maximize the likelihood.
\newblock \emph{Phys. Rev. A}, 70:\penalty0 062105, Dec 2004{\natexlab{b}}.
\newblock \doi{10.1103/PhysRevA.70.062105}.

\bibitem[Chiribella et~al.(2005)Chiribella, D'Ariano, and
  Sacchi]{Chiribella2005}
G.~Chiribella, G.~M. D'Ariano, and M.~F. Sacchi.
\newblock Optimal estimation of group transformations using entanglement.
\newblock \emph{Phys. Rev. A}, 72:\penalty0 042338, Oct 2005.
\newblock \doi{10.1103/PhysRevA.72.042338}.

\bibitem[Higgins et~al.(2007)Higgins, Berry, Bartlett, Wiseman, and
  Pryde]{Higgins2007}
B.~L. Higgins, D.~W. Berry, S.~D. Bartlett, H.~M. Wiseman, and G.~J. Pryde.
\newblock Entanglement-free heisenberg-limited phase estimation.
\newblock \emph{Nature}, 450:\penalty0 393, 2007.
\newblock \doi{10.1038/nature06257}.

\bibitem[Yurke(1986)]{Yurke86}
B.~Yurke.
\newblock Input states for enhancement of fermion interferometer sensitivity.
\newblock \emph{Phys. Rev. Lett.}, 56:\penalty0 1515--1517, Apr 1986.
\newblock \doi{10.1103/PhysRevLett.56.1515}.

\bibitem[Huelga et~al.(1997)Huelga, Macchiavello, Pellizzari, Ekert, Plenio,
  and Cirac]{Huelga:97}
S.~F. Huelga, C.~Macchiavello, T.~Pellizzari, A.~K. Ekert, M.~B. Plenio, and
  J.~I. Cirac.
\newblock Improvement of frequency standards with quantum entanglement.
\newblock \emph{Phys. Rev. Lett.}, 79:\penalty0 3865--3868, Nov 1997.
\newblock \doi{10.1103/PhysRevLett.79.3865}.

\bibitem[Banaszek et~al.(2009)Banaszek, Demkowicz-Dobrza{\'n}ski, and
  Walmsley]{Banaszek2009}
Konrad Banaszek, Rafa\l{} Demkowicz-Dobrza{\'n}ski, and Ian~A. Walmsley.
\newblock Quantum states made to measure.
\newblock \emph{Nature Photon.}, 3:\penalty0 673--676, 2009.
\newblock \doi{10.1038/nphoton.2009.223}.

\bibitem[Maccone and Giovannetti(2011)]{Maccone2011}
Lorenzo Maccone and Vittorio Giovannetti.
\newblock Quantum metrology: Beauty and the noisy beast.
\newblock \emph{Nature Phys.}, 7:\penalty0 376--377, 2011.
\newblock \doi{doi:10.1038/nphys1976}.

\bibitem[Fujiwara and Imai(2008)]{Fujiwara:08}
Akio Fujiwara and Hiroshi Imai.
\newblock A fibre bundle over manifolds of quantum channels and its application
  to quantum statistics.
\newblock \emph{Journal of Physics A: Mathematical and Theoretical},
  41\penalty0 (25):\penalty0 255304, 2008.
\newblock \doi{10.1088/1751-8113/41/25/255304}.

\bibitem[Escher et~al.(2011)Escher, de~Matos~Filho, and Davidovich]{Escher:11}
B.~M. Escher, R.~L. de~Matos~Filho, and L.~Davidovich.
\newblock General framework for estimating the ultimate precision limit in
  noisy quantum-enhanced metrology.
\newblock \emph{Nat. Phys.}, 7:\penalty0 406, March 2011.
\newblock \doi{10.1038/nphys1958}.

\bibitem[Escher et~al.(2012)Escher, Davidovich, Zagury, and
  de~Matos~Filho]{Escher:12}
B.~M. Escher, L.~Davidovich, N.~Zagury, and R.~L. de~Matos~Filho.
\newblock Quantum metrological limits via a variational approach.
\newblock \emph{Phys. Rev. Lett.}, 109:\penalty0 190404, Nov 2012.
\newblock \doi{10.1103/PhysRevLett.109.190404}.

\bibitem[Demkowicz-Dobrza{\'n}ski et~al.(2012)Demkowicz-Dobrza{\'n}ski,
  Ko{\l}ody{\'n}ski, and Gu{\c{t}}{\u{a}}]{Kolodynski:12}
R.~Demkowicz-Dobrza{\'n}ski, J.~Ko{\l}ody{\'n}ski, and M.~Gu{\c{t}}{\u{a}}.
\newblock The elusive heisenberg limit in quantum-enhanced metrology.
\newblock \emph{Nat. Commun.}, 3:\penalty0 1063, 2012.
\newblock \doi{10.1038/ncomms2067}.

\bibitem[Ko{\l}ody{\'n}ski and Demkowicz-Dobrza{\'n}ski(2013)]{Kolodynski:13}
J.~Ko{\l}ody{\'n}ski and R.~Demkowicz-Dobrza{\'n}ski.
\newblock Efficient tools for quantum metrology with uncorrelated noise.
\newblock \emph{New J. Phys.}, 15\penalty0 (7):\penalty0 073043, 2013.
\newblock \doi{10.1088/1367-2630/15/7/073043}.

\bibitem[Alipour et~al.(2014)Alipour, Mehboudi, and Rezakhani]{Al14}
S.~Alipour, M.~Mehboudi, and A.~T. Rezakhani.
\newblock Quantum metrology in open systems: Dissipative cram\'er-rao bound.
\newblock \emph{Phys. Rev. Lett.}, 112:\penalty0 120405, Mar 2014.
\newblock \doi{10.1103/PhysRevLett.112.120405}.

\bibitem[Knysh et~al.(2011)Knysh, Smelyanskiy, and Durkin]{Kn11}
Sergey Knysh, Vadim~N. Smelyanskiy, and Gabriel~A. Durkin.
\newblock Scaling laws for precision in quantum interferometry and the
  bifurcation landscape of the optimal state.
\newblock \emph{Phys. Rev. A}, 83:\penalty0 021804, Feb 2011.
\newblock \doi{10.1103/PhysRevA.83.021804}.

\bibitem[Knysh et~al.(2014)Knysh, Chen, and Durkin]{Knysh:14}
Sergey~I Knysh, Edward~H Chen, and Gabriel~A Durkin.
\newblock True limits to precision via unique quantum probe.
\newblock \emph{preprint}, arXiv:\penalty0 1402.0495[quant--ph], 2014.
\newblock URL \url{https://arxiv.org/abs/1402.0495}.

\bibitem[Preskill(2000)]{Preskill2000}
John Preskill.
\newblock Quantum clock synchronization and quantum error correction.
\newblock \emph{preprint}, arXiv:\penalty0 0010098[quant--ph], 2000.
\newblock URL \url{http://arxiv.org/abs/quant-ph/0010098}.

\bibitem[D\"ur et~al.(2014)D\"ur, Skotiniotis, Fr\"owis, and Kraus]{Dur:14}
W.~D\"ur, M.~Skotiniotis, F.~Fr\"owis, and B.~Kraus.
\newblock Improved quantum metrology using quantum error correction.
\newblock \emph{Phys. Rev. Lett.}, 112:\penalty0 080801, Feb 2014.
\newblock \doi{10.1103/PhysRevLett.112.080801}.

\bibitem[Kessler et~al.(2014)Kessler, Lovchinsky, Sushkov, and
  Lukin]{Kessler:14}
E.~M. Kessler, I.~Lovchinsky, A.~O. Sushkov, and M.~D. Lukin.
\newblock Quantum error correction for metrology.
\newblock \emph{Phys. Rev. Lett.}, 112:\penalty0 150802, Apr 2014.
\newblock \doi{10.1103/PhysRevLett.112.150802}.

\bibitem[Arrad et~al.(2014)Arrad, Vinkler, Aharonov, and Retzker]{Arrad:14}
G.~Arrad, Y.~Vinkler, D.~Aharonov, and A.~Retzker.
\newblock Increasing sensing resolution with error correction.
\newblock \emph{Phys. Rev. Lett.}, 112:\penalty0 150801, Apr 2014.
\newblock \doi{10.1103/PhysRevLett.112.150801}.

\bibitem[Ozeri(2013)]{Ozeri:13}
Roee Ozeri.
\newblock Heisenberg limited metrology using quantum error-correction codes.
\newblock \emph{preprint}, arxiv:\penalty0 1310.3432[quant--ph], 2013.
\newblock URL \url{https://arxiv.org/abs/1310.3432}.

\bibitem[Lu et~al.(2015)Lu, Yu, and Oh]{Lu2015}
Xiao-Ming Lu, Sixia Yu, and CH~Oh.
\newblock Robust quantum metrological schemes based on protection of quantum
  fisher information.
\newblock \emph{Nat. Commun.}, 6:\penalty0 7282, 2015.
\newblock \doi{10.1038/ncomms8282}.

\bibitem[Herrera-Mart\'{\i} et~al.(2015)Herrera-Mart\'{\i}, Gefen, Aharonov,
  Katz, and Retzker]{Herrera:2015}
David~A. Herrera-Mart\'{\i}, Tuvia Gefen, Dorit Aharonov, Nadav Katz, and Alex
  Retzker.
\newblock Quantum error-correction-enhanced magnetometer overcoming the limit
  imposed by relaxation.
\newblock \emph{Phys. Rev. Lett.}, 115:\penalty0 200501, Nov 2015.
\newblock \doi{10.1103/PhysRevLett.115.200501}.

\bibitem[Gefen et~al.(2016)Gefen, Herrera-Mart\'{\i}, and Retzker]{Gefen:15}
Tuvia Gefen, David~A. Herrera-Mart\'{\i}, and Alex Retzker.
\newblock Parameter estimation with efficient photodetectors.
\newblock \emph{Phys. Rev. A}, 93:\penalty0 032133, Mar 2016.
\newblock \doi{10.1103/PhysRevA.93.032133}.

\bibitem[Plenio and Huelga(2016)]{Plenio:15}
Martin~B. Plenio and Susana~F. Huelga.
\newblock Sensing in the presence of an observed environment.
\newblock \emph{Phys. Rev. A}, 93:\penalty0 032123, Mar 2016.
\newblock \doi{10.1103/PhysRevA.93.032123}.

\bibitem[Sekatski et~al.(2016)Sekatski, Skotiniotis, and D{\"u}r]{SSD:15}
P~Sekatski, M~Skotiniotis, and W~D{\"u}r.
\newblock Dynamical decoupling leads to improved scaling in noisy quantum
  metrology.
\newblock \emph{New J. Phys.}, 18\penalty0 (7):\penalty0 073034, 2016.
\newblock \doi{10.1088/1367-2630/18/7/073034}.

\bibitem[Ulam-Orgikh and Kitagawa(2001)]{Ulam2001}
Duger Ulam-Orgikh and Masahiro Kitagawa.
\newblock Spin squeezing and decoherence limit in ramsey spectroscopy.
\newblock \emph{Phys. Rev. A}, 64:\penalty0 052106, Oct 2001.
\newblock \doi{10.1103/PhysRevA.64.052106}.

\bibitem[Demkowicz-Dobrza\ifmmode~\acute{n}\else \'{n}\fi{}ski and
  Maccone(2014)]{Do14}
Rafal Demkowicz-Dobrza\ifmmode~\acute{n}\else \'{n}\fi{}ski and Lorenzo
  Maccone.
\newblock Using entanglement against noise in quantum metrology.
\newblock \emph{Phys. Rev. Lett.}, 113:\penalty0 250801, Dec 2014.
\newblock \doi{10.1103/PhysRevLett.113.250801}.

\bibitem[Chaves et~al.(2013)Chaves, Brask, Markiewicz, Ko{\l}ody{\'n}ski, and
  {Ac{\'i}n}]{Chaves:12}
R.~Chaves, J.~B. Brask, M.~Markiewicz, J.~Ko{\l}ody{\'n}ski, and A.~{Ac{\'i}n}.
\newblock Noisy metrology beyond the standard quantum limit.
\newblock \emph{Phys. Rev. Lett.}, 111:\penalty0 120401, Sep 2013.
\newblock \doi{10.1103/PhysRevLett.111.120401}.

\bibitem[Viola and Lloyd(1998)]{Viola:98}
Lorenza Viola and Seth Lloyd.
\newblock Dynamical suppression of decoherence in two-state quantum systems.
\newblock \emph{Phys. Rev. A}, 58:\penalty0 2733--2744, Oct 1998.
\newblock \doi{10.1103/PhysRevA.58.2733}.

\bibitem[Viola et~al.(1999)Viola, Knill, and Lloyd]{Viola:99}
Lorenza Viola, Emanuel Knill, and Seth Lloyd.
\newblock Dynamical decoupling of open quantum systems.
\newblock \emph{Phys. Rev. Lett.}, 82:\penalty0 2417--2421, Mar 1999.
\newblock \doi{10.1103/PhysRevLett.82.2417}.

\bibitem[Viola and Knill(2003)]{Viola:03}
Lorenza Viola and Emanuel Knill.
\newblock Robust dynamical decoupling of quantum systems with bounded controls.
\newblock \emph{Phys. Rev. Lett.}, 90:\penalty0 037901, Jan 2003.
\newblock \doi{10.1103/PhysRevLett.90.037901}.

\bibitem[Khodjasteh and Viola(2009)]{Viola:09}
Kaveh Khodjasteh and Lorenza Viola.
\newblock Dynamically error-corrected gates for universal quantum computation.
\newblock \emph{Phys. Rev. Lett.}, 102:\penalty0 080501, Feb 2009.
\newblock \doi{10.1103/PhysRevLett.102.080501}.

\bibitem[Khodjasteh et~al.(2010)Khodjasteh, Lidar, and Viola]{Viola:10}
Kaveh Khodjasteh, Daniel~A. Lidar, and Lorenza Viola.
\newblock Arbitrarily accurate dynamical control in open quantum systems.
\newblock \emph{Phys. Rev. Lett.}, 104:\penalty0 090501, Mar 2010.
\newblock \doi{10.1103/PhysRevLett.104.090501}.

\bibitem[West et~al.(2010)West, Lidar, Fong, and Gyure]{West:10}
Jacob~R. West, Daniel~A. Lidar, Bryan~H. Fong, and Mark~F. Gyure.
\newblock High fidelity quantum gates via dynamical decoupling.
\newblock \emph{Phys. Rev. Lett.}, 105:\penalty0 230503, Dec 2010.
\newblock \doi{10.1103/PhysRevLett.105.230503}.

\bibitem[Wiseman and Milburn(2009)]{Wiseman2009}
Howard~M Wiseman and Gerard~J Milburn.
\newblock \emph{{Quantum Measurement and Control}}.
\newblock Cambridge University Press, 2009.
\newblock ISBN 0521804426.
\newblock \doi{10.1017/CBO9780511813948}.

\bibitem[Chiribella(2012)]{Chiribella2012}
Giulio Chiribella.
\newblock Optimal networks for quantum metrology: semidefinite programs and
  product rules.
\newblock \emph{New Journal of Physics}, 14\penalty0 (12):\penalty0 125008,
  2012.
\newblock \doi{10.1088/1367-2630/14/12/125008}.

\bibitem[Sergeevich et~al.(2011)Sergeevich, Chandran, Combes, Bartlett, and
  Wiseman]{Sergeevich2011}
Alexandr Sergeevich, Anushya Chandran, Joshua Combes, Stephen~D. Bartlett, and
  Howard~M. Wiseman.
\newblock Characterization of a qubit hamiltonian using adaptive measurements
  in a fixed basis.
\newblock \emph{Phys. Rev. A}, 84:\penalty0 052315, Nov 2011.
\newblock \doi{10.1103/PhysRevA.84.052315}.

\bibitem[Tsang(2012)]{Tsang:12}
Mankei Tsang.
\newblock Ziv-zakai error bounds for quantum parameter estimation.
\newblock \emph{Phys. Rev. Lett.}, 108:\penalty0 230401, Jun 2012.
\newblock \doi{10.1103/PhysRevLett.108.230401}.

\bibitem[Gill and Levit(1995)]{Gill:95}
Richard~D Gill and Boris~Y Levit.
\newblock Applications of the van {Trees} inequality: a {Bayesian}
  {C}ram{\'e}r-{R}ao bound.
\newblock \emph{Bernoulli}, 1(1/2):\penalty0 59--79, 1995.
\newblock \doi{10.2307/3318681}.

\bibitem[Helstrom(1976)]{Helstrom1976}
C.~W. Helstrom.
\newblock \emph{Quantum Detection and Estimation Theory}.
\newblock Academic Press, 1976.
\newblock ISBN 0123400503.

\bibitem[Holevo(1980)]{H80}
A.~S. Holevo.
\newblock \emph{Probabilistic and Statistical Aspects of Quantum Theory}.
\newblock North-Holland Series in Statistics and Probability, 1980.
\newblock \doi{10.1007/978-88-7642-378-9}.

\bibitem[Braunstein and Caves(1994)]{BC94}
Samuel~L. Braunstein and Carlton~M. Caves.
\newblock Statistical distance and the geometry of quantum states.
\newblock \emph{Phys. Rev. Lett.}, 72:\penalty0 3439--3443, May 1994.
\newblock \doi{10.1103/PhysRevLett.72.3439}.

\bibitem[Pezz\'e and Smerzi(2009)]{Pezze2009}
Luca Pezz\'e and Augusto Smerzi.
\newblock Entanglement, nonlinear dynamics, and the {H}eisenberg limit.
\newblock \emph{Phys. Rev. Lett.}, 102:\penalty0 100401, Mar 2009.
\newblock \doi{10.1103/PhysRevLett.102.100401}.

\bibitem[L\"ucke et~al.(2014)L\"ucke, Peise, Vitagliano, Arlt, Santos, T\'oth,
  and Klempt]{Lucke2014}
Bernd L\"ucke, Jan Peise, Giuseppe Vitagliano, Jan Arlt, Luis Santos, G\'eza
  T\'oth, and Carsten Klempt.
\newblock Detecting multiparticle entanglement of dicke states.
\newblock \emph{Phys. Rev. Lett.}, 112:\penalty0 155304, Apr 2014.
\newblock \doi{10.1103/PhysRevLett.112.155304}.

\bibitem[Strobel et~al.(2014)Strobel, Muessel, Linnemann, Zibold, Hume,
  Pezz\`{e}, Smerzi, and Oberthaler]{Strobel2014}
Helmut Strobel, Wolfgang Muessel, Daniel Linnemann, Tilman Zibold, David~B.
  Hume, Luca Pezz\`{e}, Augusto Smerzi, and Markus~K. Oberthaler.
\newblock Fisher information and entanglement of non-gaussian spin states.
\newblock \emph{Science}, 345\penalty0 (6195):\penalty0 424--427, 2014.
\newblock \doi{10.1126/science.1250147}.

\bibitem[Pires et~al.(2016)Pires, Cianciaruso, C\'eleri, Adesso, and
  Soares-Pinto]{Pires2015}
Diego~Paiva Pires, Marco Cianciaruso, Lucas~C. C\'eleri, Gerardo Adesso, and
  Diogo~O. Soares-Pinto.
\newblock Generalized geometric quantum speed limits.
\newblock \emph{Phys. Rev. X}, 6:\penalty0 021031, Jun 2016.
\newblock \doi{10.1103/PhysRevX.6.021031}.

\bibitem[Taddei et~al.(2013)Taddei, Escher, Davidovich, and
  de~Matos~Filho]{Taddei2013}
M.~M. Taddei, B.~M. Escher, L.~Davidovich, and R.~L. de~Matos~Filho.
\newblock Quantum speed limit for physical processes.
\newblock \emph{Phys. Rev. Lett.}, 110:\penalty0 050402, Jan 2013.
\newblock \doi{10.1103/PhysRevLett.110.050402}.

\bibitem[Fr{\"o}wis and D{\"u}r(2012)]{Florian2012}
Florian Fr{\"o}wis and Wolfgang D{\"u}r.
\newblock Measures of macroscopicity for quantum spin systems.
\newblock \emph{New J. Phys.}, 14\penalty0 (9):\penalty0 093039, 2012.
\newblock \doi{10.1088/1367-2630/14/9/093039}.

\bibitem[Nielsen and Chuang(2010)]{MikeIke}
M.~A. Nielsen and I.~L. Chuang.
\newblock \emph{Quantum computation and quantum information}.
\newblock Cambridge university press, 2010.
\newblock \doi{10.1017/CBO9780511976667}.

\bibitem[Alicki and Lendi(1987)]{Alicki1987}
Robert Alicki and Karl Lendi.
\newblock \emph{Quantum Dynamical Semigroups and Applications}.
\newblock Springer, 1987.
\newblock \doi{10.1007/3-540-18276-4}.

\bibitem[Breuer and Petruccione(2002)]{Breuer2002}
Heinz-Peter Breuer and Francesco Petruccione.
\newblock \emph{The Theory of Open Quantum Systems}.
\newblock Oxford University Press, 2002.
\newblock \doi{10.1093/acprof:oso/9780199213900.001.0001}.

\bibitem[Matsuzaki et~al.(2011)Matsuzaki, Benjamin, and
  Fitzsimons]{Matsuzaki2011}
Yuichiro Matsuzaki, Simon~C. Benjamin, and Joseph Fitzsimons.
\newblock Magnetic field sensing beyond the standard quantum limit under the
  effect of decoherence.
\newblock \emph{Phys. Rev. A}, 84:\penalty0 012103, Jul 2011.
\newblock \doi{10.1103/PhysRevA.84.012103}.

\bibitem[Chin et~al.(2012)Chin, Huelga, and Plenio]{Chin2012}
Alex~W. Chin, Susana~F. Huelga, and Martin~B. Plenio.
\newblock Quantum metrology in non-markovian environments.
\newblock \emph{Phys. Rev. Lett.}, 109:\penalty0 233601, Dec 2012.
\newblock \doi{10.1103/PhysRevLett.109.233601}.

\bibitem[Macieszczak(2015)]{Macieszczak2015}
Katarzyna Macieszczak.
\newblock Zeno limit in frequency estimation with non-markovian environments.
\newblock \emph{Phys. Rev. A}, 92:\penalty0 010102, Jul 2015.
\newblock \doi{10.1103/PhysRevA.92.010102}.

\bibitem[Smirne et~al.(2016)Smirne, Ko\l{}ody\ifmmode~\acute{n}\else
  \'{n}\fi{}ski, Huelga, and Demkowicz-Dobrza\ifmmode~\acute{n}\else
  \'{n}\fi{}ski]{Smirne2015}
Andrea Smirne, Jan Ko\l{}ody\ifmmode~\acute{n}\else \'{n}\fi{}ski, Susana~F.
  Huelga, and Rafa\l{} Demkowicz-Dobrza\ifmmode~\acute{n}\else \'{n}\fi{}ski.
\newblock Ultimate precision limits for noisy frequency estimation.
\newblock \emph{Phys. Rev. Lett.}, 116:\penalty0 120801, Mar 2016.
\newblock \doi{10.1103/PhysRevLett.116.120801}.

\bibitem[Addis et~al.(2016)Addis, Laine, Gneiting, and Maniscalco]{Addis2016}
Carole Addis, Elsi-Mari Laine, Clemens Gneiting, and Sabrina Maniscalco.
\newblock Problem of coherent control in non-{M}arkovian open quantum systems.
\newblock \emph{Phys. Rev. A}, 94:\penalty0 052117, Nov 2016.
\newblock \doi{10.1103/PhysRevA.94.052117}.

\bibitem[Andersson et~al.(2007)Andersson, Cresser, and Hall]{Anderson:07}
E.~Andersson, J.~D. Cresser, and M.~J.~W. Hall.
\newblock Finding the {K}raus decomposition from a master equation and vice
  versa.
\newblock \emph{J. Mod. Opt.}, 54\penalty0 (12):\penalty0 1695--1716, 2007.
\newblock \doi{10.1080/09500340701352581}.

\bibitem[Brask et~al.(2015)Brask, Chaves, and Ko\l{}ody\ifmmode~\acute{n}\else
  \'{n}\fi{}ski]{Brask:15}
J.~B. Brask, R.~Chaves, and J.~Ko\l{}ody\ifmmode~\acute{n}\else \'{n}\fi{}ski.
\newblock Improved quantum magnetometry beyond the standard quantum limit.
\newblock \emph{Phys. Rev. X}, 5:\penalty0 031010, Jul 2015.
\newblock \doi{10.1103/PhysRevX.5.031010}.

\bibitem[Taminiau et~al.(2014)Taminiau, Cramer, Sar, Dobrovitski, and
  Hanson]{Taminiau2014}
T.~H. Taminiau, J.~Cramer, T.~van~der Sar, V.~V. Dobrovitski, and R.~Hanson.
\newblock Universal control and error correction in multi-qubit spin registers
  in diamond.
\newblock \emph{Nat. Nanotechnol.}, 9\penalty0 (3):\penalty0 171--176, March
  2014.
\newblock \doi{10.1038/nnano.2014.2}.

\bibitem[Waldherr et~al.(2014)Waldherr, Wang, Zaiser, Jamali,
  Schulte-Herbruggen, Abe, Ohshima, Isoya, Du, Neumann, and
  Wrachtrup]{Waldherr2014}
G.~Waldherr, Y.~Wang, S.~Zaiser, M.~Jamali, T.~Schulte-Herbruggen, H.~Abe,
  T.~Ohshima, J.~Isoya, J.~F. Du, P.~Neumann, and J.~Wrachtrup.
\newblock Quantum error correction in a solid-state hybrid spin register.
\newblock \emph{Nature}, 506\penalty0 (7487):\penalty0 204--207, February 2014.
\newblock ISSN 0028-0836.
\newblock \doi{10.1038/nature12919}.

\bibitem[Unden et~al.(2016)Unden, Balasubramanian, Louzon, Vinkler, Plenio,
  Markham, Twitchen, Stacey, Lovchinsky, Sushkov, Lukin, Retzker, Naydenov,
  McGuinness, and Jelezko]{Unden2016}
Thomas Unden, Priya Balasubramanian, Daniel Louzon, Yuval Vinkler, Martin~B.
  Plenio, Matthew Markham, Daniel Twitchen, Alastair Stacey, Igor Lovchinsky,
  Alexander~O. Sushkov, Mikhail~D. Lukin, Alex Retzker, Boris Naydenov, Liam~P.
  McGuinness, and Fedor Jelezko.
\newblock Quantum metrology enhanced by repetitive quantum error correction.
\newblock \emph{Phys. Rev. Lett.}, 116:\penalty0 230502, Jun 2016.
\newblock \doi{10.1103/PhysRevLett.116.230502}.

\bibitem[Reiter et~al.(2017)Reiter, S{\o}rensen, Zoller, and
  Muschik]{Reiter2017}
F.~Reiter, A.~S. S{\o}rensen, P.~Zoller, and C.~A. Muschik.
\newblock Autonomous {Quantum} {Error} {Correction} and {Application} to
  {Quantum} {Sensing} with {Trapped} {Ions}.
\newblock \emph{preprint}, arXiv:\penalty0 1702.08673[quant--ph], 2017.
\newblock URL \url{http://arxiv.org/abs/1702.08673}.

\bibitem[Demkowicz-Dobrza\'{n}ski et~al.(2017)Demkowicz-Dobrza\'{n}ski,
  Czajkowski, and Sekatski]{Demkowicz2017}
R.~Demkowicz-Dobrza\'{n}ski, J.~Czajkowski, and P.~Sekatski.
\newblock Adaptive quantum metrology under general {Markovian} noise.
\newblock \emph{preprint}, arXiv:\penalty0 1704.06280[quant--ph], 2017.
\newblock URL \url{http://arxiv.org/abs/1704.06280}.

\bibitem[Zhou et~al.(2017)Zhou, Zhang, Preskill, and Jiang]{zhou2017}
Sisi Zhou, Mengzhen Zhang, John Preskill, and Liang Jiang.
\newblock Achieving the {Heisenberg} limit in quantum metrology using quantum
  error correction.
\newblock \emph{preprint}, arXiv:\penalty0 1706.02445[quant--ph], 2017.
\newblock URL \url{http://arxiv.org/abs/1706.02445}.

\bibitem[Bengtsson and {\.Z}yczkowski(2006)]{Bengtsson2006}
Ingemar Bengtsson and Karol {\.Z}yczkowski.
\newblock \emph{Geometry of Quantum States: An Introduction to Quantum
  Entanglement}.
\newblock Cambridge University Press, 2006.
\newblock \doi{10.1017/CBO9780511535048}.

\end{thebibliography}

\onecolumn\newpage
\appendix

\section{Relation between the Kraus representations and the dynamical matrix}
\label{app:1}

Any Kraus operator, $K_i\!=\!\sum_k \m M_{ik} \sigma_k$, can be expressed in the operator 
basis given by the four Pauli operators in the qubit case, which we write for the vector of Kraus operators as
\be
{\bf K} = \m M \Vs.
\ee
 The action of the channel on the density matrix is expressed in the Kraus representation as
 \bea
 \cE (\rho) = \sum_i K_i \rho K_i^\dag = \sum_i( \sum_k \m M_{ik} \sigma_k) \rho ( \sum_\ell \m M_{i\ell}^* \sigma_\ell)\\ \nonumber
 = \sum_{k\ell} (\sum_i \m M_{\ell i}^\dag \m M_{ik})  \sigma_k \rho \sigma_\ell =  \sum_{k\ell} (\m M^\dag \m M)_{\ell k}\,  \sigma_k \rho \sigma_\ell. 
 \eea
 Imposing the equality with the dynamical matrix representation \eref{eq:Srep} one arrives at
 \be
 \m M^\dag \m M = \m S^T.
 \ee
From this expression it follows that all Kraus representation 
satisfying $\m M'= \m u \m M$ (with a unitary $\m u$) lead to the 
same dynamical matrix and therefore correspond to the same channel. Remark that in general $\m M$  may not be a square 
but a rectangular $n\times 4$ matrix, where $n$ is the number of Kraus operators in a given representation. 
However, any $\m M$ admits a singular value decomposition of the form
 \be
 \m M = \m u \, \m D_{n\times4} \, \m v^\dag,
 \ee 
 with $\m u$-- a unitary $n\times n$ matrix , $\m v$-- a unitary $4\times4$ matrix  and $\m D_{n\times 4}$ -- a rectangular $n\times 4$ diagonal matrix. Now $\m M' =\m D_{n\times4} \, \m v^\dag$ is also a valid Kraus representation, and so is the $4\times 4$ matrix $\m M'' = \m D_{4\times 4}\, \m v^\dag$ (where to get to $\m D_{4\times4}$ from $\m D_{n\times4}$  we either remove all the $n-4$ zero lines in the case $n>4$, or add $4-n$ zero lines in the case $n<4$). Finally, we chose the Hermitian $4\times 4$ matrix
 \be
\m M''' = \m v \, \m D_{4\times 4} \, \m v^\dag
 \ee
 to be our canonical Kraus representation for the channel.
 
\section{Infinitesimal-timestep CE bound}
\label{app:2}
Here, we prove the infinitesimal-timestep CE bound of \eqnref{eq:CE_inf} that may be applied
also to schemes that incorporate FFQC, i.e., the ones of type (d) in \figref{fig:hierarchy}. 
We essentially follow the derivation contained in the Supplementary Material of \refcite{Do14} except 
the last step, at which we slightly generalize the bound presented therein.

In particular, one should follow the reasoning up to Eq.~(S15) of \refcite{Do14}, at which a more 
general upper bound (for any real $\sqrt{x}$) may be derived via
\al{
|| \sum_k \dot K_k^\dag i A K_k + h.c.|| & = ||  i \sum_k   (\sqrt{x}\,\dot K_k^\dag)  A 
(\frac{1}{\sqrt{x}} K_k) -  (\frac{1 }{\sqrt{x}} K_k^\dag)  A (\sqrt{x}\, \dot K_k) || \\
&=||i \sum_k   (\sqrt{x}\,\dot K_k+ i \frac{1}{\sqrt{x}} K_k  \big)^\dag  A (\sqrt{x}\,\dot K_k+ i 
\frac{1}{\sqrt{x}} K_k ) - \lambda \dot K_k^\dag A \dot K_k  -\frac{1}{x}  K_k^\dag A  K_k||
}
Using the triangle inequality and the one stated in Eq.~(S13) of \refcite{Do14}, 
we arrive at a generalized version of Eq.~(S15) therein:
\be
|| \sum_k \dot K_k^\dag i A K_k + h.c.|| \leq 2 ||A|| \Big(x ||\sum_k \dot K_k^\dag  \dot K_k||+||\sum_k \dot K_k^\dag   K_k||+\frac{1}{x} \Big),
\ee
which in turn results in the generalization of Eq.~(S19):
\be
\QFI\leq 4\,k ||\alpha|| + 4\,k(k-1) ||\beta|| \left(x ||\alpha|| + ||\beta||+\frac{1}{x}\right).
\ee
By taking the limit of infinitesimal timesteps $k\!=\!t'/\d t$ and writing explicitly the time-dependences, we obtain the bound:
\be
\QFI\leq 4\frac{t'}{\d t} ||\alpha(\d t)|| + 4\frac{(t')^2}{\d t^2} ||\beta(\d t)|| \left( x ||\alpha(\d t)|| + ||\beta(\d t)||+\frac{1}{x}\right),
\label{eq:dupablada}
\ee
whose tightest form we would like determine in the $\d t\!\to\! 0$ limit. Hence, as otherwise \eqnref{eq:dupablada} must diverge, we can restrict 
to Kraus representations for which 
$\alpha(\d t)=\alpha^{(2)} \d t + \bigO(\d t^{3/2})$ and 
$\beta(\d t)=\beta^{(2)}\d t + \beta^{(3)}\d t^{3/2}+ \bigO(\d t^2)$.
As we are willing to show a linear scaling of the QFI in $t'$ due to vanishing of the second term in \eqnref{eq:dupablada}, we set $x=1/\sqrt{\d t}$ 
and compute the Taylor expansion:
\be
\QFI\leq 
4\,t'\left(||\alpha^{(2)}||+\bigO(\d t)\right) + 4\,t'^2 \left\Vert\frac{\beta^{(2)}}{\d t}+\frac{\beta^{(3)}}{\sqrt{\d t}} + \bigO(1)\right\Vert 
\left[\left(||\alpha^{(2)} ||+1\right)\sqrt{\d t}+\bigO(\d t) \right].
\ee
Thus, if we are able to find a Kraus representation for which both $\beta^{(2)}=0$
and $\beta^{(3)}=0$, the above expression does not diverge in the $\d t \!\to\!0$ limit
but rather provides the desired upper bound---\eqnref{eq:CE_inf_lin} in the main text:
\bea
\QFI\leq 4\, t' ||\alpha^{(2)}|| \qquad \text{with}\quad  \beta = \bigO(\d t^2),
\eea
which importantly imposes the SQL-like scaling due to its linearity in $t'$.

\section{Asymptotic FFQC-valid CE bounds for qubit Liouvillians}
\label{app:3}
%

\subsection{Existence of asymptotic bound for any Liouvillian but the non-parallel rank-one Pauli noise}
\label{app:3.1}

In what follows, we perform analytic minimization of the infinitesimal-timestep CE bound introduced in
\secref{asymptoticbounds} that lead to the \eqnref{eq:CE_inf_lin}. For ease of notation we denote
the matrix $\m S^{(1)}$ in \eqnref{eq:S(1)} by
\begin{equation}
\m S^{(1)}= {\small\left(
\begin{array}{c|c} s^{(1)}  &  {\bf  s}^{(1)\dag} \\
 \hline  {\bf s}^{(1)} & \bar{\m{S}}^{(1)} 
\end{array}\right)}.
\label{Smatrixsimple}
\end{equation}

Recall that our goal is to minimize over all Kraus decompositions ${\bf K}=\m M  \Vs$ of the channel 
$\mathcal{E}_{\omega,\d t}$ up to first order in $\d  t$.  Expanding the matrix $\m M$ up to first order in $\d t$, 
$\m M= \m M^{(0)} + \sqrt{\d t}\, \m M^{(1)} + \d t\, \m M^{(2)}$, and using 
\eqnref{eq:conditionsM}  the matrices $\m M^{(\ell)}$ satisfy 
\begin{align}\nonumber
\m M^{(0)} \m M^{(0)}&= \m S^{(0)}\\ \nonumber
\m M^{(0)} \m M^{(1)}+\m M^{(1)} \m M^{(0)} &= 0\\
\m M^{(1)} \m M^{(1)} +\m M^{(0)}\m M^{(2)}+\m M^{(2)} \m M^{(0)} &= \m S^{(1)}.
\label{constraintsM}
\end{align}
These constraints enforce the following structure on the matrices $\m M^{(\ell)}$
\begin{align}\nonumber \label{eq: M cons}
\m M^{(0)}&=\mathrm{diag}(1,0,0,0),\\
\m M^{(1)}&=
{\small  \left(\begin{array}{c|c}
0& {\bf 0}^T \\ \hline
 {\bf 0 } & \bar{\m M}^{(1)} =  \sqrt{\bar{ \m L}^{}}\\
\end{array}\right)},\quad
\m M^{(2)}=
{\small  \left(\begin{array}{c|c}
\frac{1}{2} s^{(1)} &  {\bf s}^{(1)\dag} \\ \hline
{\bf s }^{(1)} &  \bar{ \m M}^{(2)}
\end{array}\right)},
\end{align}
where $\bar{\m {M}}^{(2)}$ does not contribute to the first order expansion of the channel and can thus be chosen at will. 

By performing a power series expansion of the matrices $\alpha$ and $\beta$ in terms of $\d t$
\begin{align}\nonumber
\alpha &= \alpha^{(0)} +\sqrt{\d t}\, \alpha^{(1)}  +\d t\,\alpha^{(2)}  +O( \d t^{3/2}) \\
\beta &= \beta^{(0)} +\sqrt{\d t}\, \beta^{(1)}  +\d t \beta^{(2)}  +\d t^{3/2} \beta^{(3)} +O( dt^{2}),
\label{coefficientssmallorder}
\end{align}
and using \eqnref{eq:alphaM} and \eqnref{eq:betaM} with $\m h=\m h^{(0)}+\m h^{(1)} \sqrt{\d t}+ \m h^{(2)} \d t$
the minimization over all equivalent Kraus operators is equivalent to searching over Hermitian matrices $\m h^{(\ell)}$ that 
minimize the bound of \eqnref{eq:CE_inf_lin}.  

To yield a non-trivial bound a Kraus decomposition has to satisfy $\alpha^{(0)}=\alpha^{(1)}= \beta^{(0)}=\beta^{(1)}=0$.  
Expanding  \eqnref{eq:alphaM} and \eqnref{eq:betaM} in powers of $\d t$ and minimizing order by order imposes the following 
structure for $\m h^{(0)},\, \m h^{(1)}$
\begin{align}
\m h^{(0)} = {\small\left(
\begin{array}{c|c} 0 & 0 \\ \hline 
 0 & {\m H}^{(0)} 
\end{array}\right)}, \quad  \quad
\m h^{(1)} = {\small\left(
\begin{array}{c|c} 0 & {\bf h}^{(1)\dag}  \\ \hline 
 {\bf h}^{(1)} & {\m H}^{(1)} 
\end{array}\right)},
\end{align}
and the coefficients $\alpha^{(2)}$ and $\beta^{(2)}$ are given by 
\begin{align}
\label{alpha2}
&\alpha^{(2)}= \Vs ^\dag 
{\small\left(
\begin{array}{c|c} {\bf h}^{(1)^\dag} {\bf h}^{(1)}  & {\bf h}^{(1)\dag} \, {\m H}^{(0)} \bar{\m M}^{(1)}  \\ \hline 
\bar{\m M}^{(1)} {\m H}^{(0)} {\bf h}^{(1)} & \bar{\m M}^{(1)} {\m H}^{(0)}{\m H}^{(0)} \bar{\m M}^{(1)} 
\end{array}\right)} \Vs \\
\label{beta2}
&\beta^{(2)}=\ii\, \left(\frac{-1}{2} \sthree + \Vs^\dag
{\small\left(
\begin{array}{c|c} h^{(2)}_{00} & {\bf h}^{(1)\dag} \bar{\m M}^{(1)}  \\ \hline 
 \bar{\m M}^{(1)}  {\bf h}^{(1)} & \bar{\m M}^{(1)} {\m H}^{(0)} \bar{\m M}^{(1)} 
\end{array}\right)}\Vs \right). 
\end{align}
%

Moreover, the coefficient $\beta^{(3)}$ in \eqnref{coefficientssmallorder} reads
\bea\nonumber
\beta^{(3)} &=\ii\Vs ^\dag\left[ 
{\small\left(
\begin{array}{c|c} \frac{1}{2} h^{(3)}_{00} + {\bf s}^{(1)\dag}  {\bf  h}^{(1)}& 
{\bf  h}^{(1)^\dag} \bar{\m  M}^{(2)}+ ({\bf s}^{(1)\dag} {\m H}^{(0)} + {\bf h}^{(2)\dag})\bar{\m M}^{(1)}\\ \hline 
0  &\frac{1}{2} \bar{\m M}^{(1)} {\m H}^{(1)} \bar{\m M}^{(1)}+ \bar{\m M}^{(2)} {\m H}^{(0)} \bar{\m M}^{(1)} 
\end{array}\right)}+ h.c.\right]\Vs +  
i \Vs^\dag {\small\left(
\begin{array}{c|c} 0& 
(0\,\, 0\,\,\frac{1}{2})^\dag \bar{\m M}^{(1)}\\ \hline 
0  &0
\end{array}\right)}
\Vs ,\\
\label{beta3}  
\eea
where $h.c.$ stands for the Hermitian conjugate.
As none of the parameters $h^{(3)}_{00},\, {\m H}^{(1)},\,\bar{\m M}^{(2)}$, and ${\bf h}^{(2)}$ appearing in \eqnref{beta3} 
enter into the expressions for $\alpha^{(2)}$ and $\beta^{(2)}$ the norm of $\beta^{(3)}$ can be set to zero independently of all 
lower orders of coefficients for $\alpha$ and $\beta$.  Indeed, the following choice of the free parameters
\begin{align}\nonumber
h^{(3)}_{00} =& - ({\bf s}^{(1)\dag}  {\bf  h}^{(1)} + {\bf  h}^{(1)^\dag} \cdot {\bf s}^{(1)}) \\  \nonumber
{\m H}^{(1)}=&2\, {\m H}^{(0)}\\  \nonumber
\bar{\m M}^{(2)} =& -\bar{\m M}^{(1)} \\ 
\t{Re}({\bf h}^{(2)}) =& \t{Re}\left({\bf h}^{(1)} - {\m H}^{(0)}  {\bf s}^{(1)}\right) - \frac{1}{4}
{\small\left(
\begin{array}{c}
0\\0\\1
\end{array}\right)}
\end{align}
sets $\beta^{(3)}$ to zero. 

\subsubsection{Rank-one noise}
We start with a general rank-one Liouvillian $\bar{\m L}_{\bf r}^{1\text{G}}$ defined in \eqnref{eq:rank1gen}
\begin{equation}\label{rank-1 L}
\bar{\m L}_{\bf r}^{1\text{G}}= {\bf r} \,{\bf r}^\dag = \frac{\gamma}{2} {\bf v}\, {\bf v}^\dag
\end{equation}
with $ {\bf r}=(x, y, z)^T$, $\frac{\gamma}{2}= |x|^2+|y|^2+|z|^2$ and the vector ${\bf v}$ has unit length such that ${\bf v}\, {\bf v}^\dag$ is a rank-1 projector. Equation \eref{eq: M cons} implies
$\bar{\m M}^{(1)} = \sqrt{\frac{\gamma}{2}} {\bf v}\, {\bf v}^\dag$. Expressing ${\bf h}^{(1)} =  (h_r+i h_i) {\bf v}+ h_\perp {\bf v}_\perp$  in a basis containing ${\bf v}$ allows to  re-write \eqnref{beta2} as 
\begin{align}
\label{rank1 beta}
\beta^{(2)} &=\ii (-\frac{\sthree}{2} + \Vs^\dag 
{\small\left(
\begin{array}{c|c} h^{(2)}_{00} &  (h_r-\ii h_i) {\bf r}^\dag \\ \hline 
 h.c.  &  \m H_{11}^{(0)} {\bf r}\,{\bf r}^\dag
\end{array}\right)} \Vs),
\end{align}
where $\m H_{11}^{(0)} =  {\bf v}^\dag {\m H}^{(0)} {\bf v} $. We decompose the vector ${\bf r} ={\bf r}_r + \ii \,{\bf r}_i$, with ${\bf r}_r  =\t{Re}({\bf r})$ and ${\bf r}_i  =\t{Im}({\bf r})$. A little bit of algebra allows one to rewrite \eqnref{rank1 beta} as
\begin{align}\label{rank1 condition}
\beta^{(2)} &= \ii \sigma_0 (h^{(2)}_{00}+ \frac{\gamma}{2} \m H_{11}^{(0)})\nonumber  \\&+ \ii \left(-\frac{\sthree}{2} + 2 \, {\bm \sigma}^\dag 
( h_r {\bf r}_r -  h_i {\bf r}_i +  \m H_{11}^{(0)}{\bf r}_r \times {\bf r}_i)\right),
\end{align}
where we used the property $[{\bf r}_i^\dag {\bm \sigma}, {\bf r}_r^\dag {\bm \sigma}] =2 \ii ({\bf r}_i \times {\bf r}_r)^\dag {\bm \sigma}$.  We wish to set the operator $\beta^{(2)}$ to zero.  The $\sigma_0$ term can be trivially put to zero by choosing 
$h_2^{(00)}=-\frac{\gamma}{2} \m H_{11}^{(0)}$. The remaining three terms can be put to zero if the following conditions are fulfilled. Either 
${\bf r}_r$ and ${\bf r}_i$ are linearly independent---in which case the vectors 
${\bf r}_r, \,{\bf r}_i,\, {\bf r}_r \times {\bf r}_i$ span the whole vector space; or ${\bf r}_r$ and ${\bf r}_i$ are parallel to 
each other and to the vector ${\bf z}$---which corresponds to the dephasing noise. 

Hence, $\beta^{(2)}$ can be set to zero, so that \eqnref{eq:CE_inf_lin} holds for any rank-one noise except for Pauli noise which is not parallel to $\sthree$. 
For this later case we  have outlined a FFQC strategy in \secref{sec:qfi_noise} that effectively removes such noise 
at the cost of slowing down the evolution by a constant factor.

\subsubsection{Liouvillians of higher rank and CE bound}

Let us now consider a general Lindbladian, it can always be written in the diagonal form
\bea 
\bar{\m L} = {\bf r}_1{\bf r}_1^\dag \bigoplus {\bf r}_2{\bf r}_2^\dag  \bigoplus {\bf r}_3{\bf r}_3^\dag
\eea
with ${\bf r}_i^\dag {\bf r}_j =0$ for $i \neq j$ (as suggested by the direct sum). Again by ${\bf v}_j = \frac{{\bf r}_j}{|{\bf r}_j|}$ we denote the normalized vecors.
Now we show that if at least one Liouvillian $ \bar {\m L}_j ={\bf r}_j{\bf r}_j^\dag $ (e.g., $j=1$) implies SQL scaling, then also $\bar{ \m L}$
leads to SQL scaling. To see this simply pick ${\bf h}^{(1)} = (h_r + i h_i) {\bf v}_1$  and ${\m H}^{(0)} = {\m H}^{(0)}_{11} {\bf v}_1\, {\bf v}_1^\dag$,  \eqnref{beta2} implies
\begin{align}
\beta^{(2)} &=i \left( - \frac{\sthree}{2} + \Vs\dag
{\small\left(
\begin{array}{c|c} h^{(2)}_{00} & (h_r- i h_i ){\bf r}_1^\dag \\ \hline 
 h.c.  &  {\m H}^{(0)}_{11}{\bf r}_1{\bf r}_1^\dag 
\end{array}\right)} \Vs \right).
\end{align}
This expression is exactly equal to \eqnref{rank1 beta}, consequently it can be set to zero if the Liouvillian $\bar{\m L}_1$ leads to standard scaling. Accordingly any higher rank Liouvillian that contains a non-Pauli rank-1 noise in its diagonal decomposition leads to standard scaling.

\subsubsection{Rank-two noise}

Now consider a general Lindbadian of rank-two. Because of what we just showed in the previous section, the only case which can potentially lead to Heisenberg scaling is the rank-two Pauli noise in \eqnref{eq:rank2pauli}
\bea
\bar{\m L}^{2P} =\frac{1}{2}\,
R_{\bf \Omega}^T\begin{small} \left(\begin{array}{ccc}
\gamma_1 & &\\
& \gamma_2 &\\
&&0
\end{array}\right)
 \end{small} R_{\bf \Omega}= 
{\bf r}_1{\bf r}_1^\dag \bigoplus {\bf r}_2{\bf r}_2^\dag,\nonumber\\
\eea
with orthogonal real vecrtors ${\bf r}_1= \sqrt{\frac{\gamma_1}{2}} {\bf v}_1$ and ${\bf r}_2 =\sqrt{\frac{\gamma_2}{2}} {\bf v}_2$. The choice $h_{00}^{(2)}=0$,  
${\bf h}^{(1)} =  h_1{\bf v}_1+  h_2 {\bf v}_2$ (with $h_1, \, h_2 \in\mathbb{R}$),
and
\bea
\m {H}^{(0)} =
{\small\left(
\begin{array}{ccc} 0 & - \ii c& 0 \\
  \ii c & 0& 0\\
0&0&0
\end{array}\right)}
\eea
expressed in the basis $\{{\bf v}_1,\,{\bf v}_2, \,{\bf v}_3 = {\bf v}_1\times{\bf v}_2\}$ yields in \eqnref{beta2}
\bea
\beta^{(2)}= i\left(- \frac{\sthree}{2} +  2\, {\bm \sigma}^\dag ( h_1 {\bf r}_1 + h_2 {\bf r}_2 + c\,  {\bf r}_1 \times {\bf r}_2)\right).
\eea
As the three vectors $\{{\bf r}_1 , {\bf r}_2, {\bf r}_1 \times {\bf r}_2 \}$ form a 
basis, the expression above can always be set to zero. Hence, \eqnref{eq:CE_inf_lin} holds for 
any rank two noise process as well. 

\subsubsection{Rank-three noise}  
Finally let us consider the case of rank-three noise. In this case the matrix $\bar{M}^{(1)}$ is 
invertible such that $\beta^{(2)}$ in \eqnref{beta2} can be trivially set to 
zero by choosing  ${\bf h}^{(1)}= \bar{M}^{(1)-1} (0,0,1/4)^T$, $h_{00}^{(2)}=0$ and $\bar{H}_0=0$.

Thus, we have shown that even in the most general situation where the experimentalist has full quantum control any 
Liouvillian with the exception of the rank-one Pauli noise (not parallel to the generator of the evolution) yields a QFI that is upper 
bounded by, $\QFI \leq 4t||\alpha^{(2)}(h)||$, i.e., the standard scaling.

\subsection{Optimization of the bound for the exemplary noise-types}
\label{app:5}
We derive below for exemplary noise types the general bounds on QFI presented 
\eqnref{eq:exemplary_Fbounds} that encapsulate also the FFQC-assisted schemes
and show their optimality.

\subsubsection{Rank-one noise}

Let's go back to the general rank-one Liouvillian in \eqnref{rank-1 L}

\begin{equation}
\bar{\m L}_{\bf r}^{1\text{G}}= {\bf r} \,{\bf r}^\dag = \frac{\gamma}{2}\,\v{v}_1 \v{v}_1^\dag.
\end{equation}
We call ${\bf r}_\t{R} \!=\! \re({\bf r}) \!=\!(x_\t{R}, y_\t{R}, z_\t{R})^T$ and ${\bf r}_\t{I} \!=\! \im({\bf r}) \!=\!(x_\t{I}, y_\t{I}, z_\t{I})^T$, and choose the two other 
unit vectors $\v{v}_2$ and $\v{v}_3$ such that the triplet $\{ \v{v}_1, \v{v}_2, \v{v}_3\}$ forms an orthonormal basis. 
Expressing the component of the Hamiltonian matrix in this basis, we may restrict to writing:
\bea\nonumber
{\bf h}^{(1)} = {\small
\left(\begin{array}{c}
h_1^\t{R}+ \ii h_1^\t{I}\\
h_2^\t{R}+ \ii h_2^\t{I}\\
h_3^\t{R}+ \ii h_3^\t{I} 
\end{array}\right)
}\quad
\m {H}^{(0)} = 
{\small \left(\begin{array}{ccc}
\m H^{(0)}_{11}& \m H^{(0)}_{12}& \m H^{(0)}_{13}\\
\m H^{(0)*}_{12}\\
\m H^{(0)*}_{13}
\end{array}
\right)},\\
\eea
as the missing terms do not enter in $\alpha^{(2)}$ in \eqnref{alpha2}, and thus do not influence the final bound. 
Setting $\beta^{(2)}= 0$ in \eqnref{rank1 condition} uniquely specifies the parameters $h_1^\t{R}$, $h_1^\t{I}$ and $\m H^{(0)}_{11}$
\bea
\begin{cases}\label{rank1beta2}
h_1^\t{R} = \frac{(x_\t{I}^2 + y_\t{I}^2)z_\t{R} - (x_\t{I} x_\t{R} + y_\t{I} y_\t{R}) z_\t{I}}{4 (|{\bf r}_\t{R}|^2 |{\bf r}_\t{I}|^2 - ({\bf r}_\t{R}^\dag {\bf r}_\t{R})^2) } \\
h_1^\t{I} = \frac{(x_\t{I} x_\t{R} + y_\t{I} y_\t{R}) z_\t{R}- (x_\t{R}^2+y_\t{R}^2)z_\t{I} }{4 (|{\bf r}_\t{R}|^2 |{\bf r}_\t{I}|^2 - ({\bf r}_\t{R}^\dag {\bf r}_\t{R})^2) } \\
\m H_{11}^{(0)}=\frac{x_\t{R} y_\t{I} - x_\t{I} y_\t{R}}{4 (|{\bf r}_\t{R}|^2 |{\bf r}_\t{I}|^2 - ({\bf r}_\t{R}^\dag {\bf r}_\t{R})^2) }.
\end{cases}
\eea 
The rest of the parameters are free and have to be chosen such that $||\alpha^{(2)}||$ is minimal. To this end we rewrite \eqnref{alpha2} 
\bea
\alpha^{(2)} &=(
{\bf h}^{(1)\dag}{\bf h}^{(1)} +\frac{\gamma}{2} \v{v}_1^\dag \m {H}^{(0)}\m {H}^{(0)} \v{v}_1)\,  \sigma_0\nonumber\\
&+({\bf h}^{(1)\dag}\m {H}^{(0)}  \v{v}_1) \,{\bf r}^\dag {\bm \sigma} + ( \v{v}_1^\dag \m {H}^{(0)} {\bf h}^{(1)}) \, {\bm \sigma}^\dag {\bf r}\nonumber\\
&+ (2 \v{v}_1^\dag \m {H}^{(0)}\m {H}^{(0)} \v{v}_1 )\, {\bm \sigma}^\dag ({\bf r}_\t{R} \times {\bf r}_\t{I}).
\eea

A direct way to obtain an upper-bound is to set all the free parameters to zero, this yields 
\bea
\alpha|_0^{(2)} = \sigma_0 (|h_1^\t{R}|^2+ |h_2^\t{I}|^2+ |\v r|^2 (\m H_{11}^{(0)})^2)\qquad \\
+  \m H_{11}^{(0)} {\bm \sigma}^\dag \left( (h_1^\t{R}+\ii h_1^\t{I}) \v r +(h_1^\t{R}- \ii h_1^\t{I}) \v r^* + 2\m H_{11}^{(0)} \v r_\t{R} \times \v r_\t{I} \right) \nonumber
\eea
and 
\bea\label{eq:zero set}
||\alpha|_0^{(2)}|| 
&=& 
\frac{|x|^2 + |y|^2 + 2|x_\t{I} y_\t{R} - x_\t{R} y_\t{I}|}{16(|\v r_\t{R}|^2 |\v r_\t{I}|^2 - (\v r_\t{R}\cdot \v r_\t{I})^2 )}\nonumber\\
&=& 
\frac{\max\{ |x+\ii y|,|x-\ii y|\}^2}{16(|\v r_\t{R}|^2 |\v r_\t{I}|^2 - (\v r_\t{R}\cdot \v r_\t{I})^2 )}.
\eea
This holds as an upper bound for any $\v r$, but might not be optimal in general. Now consider the deviation from this bound in case where the free parameters are not set to zero, and denote $h_1=h_1^\t{R}+\ii h_1^\t{I}$, $\v h = (h_2^\t{R}+\ii h_2^\t{I}, h_3^\t{R}+\ii h_3^\t{I})^T$ and $\v H = (\m H_{12}^{(0)}, \m H_{13}^{(0)})^T$. One easily obtains
 \bea
||\alpha^{(2)}|| - ||\alpha|_0^{(2)}|| = |\v h|^2 + |\v H|^2 \big(|\v r|^2 + 2  |\v r_\t{R} \times \v r_\t{I}|\big)\nonumber\\
+\left|   (\v H^\dag \v h+ \m H^{(0)}_{11} h_1) \,\v r+(\v h^\dag \v H+ \m H^{(0)}_{11} h_1^*) \,\v r^* \right|\nonumber\\
-\left|   ( \m H^{(0)}_{11} h_1) \,\v r+( \m H^{(0)}_{11} h_1^*) \,\v r^* \right|.
\eea
In the case where one of the coefficients of $\v r$ is zero the only negative term in the expression above vanishes making $||\alpha|_0^{(2)}||$ the optimal bound. As follows from \eqnref{rank1beta2}, if $x=0$ (up to a rotation around $\v z$) this happens because $\m H_{11}^{(0)}=0$, while the case $z=0$ implies $h_1=0$. In general, the bound in \eqnref{eq:zero set} could be potentially improved.

\subsubsection{Pauli noise in the z-direction}
This case corresponds to the much studied case of dephasing 
noise \cite{Huelga:97, Escher:11,Kolodynski:12,Kolodynski:13, Al14}. As 
the noise commutes with the evolution our FFQC can 
not detect it and therefore our FFQC strategy cannot improve the metrological performance.
Nevertheless we treat it here for the sake of completeness.  

The Liouvillian is given by 
\bea
\bar{ \m L}_{\v z}^{1\t P} = 
{\small 
\left(\begin{array}{ccc}
0&\\
& 0\\
& & \frac{\gamma}{2}
\end{array}\right)}= \v r \, \v r^\dag,
\eea
with $\v r =(0,0,\sqrt{\gamma/2})$.
It is easy to see from \eqnref{rank1 condition} that $\beta^{(2)} =0$ implies 
$4 \sqrt{\gamma / 2} \, h_r= 1$. Whereas the norm of $\alpha^{(2)}$ in \eqnref{alpha2} is given by 
\bea
||\alpha^{(2)}|| = |\v h^{(1)}|^2 + \v r^\dag \m H^{(0)} \m H^{(0)} \v r\nonumber\\+ |\v  h^{(1)\dag} \m H^{(0)} \v r + \v r \,\m H^{(0)} \v  h^{(1)}|.
\eea
Which is minimized by 
\bea
||\alpha^{(2)}||= h_r^2 = \frac{1}{8 \gamma}
\eea for ${\bf h}^{(1)}= h_r \frac{\v r}{|\v r|}$ and $\m {H}^{(0)}=0$.

\subsubsection{Rank-two Pauli noise}
Finally, let us consider the case of rank two Liouvillian noise.  For convenience we perform a basis rotation so that the 
Liouvillian is given by X-Y Pauli noise
\bea
\bar{\m L}^{2\text{P}} =
\frac{1}{2}{\small 
\left(\begin{array}{ccc}
\gamma_1\\
& \gamma_2\\
& & 0
\end{array}\right)},
\eea  
whereas the evolution is given by $\frac{\omega}{2} (s_\theta c_\varphi,\, s_\theta s_\varphi, \, c_\theta)^T  {\bm \sigma} $. In 
the rotated basis, the relevant elements in the series expansion of the Hermitian operator $\m h$ read:
\bea
{\bf h}^{(1)} &=& {\small
\left(\begin{array}{c}
h_1^{r}+ \ii h_1^i\\
h_2^{r}+ \ii h_2^i\\
h_3^{r}+ \ii h_3^i 
\end{array}\right)},\\
\m{H}^{(0)} &=& 
{\small \left(\begin{array}{ccc}
\m H^{(0)}_{11}&\m R^{(0)}_{12}- i \m I^{(0)}_{12}& \m H^{(0)}_{13}\\ 
\m R^{(0)}_{12}+i \m I^{(0)}_{12}& \m H^{(0)}_{22}& \m H^{(0)}_{23}\\
\m H^{(0)*}_{13} &\m  H^{(0)*}_{23}&\m  H_{33}^{(0)}
\end{array}
\right)}.
\eea
The requirement that $\beta^{(2)}=0$ imposes the following conditions 
\bea
\begin{cases}
 h_1^r =-\frac{ s_\theta c_\varphi }{4  \sqrt{\gamma_1/2}} \\
 h_2^r =-\frac{s_\theta s_\varphi }{4 \sqrt{\gamma_2/2}} \\
\m I_{12}^{(0)} =-\frac{c_\theta}{2 \sqrt{\gamma_1 \gamma_2}} .
\end{cases}
\eea

For $\alpha^{(2)}$ one gets a cumbersome expression, whose exact form is not important. It is sufficient to notice that it can be 
written as
\bea\nonumber
\alpha^{(2)}= \frac{\sigma_0}{2} \Big( 2 {\bf h}^{(1)\dag} {\bf h}^{(1)} +\gamma_1(|\m H^{(0)}_{11}|^2+ \m R^{(0)2}_{12} + \m I^{(0)2}_{12} + 
|\m H^{(0)}_{13}|^2) \\ \nonumber
+\gamma_2(|\m H^{(0)}_{22}|^2+ \m R^{(0)2}_{12} + \m I^{(0)2}_{12} + |\m H^{(0)}_{23}|^2) \Big)+ f({\bf h}^{(1)}, {\m H}^{(0)}) \,{\bm\sigma}^\dag \hat {\v{n}},\\ 
\eea
where $f({\bf h}^{(1)}, \bar{H}^{(0)})=0$ when all the parameters except $h_1^r$,  $h_2^r$ and $\m  I^{(0)}_{12}$ are set to zero. 
Since this is also the choice that minimizes the pre-factor of $\sigma_0$ in the expression above, it is the optimal choice for the norm of the whole operator. This implies the optimal bound
\bea
||\alpha^{(2)}||=(h_1^r)^2+(h_2^r)^2+ \frac{\gamma_1+\gamma_2}{2} \m  I^{(0)2}_{12}\nonumber\\
=\frac{ c_\theta^2 (\gamma_1+\gamma_2) + s_\theta^2 (\gamma_1 s_\varphi^2 + \gamma_2 c_\varphi^2)}{8 \gamma_1 \gamma_2}.
\eea

\section{Analysis of the X-Y noise}
\label{app:4}
The full master equation incorporating the X-Y noise introduced in \eqnref{eq:xynoise} reads
\bea
\frac{\d \rho_\omega(t)}{\d t} &=& - \ii \frac{\omega}{2}[\sigma_3, \rho_\omega(t)]
\label{eq:XYnoise_app}
+\frac{\gamma}{2}
\left[p\, \sigma_1 \rho_\omega(t) \sigma_1 +(1-p) \sigma_2 \rho_\omega(t) \sigma_2-\rho_\omega(t)
\right]
\eea
and can be integrated analytically, e.g., by methods described in~\cite{Anderson:07},
to explicitly derive the probe dynamics.

\subsection{Simple FFQC strategy}
\label{app:XYffqc}

Let us carefully explain the sequential FFQC strategy used in \secref{sec:XYnoise} and \secref{sec:Finite_sequential}, which 
is inspired by the strategy of \secref{sec:rank1pauli} for perpendicular dephasing~\cite{Kessler:14}. 
Recall that neglecting the higher orders in $\d t$ the probe-plus-ancilla state after an evolution of 
infinitesimal duration can follow two possible branches, see \eqnref{eq:qubit_ancilla_int_dt}:
\bea
\varrho(t+\d t) = \varrho_\t{C}(t+\d t) + \varrho_\t{E}(t+\d t).
\eea
Either no error happened $\varrho_\t{C}(t+\d t) =(1-\frac{\gamma}{2} \d t) \varrho(t) -\ii\,  \frac{\omega}{2} \d t [\sigma_3\otimes\eins,\varrho(t)]$
and the parameter was imprinted on the state or with probability 
$\tfrac{\gamma}{2}\d t$ an error occurred $\varrho_\t{E}(t+\d t) = \frac{\gamma}{2} \d t \big(p\, (\sone\otimes \eins) \varrho(t) (\sone\otimes \eins) + (1-p)(\stwo\otimes \eins) \varrho(t) (\stwo\otimes \eins) \big)$. If in addition the state at time $t$ belongs to  the \emph{code space} $\varrho \in {\cB(\mathcal{H}_\t{C})}$ with $\mathcal{H}_\t{C} =\t{span}(\ket{00},\ket{11})$ and $\mathcal{H}_\t{E} = \t{span}(\ket{01},\ket{10})$, the two branches of the infinitesimal 
evolution can be distinguished in a non-demolition manner by projecting onto the code the error spaces 
via a parity measurement (since $\varrho_\t{C}(t+\d t)\in \cB(\mathcal{H}_\t{C})$ and 
$\varrho_\t{E}(t+\d t)\in \cB(\mathcal{H}_\t{E})$). If the state happens to be in the correct branch we simply leave it alone, wheres if an error is detected our strategy consists of mapping the error branch back into the code space by correcting for the most probable error term that we assume to be $\stwo\otimes \eins$ ($0\!<\!p\!\leq\!1/2$) without loss of generality. Thus, after the correction step the error branch 
reads $\bar{\varrho}_E(t+\d t) = (\stwo\otimes \eins) \varrho_\t{E}(t+\d t) (\stwo\otimes \eins)$ with 
\al{
\label{fifty}
\bar{\varrho}_\t{E}(t+\d t) & = \frac{\gamma}{2} \d t \big( (1-p)\varrho(t)
+ p (\sthree\otimes \eins) \varrho(t) (\sthree\otimes \eins) \big)+O(\d t^2). 
}
Note that here we assume that the control steps are performed fast enough, so that \eqnref{fifty} holds up to first order in $\d t$\footnote{Note that the control steps need not be ultra fast, just sufficiently fast so that the first order approximation holds.  The duration of these control-pulses depends on the strength of the decoherence process.}.
Importantly, after the correction the final state belongs to the code space again $\bar{\varrho}(t+\d t) \in \cB(\mathcal{H}_\t{C})$, 
and we can continue running the strategy. Another important point is that we keep track of the number of errors that have been detected before a time $t$ in a classical error register. This can be made explicit by introducing an error register mode R initialized in the state $\ket{0}_\t{R}$ for each probe qubit, and a Hermitian operator that increases its count by one $R_+\ket{m}_\t{R}=\ket{m+1}_\t{R}$. For example the register can be modelled by a bosonic mode $a$ with $R_+= \frac{1}{a^\dag a} a^\dag$. Such a FFQC strategy leads to an effective evolution of the probe+register system (the ancilla does not play a role at this stage) given by the master equation 
\be
\frac{\d}{\d t}\varrho_\omega(t) = -\ii \frac{\omega}{2}[\sigma_3\otimes\eins_\t{R},\varrho_\omega(t) ] +\frac{\gamma}{2}\left(-\varrho_\omega(t) +\frac{\gamma}{2}(\eins \otimes R_+)\big( (1-p)\varrho_\omega(t)
+ p (\sthree\otimes \eins_\t{R}) \varrho_\omega(t) (\sthree\otimes \eins_\t{R})\big)(\eins \otimes R_+)\right).
\ee

For a single probe initialized in a state $\ket{+}=\frac{1}{\sqrt{2}}\binom{1}{1}$ (or $\frac{\ket{00}+\ket{11}}{\sqrt{2}}$ if we remember the ancillary qubit) the final state resulting from such an effective evolution after time $t$ reads
\be
\varrho_\omega(t)= \sum_{m=0}^\infty p(m;t)\,\bar \varrho_m(t)  \otimes \proj{m}_\t{R},
\ee
where the total number of errors $m$  detected during time $t$ 
follows the Poissonian distribution $p(m;t)\!=\!\e^{-\gamma t/2}(\gamma t/2)^m/m!$, and the state of the probe conditional on the occurrence of $m$ errors is
\be
\bar \varrho_m(t) =\frac{1}{2}\left(\begin{array}{cc}1 & (1-2p)^m \e^{\ii \omega t}\\ (1-2p)^m \e^{-\ii \omega t}&1\end{array}
\right).
\label{eq:effective stateXY}
\ee

\subsection{Description of the probe channel}
In particular, one may derive the \emph{Choi-Jamiolkowski (CJ) matrix}
\cite{Bengtsson2006} representing the probe map $\cE_{\omega,t}^\t{X-Y}$,
which is defined then as $\m{P}(\omega,t)\!:=\!\cE_{\omega,t}^\t{X-Y}\otimes\cI[I]$
with $\ket{I}\!=\!\ket{00}+\ket{11}$ and reads 
\be
\m{P}(\omega,t)=2\left(
\begin{array}{cccc}
\mathrm{C}_{\frac{\gamma}{2}} & 0 & 0 & \mathrm{C}_{\Omega}-\ii\tilde{\omega}\mathrm{S}_{\Omega}\\
0 & \mathrm{S}_{\frac{\gamma}{2}} & -\tilde{\gamma}\mathrm{S}_{\Omega} & 0\\
0 & -\tilde{\gamma}\mathrm{S}_{\Omega} & \mathrm{S}_{\frac{\gamma}{2}} & 0\\
\mathrm{C}_{\Omega}+\ii\tilde{\omega}\mathrm{S}_{\Omega} & 0 & 0 & \mathrm{C}_{\frac{\gamma}{2}}
\end{array}
\right),
\label{eq:CJmat}
\ee
where
\bea
\mathrm{C}_{x} := \frac{1}{2}\e^{-\frac{\gamma t}{2}}\cosh(xt),
&\quad&
\mathrm{S}_{x} :=  \frac{1}{2}\e^{-\frac{\gamma t}{2}}\sinh(xt),
\nonumber\\
\mu:=\frac{\gamma(1-2p)}{2},
&\quad&
\Omega:=\sqrt{\mu^2-\omega^2},
\nonumber\\
\tilde{\gamma} := \frac{\mu}{\Omega},
&\quad&
\tilde{\omega} := \frac{\omega}{\Omega}.
\eea
For $t\!>\!0$ the CJ-matrix \eref{eq:CJmat} is of rank four,
so the probe channel, $\cE_{\omega,t}^\t{X-Y}$, is 
\emph{full-rank} always possessing four Kraus operators in \eqnref{eq:Krep}. 
By performing the spectral decomposition of the CJ-matrix,
$\m{P}(\omega,t)\!=\!\sum_{i=1}^4\lambda_{i}\proj{e_{i}}$,
we obtain the \emph{CJ-canonical Kraus representation}
of channel $\cE_{\omega,t}^\t{X-Y}$ after identifying
$\sqrt{\lambda_{i}}\ket{e_{i}}\!=\!K_i\otimes\eins\ket{I}$
for all $i$, where
\bea
&
K_{1}=\sqrt{\mathrm{S}_{\frac{\gamma}{2}}-\tilde{\gamma}\mathrm{S}_{\Omega}}\,\sigma_{1},
\; 
K_{2}=-\ii\sqrt{\mathrm{S}_{\frac{\gamma}{2}}+\tilde{\gamma}\mathrm{S}_{\Omega}}\,\sigma_{2}, 
&
\label{eq:Krep_XY}
\\
 & K_{3}=\sqrt{\mathrm{C}_{\frac{\gamma}{2}}-\sqrt{\frac{\e^{-\gamma t}}{4}+\left(\tilde{\gamma}\mathrm{S_{\Omega}}\right)^{2}}}
\left(
\begin{array}{cc}
-\frac{\sqrt{\frac{\e^{-\gamma t}}{4}+\left(\tilde{\gamma}\mathrm{S_{\Omega}}\right)^{2}}}{\mathrm{C}_{\Omega}+\ii\tilde{\omega}\mathrm{S}_{\Omega}} & 0\\
0 & 1
\end{array}
\right),
\nonumber \\
 & K_{4}=\sqrt{\mathrm{C}_{\frac{\gamma}{2}}+\sqrt{\frac{\e^{-\gamma t}}{4}+\left(\tilde{\gamma}\mathrm{S_{\Omega}}\right)^{2}}}\left(\begin{array}{cc}
\frac{\sqrt{\frac{\e^{-\gamma t}}{4}+\left(\tilde{\gamma}\mathrm{S_{\Omega}}\right)^{2}}}{\mathrm{C}_{\Omega}+\ii\tilde{\omega}\mathrm{S}_{\Omega}} & 0\\
0 & 1
\end{array}\right).
\nonumber
\eea

On the other hand, we may rewrite the action of 
the probe channel, $\cE_{\omega,t}^\t{X-Y}$, in the Pauli basis 
and thus derive its corresponding \emph{dynamical matrix} 
$\m{S}$ defined in \eqnref{eq:Srep}:
\be
\m{S}(\omega,t)
=
\left(
\begin{array}{cccc}
\mathrm{C}_{\frac{\gamma}{2}}+\mathrm{C}_{\Omega} & 0 & 0 & \ii\tilde{\omega}\mathrm{S}_{\Omega}\\
0 & \mathrm{S}_{\frac{\gamma}{2}}-\tilde{\gamma}\mathrm{S}_{\Omega} & 0 & 0\\
0 & 0 & \mathrm{S}_{\frac{\gamma}{2}}+\tilde{\gamma}\mathrm{S}_{\Omega} & 0\\
-\ii\tilde{\omega}\mathrm{S}_{\Omega} & 0 & 0 & \mathrm{C}_{\frac{\gamma}{2}}-\mathrm{C}_{\Omega}
\end{array}
\right).
\label{eq:SmatXY}
\ee

In the special case of \emph{balanced X-Y noise}, for which $p\!=\!1/2$,
all the above expressions dramatically simplify ($\mu\!=\!\tilde\gamma\!=\!0$, $\Omega\!=\!\ii\omega$, $\tilde\omega\!=\!-\ii$), 
as the Liouvillian part of \eqnref{eq:XYnoise_app} yields a noisy channel that commutes with parameter encoding, 
or in other words, the channel $\cE_{\omega,t}^\t{X-Y}$ becomes then
\emph{phase-coviariant} with respect to rotations generated by $\sigma_3$ \cite{Smirne2015}.

{
\subsection{Parallel CE bound}
We follow the methods of \refcite{Kolodynski:13} summarised in \secref{sec:bound_opt},
in order to compute the parallel CE bound \eref{eq:CE_par} for the X-Y noise 
and the scenario depicted \figref{fig:XY_schemes}(b):
\be
\QFI^{\t{(b)}\uparrow}_\t{X-Y}(N,t)
:=
4\,\min_{\m{h}(t)}\left\{N || \alpha(t) || + N(N-1)  || \beta(t)||^2 \right\},
\label{eq:CE_boundXY}
\ee
which then allows to determine an upper bound on the maximal QFI rate
per probe, defined in \eqnref{eq:freq_par_CRB}:
\be
\mathfrak{f}_N^{\t{(b)}\uparrow}
:= 
\max_t\,
\frac{\QFI^{\t{(b)}\uparrow}_\t{X-Y}(N,t)}{N\,t}.
\label{eq:QFIrateXY}
\ee
For given $\gamma$, $\omega$, $N$ and $t$, we substitute into \eqnref{eq:CE_boundXY} 
the canonical Kraus operators of \eqnref{eq:Krep_XY}, $\v{K}(t)\!=\!(K_1,\dots,K_4)^T$, 
and their shifted derivatives, $\dot{\v{K}}(t)-\ii\m{h}(t)\v{K}(t)$, so that the minimisation 
can be efficiently performed by means of semi-definite programming (SDP) \cite{Kolodynski:13}.
In order to compute $\mathfrak{f}_N^{\t{(b)}\uparrow}$, we repeat such procedure for various $t$,
while numerically optimising \eqnref{eq:QFIrateXY} over $t$ that then yields $t_\t{opt}^{\t{(b)}\uparrow}$. 

However, in the special single-qubit ($N\!=\!1$) and asymptotic cases ($N\!\to\!\infty$),
we are able to determine the analytic expressions for $\QFI^\t{(b)}_\uparrow(t)$ in 
\eqnref{eq:CE_boundXY} after correctly choosing an analytic ansatz form (motivated by the 
numerical SDPs) for the Hermitian matrix $\m{h}(t)$ and explicitly performing the 
minimisation in \eqnref{eq:CE_boundXY}. Although such expressions cannot 
be claimed to be derived in a fully analytic manner, their correctness can be 
verified numerically to arbitrary precision thanks to the SDP formulation.

Nevertheless, due to their cumbersome form we present below only their
simplified versions for the special case of balanced X-Y noise.
On the other hand, in case of the asymptotic ($N\!\to\!\infty$) regime,
we provide a compact derivation of slightly loosened CE bound \eref{eq:CE_boundXY},
which possesses much simpler form but nonetheless approximates \eqnref{eq:CE_boundXY}
to negligible precision, what we have numerically verified.
 
\subsection{Parallel CE bound for $p\!=\!1/2$}
In the case of balanced X-Y noise we explicitly compute the CE bound \eref{eq:CE_boundXY} 
thanks to the recently derived analytic generalisation of the SDP-formulation of 
\eqnref{eq:CE_boundXY} for \emph{phase-covariant}, \emph{unital} qubit maps \cite{Smirne2015}:
\be
\left.\QFI^{\t{(b)}\uparrow}_\t{X-Y}(N,t)\right|_{p=\frac{1}{2}}
=
\frac{N^2 t^{2}}{1+2N\,\e^{\gamma t}S_{\frac{\gamma}{2}}},
\label{eq:CEboundXYp05}
\ee
which is obtained after choosing as the channel Kraus representation
the canonical one of \eqnref{eq:Krep_XY}, and optimally setting in \eqnref{eq:CE_boundXY}:
\be
\left.\m{h}_\t{opt}(t)\right|_{p=\frac{1}{2}}
=
\frac{t}{2\kappa}
\left(
\begin{array}{cccc}
0 & 1-N & 0 & 0\\
1-N & 0 & 0 & 0\\
0 & 0 & -\kappa & \kappa-1\\
0 & 0 & \kappa-1 & -\kappa
\end{array}
\right)
\ee
with $\kappa\!=\!1+2N\,\e^{\gamma t}S_{\frac{\gamma}{2}}$.

Thanks to analytic form of \eqnref{eq:CEboundXYp05},
we explicitly compute the corresponding upper bound on maximal QFI rate
\eref{eq:QFIrateXY}:
\be
\left.\mathfrak{f}^{\t{(b)}\uparrow}_N\right|_{p=\frac{1}{2}}
= 
\frac{2}{\gamma}\;\frac{N\left(1+W\left(\frac{2-N}{\e N}\right)\right)}{2-\left(1-\e^{1+W\left(\frac{2-N}{\e N}\right)}\right)N}
\label{eq:QFIrateXYoptp05}
\ee
which occurs at $t_\t{opt}^{\t{(b)}\uparrow}(N)\!=\!\frac{1+W\!\left(\frac{2-N}{\e N}\right)}{\gamma}$
with $W(x)$ representing the Lambert function.

Note that for $N\!=\!2$, $\left.\mathfrak{f}^{\t{(b)}\uparrow}_{N=2}\right|_{p=\frac{1}{2}}\!\!=\!2/(\gamma \e)$,
so that the bound \eref{eq:QFIrateXYoptp05}
coincides with the QFI rate attained 
by our proposed sequential strategy of \figref{fig:XY_schemes}(a2), i.e., $\mathfrak{f}^\t{(s)}$ in \eqnref{eq:f_seq}. 
As for any smaller $p$ (more noise
asymmetry), the sequential strategy can perform only better in comparison to the 
parallel schemes (b) of \figref{fig:hierarchy}---see also \figref{fig:Nth}---this proves that 
for any $p$ there always exist a non-trivial range of $N\!>\!1$
for which the sequential protocol of \figref{fig:XY_schemes}(a2) 
outperforms all the parallel schemes.
}

\subsection{Asymptotic parallel CE bound}
In the asymptotic regime of $N\!\to\!\infty$
and fixed $t$, it is always optimal to set $\beta(t)\!=\!0$
in \eqnref{eq:CE_boundXY} if possible, so that the parallel CE bound 
\eref{eq:CE_boundXY} then reads
\be
\QFI^{\t{(b)}\uparrow}_\t{X-Y}(N\!\to\!\infty,t)
=
4N\,\min_{\m{h}(t)\,\t{s.t.}\,\beta(t)=0} || \alpha(t) ||.
\label{eq:CE_boundXYas}
\ee

We determine an upper-bound on the r.h.s.~of 
\eqnref{eq:CE_boundXYas} making use of the $\m{M}$-matrix
formulation introduced in \secref{sec:bound_opt} and the alternative 
expressions for $\alpha$ and $\beta$ of \eqnsref{eq:alphaM}{eq:betaM}.

Basing on the dynamical matrix $\m S$ stated \eqnref{eq:SmatXY},
we derive the $\m M$ matrix following \eqnref{eq:conditionsM}
and choosing $\m M\!=\!\m M^\dag\!=\!\sqrt{\m S}^T$:
\be
\m M(\omega,t)=\left(\begin{array}{cccc}
\varepsilon+\zeta&0&0 & - \ii \chi\\
0& \Delta&0&0\\
0&0&\Gamma&0\\
\ii \chi&0 &0& \varepsilon-\zeta
\end{array}\right),
\ee
with 
\bea
&\varepsilon^2 = \frac{\Theta_+}{2}, 
\qquad 
\zeta^2 = \frac{{C_\Omega}^2}{2\Theta_-},
\qquad 
\chi^2 = \frac{{\tilde{\omega}\mathrm{S}_{\Omega}}^2}{2\Theta_-},\\
&\Delta^2= S_\frac{\gamma}{2} -  \tilde{\gamma}\mathrm{S}_{\Omega},
\qquad \Gamma= S_\frac{\gamma}{2}+  \tilde{\gamma}\mathrm{S}_{\Omega},&
\eea
and $\Theta_\pm\!=\!{C_\frac{\gamma}{2}}\pm\sqrt{{C_\frac{\gamma}{2}}^2-{C_\Omega}^2-{\tilde{\omega}\mathrm{S}_{\Omega}}^2}$.

We inspect \eqnref{eq:betaM} and, in order to satisfy the asymptotic condition
$\beta(t)\!=\!0$, we set following an educated guess:
\be
\m h 
\;=\; 
\frac{\dot \zeta \chi-\zeta \dot \chi}{\Delta\,\Gamma}
\left(\begin{array}{ccc}
0 & 0\;0 & 0\\
\begin{array}{c}
0\\
0
\end{array} & \sigma_{2} & \begin{array}{c}
0\\
0
\end{array}\\
0 & 0\;0 & 0
\end{array}\right),
\ee
which then after evaluating  $\alpha_{\omega, \gamma, p}^\t{(b)}(t) \!:=\!||\alpha(t)||$ according to \eqnref{eq:alphaM} 
yields an upper bound on \eqnref{eq:CE_boundXYas}:
\be
\QFI^{\t{(b)}\uparrow}_\t{X-Y}(N\!\to\!\infty,t)
\le
4\,N\,\alpha_{\omega, \gamma, p}^\t{(b)}(t),
\ee
where
\bea
\alpha_{\omega, \gamma, p}^\t{(b)}(t)&=&
2 (\dot{\varepsilon}^2+\dot{\zeta }^2+\dot{\chi }^2)+\nonumber\\
&&+\dot{\Delta }^2+\dot{\Gamma }^2+\frac{(\Delta^2+\Gamma^2)(\zeta \dot{\chi }-\dot{\zeta } \chi)^2 }{\Delta^2\Gamma^2}
\eea
is of complex but importantly of fully analytic form.

\section{Ziv-Zakai precision bound and the QFI}
\label{app:6}
Consider the parameter $\omega$ with prior distribution $p_0(\omega)$. The state  at time $t$ is denoted 
as $\rho_{\omega,t}$.
The Ziv-Zakai bound on the \aMSE of the parameter reads \cite{Tsang:12}
\be
\langle\delta \omega^2\rangle \geq \int_0^\infty \!\!\!\!d\tau \tau \int_{-\infty}^\infty \!\!\!\!d\omega \min[p_0(\omega), p_0(\omega+\tau)] \text{Pr}_e(\omega, \omega +\tau),\nonumber\\
\ee
where $ \text{Pr}_e(\omega, \omega +\tau)$ is minimum error probability of the binary hypothesis testing between $\omega$ and $\omega+\tau$. Quantum mechanically it is given by the trace distance \cite{Tsang:12}
\bea
 \text{Pr}_e(\omega, \omega +\tau)\geq \frac{1}{2}(1-D[\rho_{\omega,t}, \rho_{\omega+\tau,t}]) \nonumber\\:=\frac{1}{2}(1-\frac{1}{2}||\rho_{\omega,t}-\rho_{\omega+\tau,t}||_1).
\eea

We use the triangle inequality to decompose the trace distance in elementary steps
\be
D[\rho_\omega^T, \rho_{\omega+\tau}^T]\leq \min\Big(1, \int_{\omega}^{\omega+\tau}\!\!\!\!  D[\rho_{\tau',t}, \rho_{\tau'+d\tau',t}] d\tau' \Big),
\ee
and the local expansion of the trace distance $ D[\rho_{\tau',t}, \rho_{\tau'+d\tau',t}]\leq d\tau' \frac{\sqrt{\QFI(\rho_{\tau',t})}}{2}$ we rewrite the error probability as
\be
\text{Pr}_e(\omega, \omega +\tau) \geq \frac{1}{2}(1- \min\!\Big(1, \int_{\omega}^{\omega+\tau}\!\!\!\! d\tau' \frac{\sqrt{\QFI(\rho_{\tau',t})}}{2}\Big)
\ee
  Now we are in position to plug in the bound for the QFI, as we showed for the SQL Liouvillians the QFI of the state at time $t$ satisfies $\QFI_\mathcal{L}\leq 4 \alpha_\mathcal{L}t$, such that the error probability fulfills
\be
\text{Pr}_e(\omega, \omega +\tau) \geq \frac{1}{2} \max (0, 1- \tau \sqrt{\alpha_\mathcal{L} t }).
\ee
And the Ziv-Zakai bound reads
\be
\langle\delta \omega^2\rangle \geq \frac{1}{2} \int_0^\frac{1}{\sqrt{\alpha_\mathcal{L}t}} \!\!\!\!d\tau \tau (1- \tau \sqrt{\alpha_\mathcal{L} t } )\nonumber \\ \int_{-\infty}^\infty \!\!\!\!d\omega  \min[p_0(\omega), p_0(\omega+\tau)]
\ee
Which in the case $t\gg 1$ leads to 
\be
\langle\delta \omega^2\rangle \geq \frac{1}{12 \alpha_\mathcal{L} t},
\ee
for any prior which respects the regularity condition $\lim_{\tau\to 0}\int_{-\infty}^\infty \d\omega\;  \min[p_0(\omega), p_0(\omega+\tau)] \to 1$.

\section{Strategy for noiseless frequency estimation}
 \label{app:7}

In the context of phase alignment it is known \cite{Berry00} that the state minimizing the cost-function $\sin^2(\frac{\varphi}{2})$
 is given by
\bea
\label{eq:Berry-Wisemann}
\ket{\Psi}=\sqrt{\frac{2}{M+2}}\sum_{n=0}^{N} \sin(\frac{\pi(n+1)}{N+2})\ket{n},
\eea 
where $\ket{n}$ is the symmetric eigenstate of the total spin $J_z =\frac{1}{2}\sum_{k=1}^N \sigma_3^{(k)}$ corresponding to the eigenvalue $\frac{2n-N}{2}$. We call the rotated state  $\ket{\Psi_\varphi}=e^{\ii \varphi J_z}\ket{\Psi}$. For the measurement given by the covariant POVM $\{\ketbra{\theta}{\theta}\}$ with $\ket{\theta}=(N+1)^{-1/2}\sum e^{\ii \theta n}\ket{n}$ and the state $\ket{\Psi_\varphi}$ we denote the probability to observe an outcome $\theta$ by 
\bea\label{eq:prob_def}
p(\theta-\varphi) = p(\theta|\varphi):= |\bracket{\theta}{\Psi_\varphi}|^2.
\eea 
It has been shown in \cite{Berry00} that the probability $p(\delta)$ in \eqnref{eq:prob_def} yields a cost
\bea
C= \int_{-\pi}^{\pi} \sin^2(\frac{\delta}{2}) p(\delta) d\delta = \frac{\pi^2}{4 N^2}
\label{eq:C_cost}
\eea
in the regime of large $N$. Note that the cost-function $\sin^2(\frac{\delta}{2})$ is appropriate in the context of phase estimation where the estimated parameter is defined a circle $\varphi \in [-\pi, \pi)$. But in the context of frequency estimation the parameter
of interest is defined on the whole line $\omega \in (-\infty, \infty)$ (similar for the resulting phase $\varphi =\omega t$), and the appropriate cost-function is the variance (MSE). We show below how  it can be lower bounded from the value of $C$.

Using the fact that $\delta^2 \leq  \sin^2( \frac{\delta}{2} ) \pi^2 $ on the interval $[-\pi,\pi]$ we get a bound on the second moment.
\be
V=\int_{-\pi}^{\pi} \delta^2 p(\delta) d\delta \leq\pi^2 C = \frac{\pi^4}{4 N^2}.
\ee
On the  other hand, one can also upper-bound the tail probabilities from the cost $C$
\be
T(r) = \int_{r \pi}^\pi p(\delta) d\delta \leq \frac{C}{2 \sin^2(r \pi/2)}
\ee
 
 The average mean square error over all possible outcomes of the covariant POVM is given by
 \bea
 \langle\delta \varphi^2\rangle = \int_{-\pi}^{\pi} d\theta \,p(\theta)\int d\varphi (\varphi - \tilde \varphi_\theta)^2 p(\varphi|\theta)\\
 =  \int_{-\pi}^{\pi} d\theta \,\int d\varphi (\varphi - \tilde \varphi_\theta)^2 p(\theta-\varphi) p_0(\varphi),
 \eea
where $p_0(\varphi)$ is the prior knowledge and $\tilde \varphi_\theta$ is the optimal estimator of the phase given the measurement outcome $\theta$. Choosing an explicit estimator $\tilde \varphi_\theta=\theta$ (which might not be optimal for the MSE) and assume
\bea\label{eq:prior}
p_0(\varphi) = \begin{cases}
\frac{1}{2 \pi r} \quad &|\varphi|\leq r \pi\\
 0 \quad &|\varphi| > r \pi.
\end{cases} 
\eea
This implies
\bea
\langle \delta \varphi^2 \rangle =  \int_{-\pi}^{\pi} d\theta \,\int d\varphi (\varphi - \theta)^2 p(\theta-\varphi) p_0(\varphi)\\
= \frac{1}{2\pi r}\int_{-\pi}^{\pi} d\theta \,\int_{-r \pi}^{r \pi} d\varphi (\varphi - \theta)^2 p(\theta-\varphi).
\eea
Now we can change the integration variables to $\Sigma= \varphi+\theta$ and $\Delta= \varphi-\theta$. The domain of integration then is a rotated rectangle, but since the integrand is positive we can enlarge the domain to the minimal square that contains it without lowering the integral. This implies
\bea
\langle  \delta \varphi^2 \rangle &\leq  \frac{1}{2\pi r} \int_{-(1+r)\pi}^{(1+r)\pi} \int_{-(1+r)\pi}^{(1+r)\pi} \frac{d\Sigma d\Delta}{2}   \Delta^2 p(\Delta)\nonumber\\
 &= \frac{r+1}{2 r}\int_{-(1+r)\pi}^{(1+r)\pi}   \Delta^2 p(\Delta)  d\Delta \nonumber\\ &= \frac{r+1}{2 r}\left( V
 + 2 \int_{\pi}^{\pi(1+r)} \Delta^2 p(\Delta)  d\Delta \right). 
\eea
Finally, because $p(\Delta)$ is periodic, the second term in brackets in the last expression is upper-bounded by 
 \al{
\int_{\pi}^{\pi(1+r)} \Delta^2 p(\Delta)  d\Delta & \leq   \pi^2 (1+r)^2 \int_{(1-r)\pi}^{\pi}  p(\Delta)  \,\d \Delta\; =  \pi^2(1+r)^2 \,T(1-r)\\
& \leq C\,\frac{ \pi^2(1+r)^2 }{2 \sin^2((1-r)\pi/2)},
 }
which yields
\be
\langle \delta \varphi^2 \rangle \leq C \,\frac{(r+1)\pi^2}{2 r} \left(1+  \frac{(1+r)^2}{\sin^2((1-r )\pi/2)} \right).
\ee
Now, minimizing the above bound over $r$ and substituting for the cost C according to \eqnref{eq:C_cost}, we obtain
\be
\langle \delta \varphi^2 \rangle \leq  C\,\kappa\,\pi^2 \leq  \frac{\pi^4 \kappa}{4}\frac{1}{N^2},
\ee
which is attained for $r_\t{opt}\approx 0.33$ that gives $\kappa\approx 6.74$. 

It remains to show how the above bound applies 
within the frequency estimation scenario. Let us assume the prior distribution of the frequency parameter to also have an interval form:
\be
p_0(\omega)=
\begin{cases}
\frac{1}{2\omega_0} \quad &\text{if} \quad |\omega|\leq \omega_0\\
0   \quad &\text{if} \quad |\omega| >\omega_0.
\end{cases}
\ee
Remark that if the interval is not centred at zero, it can be always shifted be means of FFQC. Given a total resource $t'=N\,t$, we can construct a circuit that is equivalent to running $N= \frac{t' \omega_0}{\pi r_\t{opt}}$ parallel qubits prepared in the state
\eref{eq:Berry-Wisemann} for a time $t = \frac{r_\t{opt} \pi}{\omega_0}$. This choice maps the frequency estimation task on the estimation of the phase $\varphi = \omega t=\omega \frac{r_\t{opt} \pi}{\omega_0}$ for the prior defined in \eqnref{eq:prior} with $r=r_\t{opt}$, which we have solved above. Consequently, the MSE for the frequency reads
\be
\langle \delta \omega^2 \rangle= \frac{\langle\delta\varphi^2\rangle}{t^2} \leq \frac{\pi^4 \kappa}{4}\frac{1}{t^2\, N^2} = \frac{\pi^4 \kappa}{4}\frac{1}{(t')^2}.
\ee
However, in contract to the case of phase estimation, note that if the evolution is slowed down---as it is for the rank-one Pauli noise---the time $t$ (and, hence, $t'$) has to be 
set longer, what affects the frequency MSE above.

\end{document}